\documentclass{aa}
\usepackage{graphicx}
\usepackage{txfonts}
\usepackage{natbib}
\usepackage[caption=false]{subfig}
\usepackage[linktocpage]{hyperref}
\usepackage{xcolor}
\usepackage{threeparttable}
\usepackage{etoolbox}
\makeatletter
\patchcmd\@combinedblfloats{\box\@outputbox}{\unvbox\@outputbox}{}{%
   \errmessage{\noexpand\@combinedblfloats could not be patched}%
}%
 \makeatother

\begin{document}

   \title{Assembly of spheroid-dominated galaxies in the EAGLE simulation}

    \author{M. S. Rosito \inst{1}, P. B. Tissera \inst{2,3}, S. E. Pedrosa \inst{1}, Y. Rosas-Guevara \inst{2,4,5}} 

   \institute{Instituto de Astronom\'{\i}a y F\'{\i}sica del Espacio, 
CONICET-UBA, Casilla de Correos 67, Suc. 28, 1428, Buenos Aires, Argentina.
         \and
             Departamento de Ciencias F\'{\i}sicas, Universidad Andr\'{e}s 
Bello, 700 Fernandez Concha, Santiago, Chile. 
\and Corresponding Investigator, IATE-CONICET, Laprida 927, C\'ordoba, Argentina.
\and  Centro de Estudios de F\'isica  del Cosmo de Arag\'on, Plaza San
Juan 1, Planta 2, E-44001 Teruel, Spain.
\and Donostia International Physics Center (DIPC), Manuel Lardizabal pasealekua 4, 20018 Donostia, Basque Country, Spain.\\
	}


\abstract
   {Despite the insights gained in the last few years, our knowledge about the formation and evolution scenario for the spheroid-dominated galaxies is still incomplete.
   New and more powerful cosmological simulations have been developed that together with more precise observations open the possibility of more detailed study of the formation of early-type galaxies (ETGs). }
   {The aim of this work is to analyse the assembly histories of ETGs in a $\Lambda$-CDM cosmology,  focussing on the archeological approach given by the mass-growth histories.
  }
   {We inspected a sample of dispersion-dominated galaxies selected from the largest volume simulation of the  EAGLE project. This simulation includes a variety of physical processes such as radiative cooling, star formation (SF), metal enrichment, and stellar and active galactic nucleus (AGN) feedback.
   The selected sample comprised 508  spheroid-dominated galaxies classified according to their dynamical properties.
   Their surface brightness profile, the fundamental relations, kinematic properties, and stellar-mass growth histories are estimated and analysed.
   The findings are confronted with recent observations.}
   { The simulated ETGs are found to globally reproduce the fundamental relations of ellipticals. All of them have an inner disc component where residual younger stellar populations (SPs) are detected.
   A fraction of this inner-disc correlates with bulge-to-total ratio.
   We find a  relation between kinematics and shape that implies that dispersion-dominated galaxies with low $V/\sigma_L$ (where $V$ is the average rotational velocity and $\sigma_L$ the one dimensional velocity dispersion) tend to have ellipticity smaller than $\sim 0.5$ and are dominated by old stars.
  On average, less massive galaxies  host  slightly younger stars. More massive spheroids show coeval SPs while for less massive galaxies (stellar masses lower than $\sim 10^{10}{\rm M_{\odot}}$), there is a clear trend to have rejuvenated inner regions, showing an age gap between the inner and the outer regions up to $\sim 2$ Gyr, in apparent contradiction with observational findings. 
We find evidences suggesting that both the existence of the disc components with SF activity in the inner region and the accretion of satellite galaxies in outer regions could contribute to the outside-in formation history in galaxies with low stellar mass.
On the other hand, there are non-negligible uncertainties in the determination of the ages of old stars in observed galaxies.
Stronger supernova (SN) feedback and/or the action of AGN feedback for galaxies with stellar masses lower than $10^{10}{\rm M_{\odot}}$ could  contribute to prevent the SF in the inner regions.
}
   {}


  \authorrunning{M. S. Rosito et al.}
  \maketitle

\section{Introduction}

Once thought as simple systems formed in a monolithic collapse, early-type galaxies (ETGs)
are currently considered complex structures likely formed by the combination of diverse physical processes such as infall and collapse, major and/or minor mergers \citep[e.g.][and references therein]{Burkert2004, Bournaud2007, Gonzalez2009, Zavala+2012, Tissera2012form,perez2013,AvilaResse2014, Dubois2016, RG2016}.
The development of integral field spectroscopy (IFS) has allowed a deeper analysis of the SPs and interstellar medium (ISM) of galaxies and particularly of ETGs. 
Recent multi-object surveys such as SAMI \citep{SAMI2012} and MaNGA \citep{Bundy2015, SDSS+2016}, and single-object ones like  ALTAS$^{3\mathrm{D}}$ \citep{AtlasI} and CALIFA \citep{CALIFA2012}
 provide more detailed information that improve our understanding of  the astrophysical properties and fundamental relations of galaxies in a wide range of stellar masses and morphologies.

The construction of stellar-mass growth histories (MGHs) for a significant number of galaxies has been made possible by  IFS surveys. The MGHs provide archaeological estimations of  galaxy assembly.  
Recent observational works  find  a significant fraction of galaxies in the Local Universe to be consistent with  an inside-out formation history as they exhibit negative age gradients \citep[][]{Wang2011, Lin2013, Li2015}.
This implies  that the star formation (SF) in the inner regions occurs at  earlier times than in outer regions of the galaxies. 
For the ETGs, it is not yet clear how they assembled and/or quenched their SF activity as a function of radius. Supernova (SN) and active galactic nucleus (AGN) feedback can contribute to
regulate the SF activity in galaxies of different masses. Recent results by \citet{Argudo2018} suggest  that AGN feedback might even act  in galaxies with stellar masses down to
$10^{10}$ M$_{\odot}$. Further, there is new evidence that suggests the action of AGN feedback in dwarf galaxies \citep{Manzano2019}.
 
Evidence for both 
 inside-out and outside-in scenarios for ETGs has been observationally reported \citep[][]{SB2007}.
\citet{IbarraMedel2016} study galaxies from the MaNGA survey \citep{Bundy2015, SDSS+2016} 
finding an inside-out scenario  in star-forming and late-type galaxies.
In less massive systems, there is a larger variety of behaviours in the observed MGHs. Early-type galaxies  are detected as being consistent with a weak inside-out formation at later epochs.
At early epochs, a slight trend from outside-in  formation is found. However, the main caveat to studying ETG assembly is the determination of ages for SPs
 older than $\sim 10$ Gyr. 
Another effect to be considered in the formation of ETGs is the contribution of old stars acquired by satellite accretion \citep[e.g.][]{Genel2018}. The accretion of
satellites could add older stars to the outer regions helping to establish an outside-in formation scenario. However, if the accreted SPs are younger than the main SPs of the principal galaxy then the opposite scenario could be set. From an observational point of view, this is yet not clearly established. 

The current cosmological scenario for the formation of the structure  shows that  different morphologies can be  associated with a variety of formation histories at a given stellar mass
\citep{derossi2015,Trayford2019}.
In particular, mergers have been shown to be able to change galaxy morphology drastically \citep[e.g.][]{mk1996}. 
During the early stages of the interactions, these processes may  trigger tidal fields that drive gas inflows, producing starbursts that can feed the spheroidal component \citep{Hernquist1989, sillero2017}. These processes can modify the metallicity distributions, contributing to weaken the metallicity gradients as well as triggering SF \citep[e.g.][]{rupke2010,perez2013,tissera2016,taylor2017,bustamente2018}. Galactic winds might be generated by stellar or AGN feedback that could transport material out of the galaxy, modifying the morphologies, the
SF activity, and the chemical abundances \citep[e.g.][]{gibson2013,Genel2015, Dubois2016,tissera2019}.
\cite{Naab2013} proposes a  two-phase assembly history: first the action of dissipative processes that triggered in situ SF at high redshift and second,  the action of dry accretion of  nearby galaxies.
Dry mergers have also been proposed as a mechanism responsible for the formation of ETGs, principally of the outskirts, because they contribute with old stars and small amount of gas \citep{Genel2018}. While SN feedback is a crucial process to regulating the transformation of gas into stars in galaxies, AGN feedback is also
required at high stellar masses where SN feedback is not that efficient. 

By analysing data from the Illustris Project \citep{Illustris2014a, Illustris2014b, Genel2014}, \cite{RG2016} quantify the fraction of ex-situ mass coming from major or minor mergers, finding that half of the ex-situ mass comes, on average, from major mergers.
For galaxies with stellar masses $\sim 10^{10}-10^{11}$ M$_{\odot}$, the ratio between the accreted mass over the total mass shows a non-negligible scatter.
They conclude that features such as morphology and halo formation time, together with the merger histories affect the fraction of ex-situ mass 
(e.g., at a fixed stellar mass spheroid galaxies with late halo formation have higher fractions).
Consistently with studies of simulations of galaxy mergers, they find that in-situ SF lies in the centre of the galaxy, while stars accreted during mergers are found in the outer regions.
\cite{Clauwens2018} study the origin of the different morphologies in central galaxies from the EAGLE simulation \citep{Schaye2015, Crain2015}. 
They describe how  the disc and the bulge components of galaxies form. They distinguish three phases of galaxy formation. 
When $M_{\mathrm{Star}} < 10^{9.5} \ \mathrm{M_{\odot}}$, galaxy evolution would be dominated by random motions, growing disorderly.  
In this phase, SF occurs  mostly in situ, possible fed by wet mergers.
When the stellar mass of galaxies is in the range $10^{9.5} \ \mathrm{M_{\odot}} < M_{\mathrm{Star}} < 10^{10.5} \ \mathrm{M_{\odot}}$, there is also in situ SF, but in systems with disc morphologies.
Finally, at higher masses, there is a transformation to more spheroid-dominated galaxies where the spheroid formation is mainly due to accretion of ex-situ stars at large radii. 
Furthermore, \citet{Trayford2019}  analyse the morphological evolution of galaxies in the EAGLE simulation, studying the  changes produced by  mergers, accretion and secular evolution in 
galaxies as a function of redshift.
They conclude that the stellar-mass fraction of spheroids increases steadily towards $z\sim 0$ and that galaxies that have mergers with mass ratios larger than $1:10$ tend to
change their morphology from disc-dominated to spheroidal-dominated.

Recently, \citet{Rosito2018} investigate a sample of field ETGs  selected from  Fenix project \citep{Pedrosa2015}. These authors
report the simulated ETGs to be able to reproduce the size-mass relation, fundamental plane (FP) and the Faber-Jackson relation (FJR) and to have formed in an inside-out fashion, in general. All the analysed dispersion-dominated galaxies
have small disc components and $\sim 60$ per cent of them can be classified as pseudo-bulges (assuming S\'ersic index lower than 2). These authors find that the spheroidal galaxies have slightly bluer colours than expected, consistent with having been recently rejuvenated. Compared to
observations, the fraction of rejuvenated simulated spheroids is larger, suggesting the need for a stronger quenching mechanism. However, the MGHs obtained by \citet{Rosito2018} are in global agreement with the observational trends by \cite{IbarraMedel2016} that show an inside-out formation history for
low-mass galaxies ETGs. For massive ETGs, all SPs seem to be coeval with a slight trend to inside-out formation.  This study is based on a small number sample of galaxies (18) restricted
to a typical field region of the universe, and hence it does not grant a statistical analysis which allows for a large variety of assembly histories.

In this paper, we  analyse the  dispersion-dominated galaxies identified in the 100 Mpc cubic volume simulation of the EAGLE project \citep{Crain2015,Schaye2015}.
Hence, this galaxy sample allows us to explore a larger variety of assembly histories in comparison with \cite{Rosito2018}, though the detailed modelling of subgrid
physics differs.
The selected EAGLE spheroid-dominated galaxies (hereafter, E-SDGs) sample comprises 508 members resolved
with more than 10000 stellar particles.
In order to validate the E-SDGs, first we estimate their structural and fundamental relations and compare them with observations from ATLAS$^{3\mathrm{D}}$ \citep{Cappellari2013, CappellariAtlasXX}. 
We also calculate the shapes and kinematic properties. 
Then, the MGHs as well as the metallicity properties are analysed as a function of stellar galaxy mass and  the bulge-to-total mass ratio.

This paper is organised as follows. In Section \ref{sec:eagle}, the main aspects of the EAGLE project are summarised and we characterise the simulated galaxies via  morphological decomposition, the analysis of the surface density profile and the  scaling relations. In Section \ref{sec:sk}, we discuss the relation between shape and kinematics. Section \ref{sec:mgh} describes the MGHs.
Finally, Section \ref{sec:conclusion} summarises the results.

\section{The EAGLE simulation}
\label{sec:eagle}

We analyse galaxies selected from the 100 Mpc sized box reference run of the EAGLE project,  a suite of hydrodynamical simulations that follows the formation of structure in cosmological representative volumes, all of them consistent with  the current
favoured $\Lambda$-CDM cosmology \citep{Crain2015,Schaye2015}.
These simulations include: radiative heating and cooling \citep{wiersma2009}, stochastic SF \citep{Schaye2008} , stochastic stellar feedback \citep{DV2012} and  AGN feedback \citep{Rosas2015}. 
The AGN feedback is particular important for the evolution of SF activity in massive ETGs. An Initial Mass Function (IMF) of \cite{Chabrier2003} is used.
A more detail description of the code and the simulations is given by  \citet{Crain2015} and \cite{Schaye2015}.

 The adopted cosmological parameters are: $\Omega_m=0.307$, $\Omega_{\Lambda}=0.693$, $\Omega_b=0.04825$, $H_0=100~h \ \mathrm{km \ s^{-1} \ Mpc^{-1}}$, with $h=0.6777$ \citep{Planck2014a, Planck2014b}.
The 100 Mpc sized box reference simulation, so called L100N1504,  is represented by $1504^3$ dark matter particles and the same initial number of gas particles, with an initial mass of  $9.70 \times 10^6$ M$_{\odot}$ and $1.81 \times 10^6$ M$_{\odot}$, respectively \citep{Schaye2015}.
A maximum gravitational softening of $0.7$ kpc is adopted.

We use the publicly available database by \citet{mcalpine2016}.

\subsection{The selected sample of galaxies}
\label{sec:characterization}

In this work,  we use the sample of 7482 central galaxies selected  by \cite{tissera2019}
from the EAGLE galaxy catalog of  L100N1504.  
To diminish numerical resolution artefacts, we only analyse those galaxies resolved with more than 10000 stellar particles within $1.5 \ R_{\rm{opt}}$ \footnote{The optical radius, $R_{\rm opt}$, is defined as the one that encloses $\sim  80$ per cent of the baryonic mass (gas and stars) of the galaxy \citep{tissera2000}. This definition allows a determination of a characteristic radius that adapts to the size and mass distribution  of a given galaxy.}
Thus, the selected EAGLE galaxy subsample has 1721 members with masses in the range $\sim [0.22-16.7] \times 10^{10} \ \mathrm{M}_{\odot}$,
where the total stellar mass is defined as the sum of the bulge and disc stellar-mass components (defined as below) within 1.5 $R_{\mathrm{opt}}$.
\footnote{There is a difference between our definition of stellar mass and the one used in \cite{Schaye2015}.
In the latter case, the stellar mass is defined as the sum of masses of all star particles within a radius of 30 kpc.}

\begin{figure}
  \centering
\includegraphics[width=0.45\textwidth]{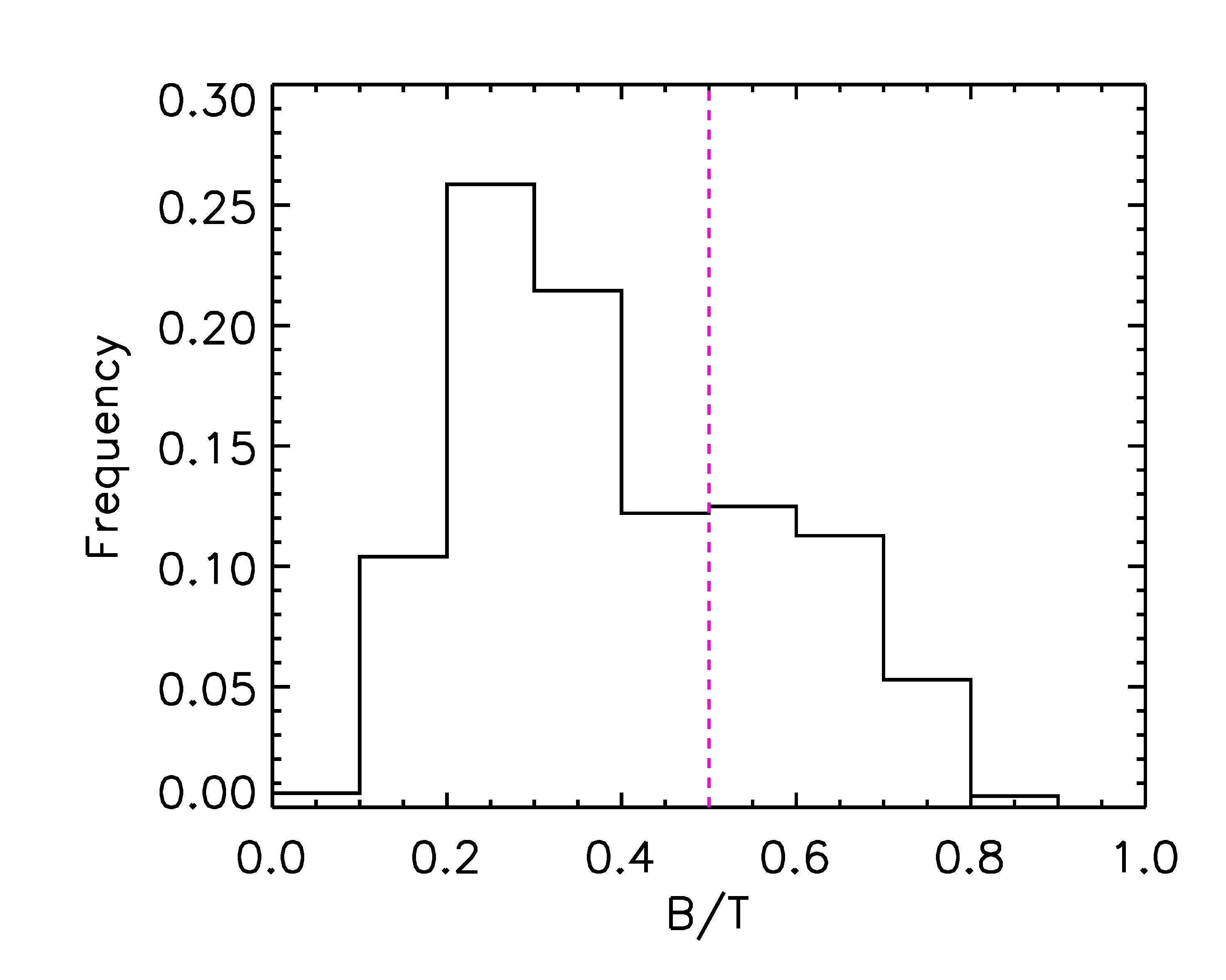}
  \caption{Distribution of $B/T$ ratios for selected
     EAGLE  galaxies. Only those galaxies with $B/T > 0.5$
    are classified
    as SDGs (dashed magenta line).}
   \label{fig:BT}
\end{figure}

The morphological classification is performed by applying the method described by \citet{tissera2012}.
For each stellar particle, we use  the parameter $\epsilon=J_z/J_{z,max}(E)$ where $J_z$ is the angular momentum component in the direction of the total angular momentum and $J_{z,max}$ is the maximum $J_z$ over all particles at a given binding energy ($E$).
Those particles with $\epsilon > 0.5$ are associated with the disc component, whereas the rest of them are  considered  part of the spheroid component.
In order to discriminate between the bulge (also called spheroid) and the stellar halo, the particle binding energy is used so that  the most bounded particles are assigned to the bulge. We assume the minimum energy of the particles at half of $R_{\mathrm{opt}}$ as the $E$  threshold \citep{tissera2012}.

This decomposition allows the definition of the bulge-to-total mass ratios ($B/T$ that takes into account only the bulge and disc components),  which are used to select galaxies according to their morphology.
In Fig. \ref{fig:BT}, we show the distribution of $B/T$ ratios for our EAGLE sample. 
Following previous works \citep{Pedrosa2015,Rosito2018, tissera2019}, we adopt $B/T=0.5$ to separate disc-dominated  from spheroid-dominated galaxies (DDGs and SDGs, respectively).

The SDG sample comprises  508 members  and  will be hereafter called E-SDGs. As can be seen from  Fig. \ref{fig:BT}, all simulated galaxies have a stellar disc component,  regardless of the size of spheroidal system.
The total stellar masses, defined by adding the bulge and the disc components, are in the range $[0.4, 14] \times 10^{10} \ \mathrm{M}_{\odot}$.

In order to analyse the scaling relations, we also estimate the stellar projected half-mass radius, $R_{\mathrm{hm}}$. The projection is peformed onto the $xy$-plane, where $z$ is the direction of the total angular momentum. We also obtain the three dimensional half-mass radius, hereafter called $R^{3D}_{\rm hm}$.

For the SPs in each E-SDG,  we measure the specific star formation rate (sSFR) calculated with star particles younger than 0.5 Gyr, the mass-weighted stellar age, the velocity dispersion ($\sigma$), and the average rotational velocity ($V$) within the  $R_{\mathrm{hm}}$. The median chemical abundances: [Fe/H], [O/Fe] and [O/H] are estimated within $R^{3D}_{\rm hm}$.

\subsubsection{Surface-mass densities}

The projected stellar-mass surface density profiles on the $xy$-plane for both disc and spheroid components are estimated.
S\'ersic profiles \citep{Sersic1968}  are fitted for the bulges,
obtaining the central surface
brightness ($I_0$), the scale radius ($R_{\rm b}$) and the so-called S\'ersic index ($n$) according to 

\begin{equation*}
 I(R)=I_{0}\exp{\big({-(R/R_{\rm b})^{(1/n)}}\big)}
\end{equation*}
The surface density profiles of the discs are fitted with an
exponential profile ($n=1$) with a scale-length $R_{\rm d}$.
We consider the projected stellar-mass surface density as a proxy of the luminosity surface brightness, which is equivalent to adopting a mass-to-light ratio $M/L=1$. This ratio is close to the observed one for the infrared bands and is consistent with that adopted  by  \citet{Rosito2018}.
In the case of the spheroidal components, the S\'ersic profile is fitted within the radial range defined by one gravitational softening and the radius that encloses 90 per cent of the total spheroid mass (to avoid numerical noise in the inner and very outer regions). The  exponential fit to the discs  is performed within the latter and the $R_{\mathrm{opt}}$.

\begin{figure}
  \centering
  \includegraphics[width=0.37\textwidth]{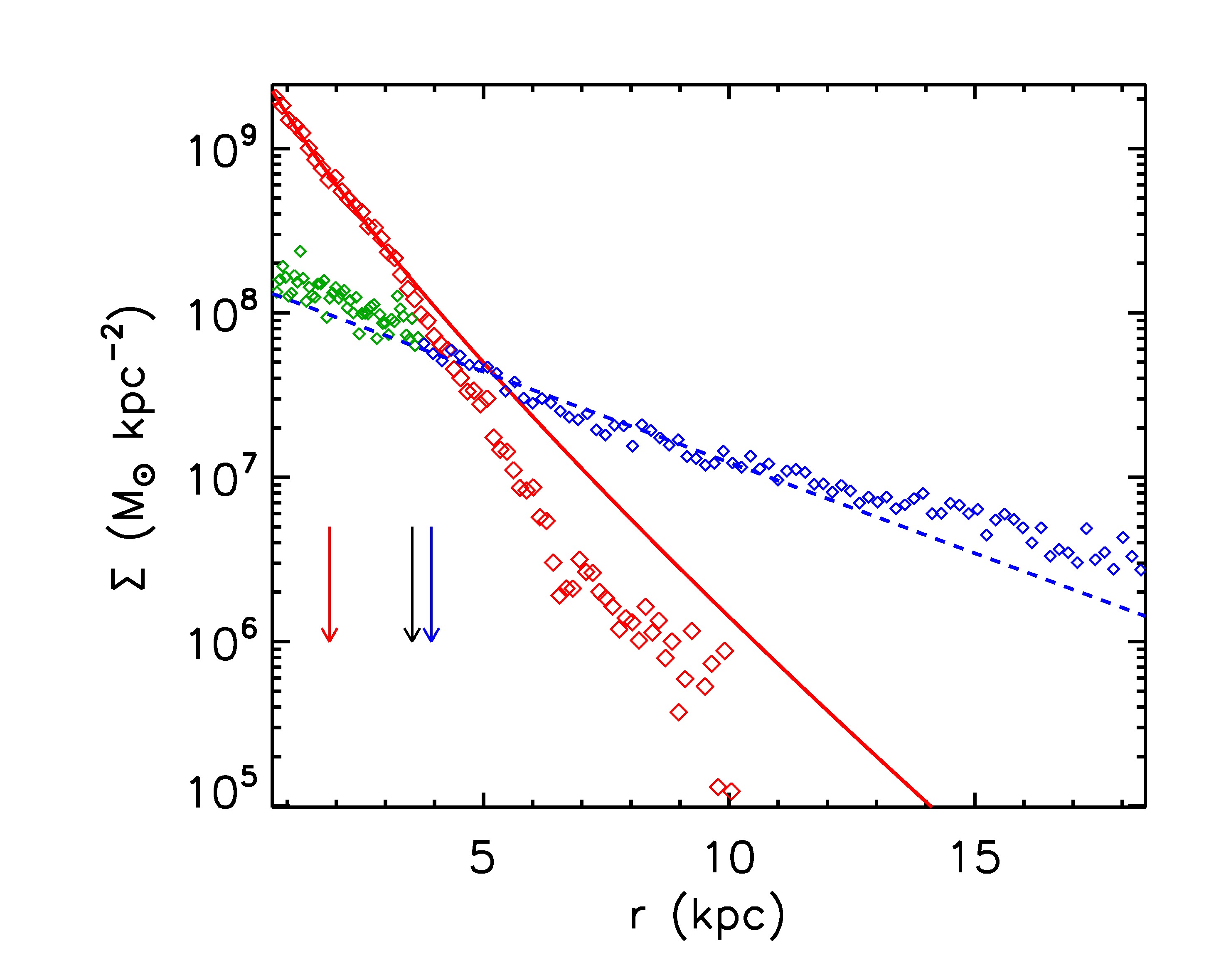} \\
  \includegraphics[width=0.37\textwidth]{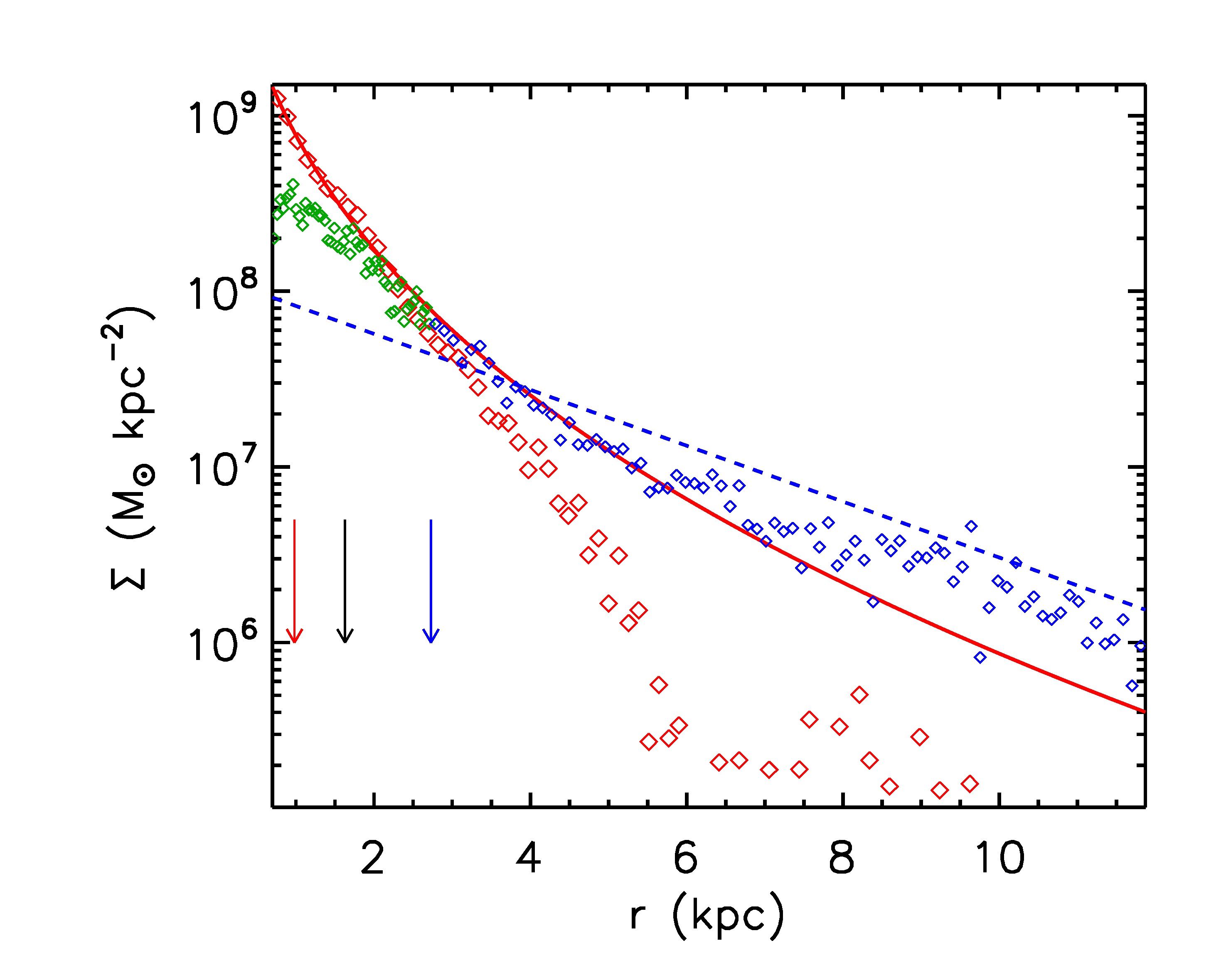} \\
  \caption{Projected stellar-mass surface density profiles for bulge (red diamonds) and disc components (blue diamonds) of two typical simulated ETGs. 
 The regions of the discs that  co-exist with spheroid components are highlighted (green diamonds). The best-fitted S\'ersic profiles for the spheroid components (solid red lines) and the exponential profiles for the  discs (dashed blue lines) are also included.
  The bulge effective radius \citep[$R_{\mathrm{eff}}$, calculated by using equation 6 in ][]{Saiz01},
 the $R_{\mathrm{hm}}$ and the $R_{\mathrm{d}}$ are depicted with red, black and blue arrows, respectively.
In the upper panel, the inner disc follows the exponential profile ($n \sim 1.25$) while in the lower panel, the same component follows the bulge profile ($n \sim 2.78$). A
variety of behaviours is detected, suggesting different contributions of processes such as collapse, mergers and secular evolution.}
  \label{fig:profiles}
\end{figure}

In Fig. \ref{fig:profiles},  we show two examples of simulated
galaxies with different surface density profiles.  As can be seen,
the profiles are well-reproduced by the S\'ersic law and an exponential profile for the spheroid and disc, respectively.
In the central regions, the bulge and disc components co-exist, although different behaviours are identified as can be seen from  Fig. \ref{fig:profiles}.  In general, the discs either extend into the bulge region, following the disc exponential profiles  (upper panel) or increases their surface densities so that they might reach that of the bulge (lower panel). In this case, they could also be interpreted as part of the bulge since their contribution is small.
The different characteristics of the  co-existence of  the discs and the spheroids reflect the variety of assembly histories \citep[e.g.][]{Trayford2019}. 

\begin{figure}[!ht]     
  \centering
  \includegraphics[width=0.45\textwidth]{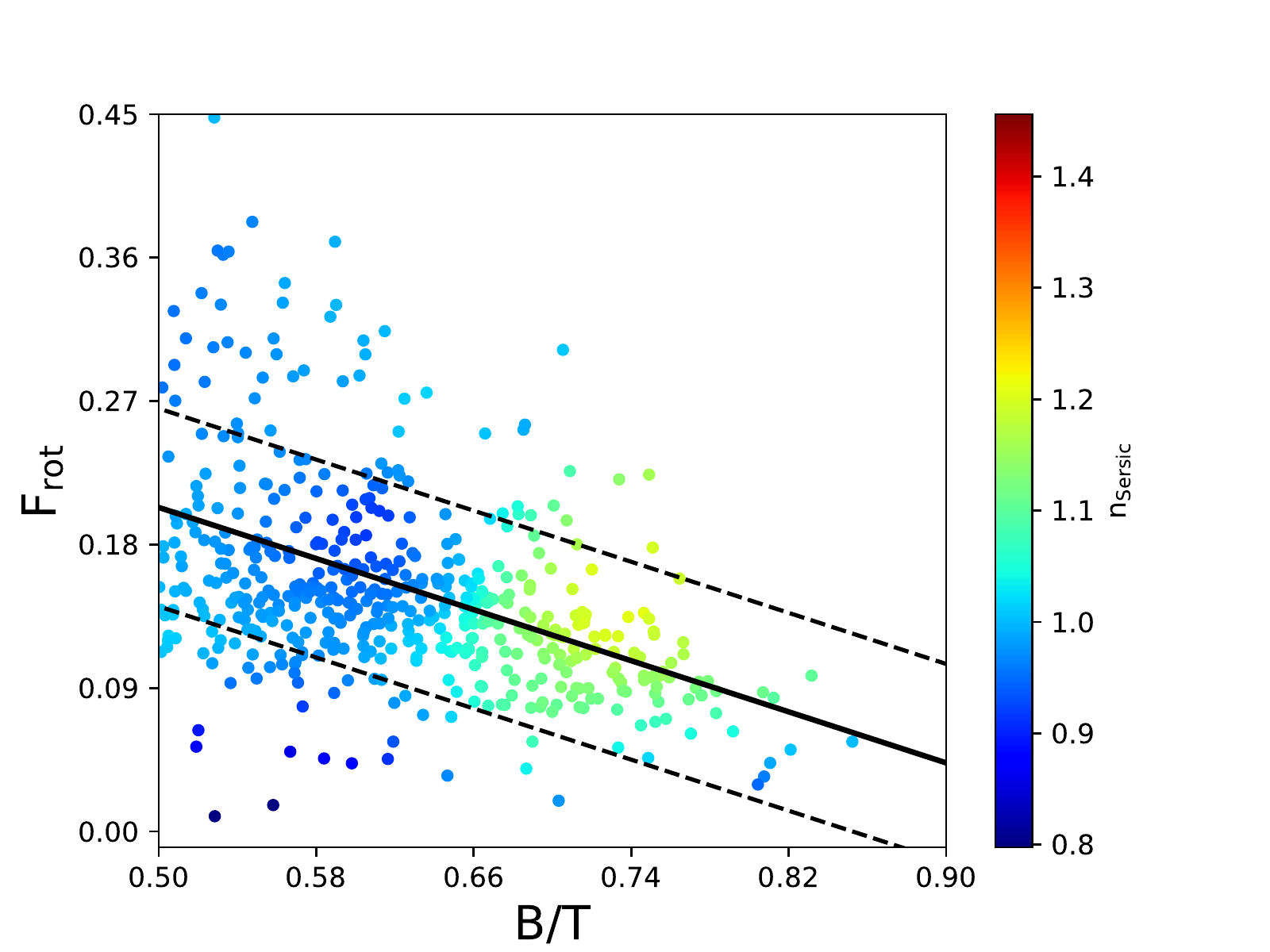}
  \caption{Stellar-mass fractions $F_{\rm rot}$ of the discs that co-exist with
      the spheroidal components as a function of the $B/T$ ratio. 
      Symbols are coloured according to $n$.  
      A linear
    regression fit is included (solid black line) along with its 1$\sigma$ dispersion (dashed black lines).}
  \label{fig:fracfracs}
\end{figure}

 Following \cite{Rosito2018}, we quantified the fraction of the disc stellar mass that co-exists with the bulge component by
estimating the stellar mass with $\epsilon > 0.5$ and binding energy lower than
the energy threshold adopted to define the bulges ($F_{\rm rot}$).
In  Fig. \ref{fig:fracfracs}, we show  $F_{\rm rot}$ as a function of $B/T$. In order to obtain smoothed distributions, we use the Python
implementation \citep{CappellariAtlasXX} of the two-dimensional
Locally Weighted Regression \citep{Cleveland88} method.
This method generalizes the polynomial regression and has the advantage that it is not necessary to specify a function to fit a model to the data sample, being notably simply to implement.
By smoothing our plots, the tendencies can be more clearly appreciated.
We must bear in mind, however, that some colours may be affected by this method.
We applied this method in all our scatter plots, hereafter. In Appendix \ref{app:plots} we include the figures without smoothing the distributions. We fix the colour-bar limits to the first and third quartile of the variable according to which we colour the symbols in the smoothed plots, except for the ones in Appendix \ref{app:all}.

Most of the  E-SDGs ($\sim 85$ per cent) have $n <2$ ($\sim 53$ per cent of the E-SDGs have $n > 1$) and will be classified as pseudo-bulges.
These systems are expected to have been  formed
by a significant contribution of  secular evolution \citep[see][for a recent review]{Kormendy2016}. We note that 
other authors assumes different definitions taking into account, for example, the inner shape or just the resulting product of secular evolution in the inner part of the galactic disc \citep[e.g.][]{Falcon2012}. 
We decide to use $n=2$ for the sake of simplicity.

As can be seen from Fig. \ref{fig:fracfracs}, E-SDGs with large  $B/T$ ratios tend to have small $F_{\rm rot}$.
This is the expected behaviour considering that more massive galaxies have larger probability to have experienced a higher rate of mergers \citep{RG2015},  producing more classical bulges.
There is only a weak trend to have slightly higher $n$ for these galaxies.
On the other hand, larger inner discs ($F_{\rm rot} >0.2$) are found in the more discoidal E-SDGs that tend to have $n \leq 1$, as expected  if bulges formed from secular evolution or wet mergers \citep{Kormendy2016}.
As can be seen from this figure, at a given $B/T$ ratio, there is a large variation of $F_{\rm rot}$, suggesting different assembly histories (i.e. different contributions of collapse, wet and dry mergers, and secular evolution for example). 
There is no clear trend for galaxies with $n>2$  as can be seen from Fig. \ref{fig:fracfrac}, where the distribution has not been smoothed.
The Spearman coefficient for the relation is  $-0.47$. 
The linear regression  yields an slope of $-0.40 \pm 0.04$ (this error is calculated by a bootstrap method).
The relation suggests that those galaxies that have larger disc
components are able to extend these discs all the way to the central
region \citep{Rosito2018}. 

\begin{figure}[!ht]     
  \centering
  \includegraphics[width=0.45\textwidth]{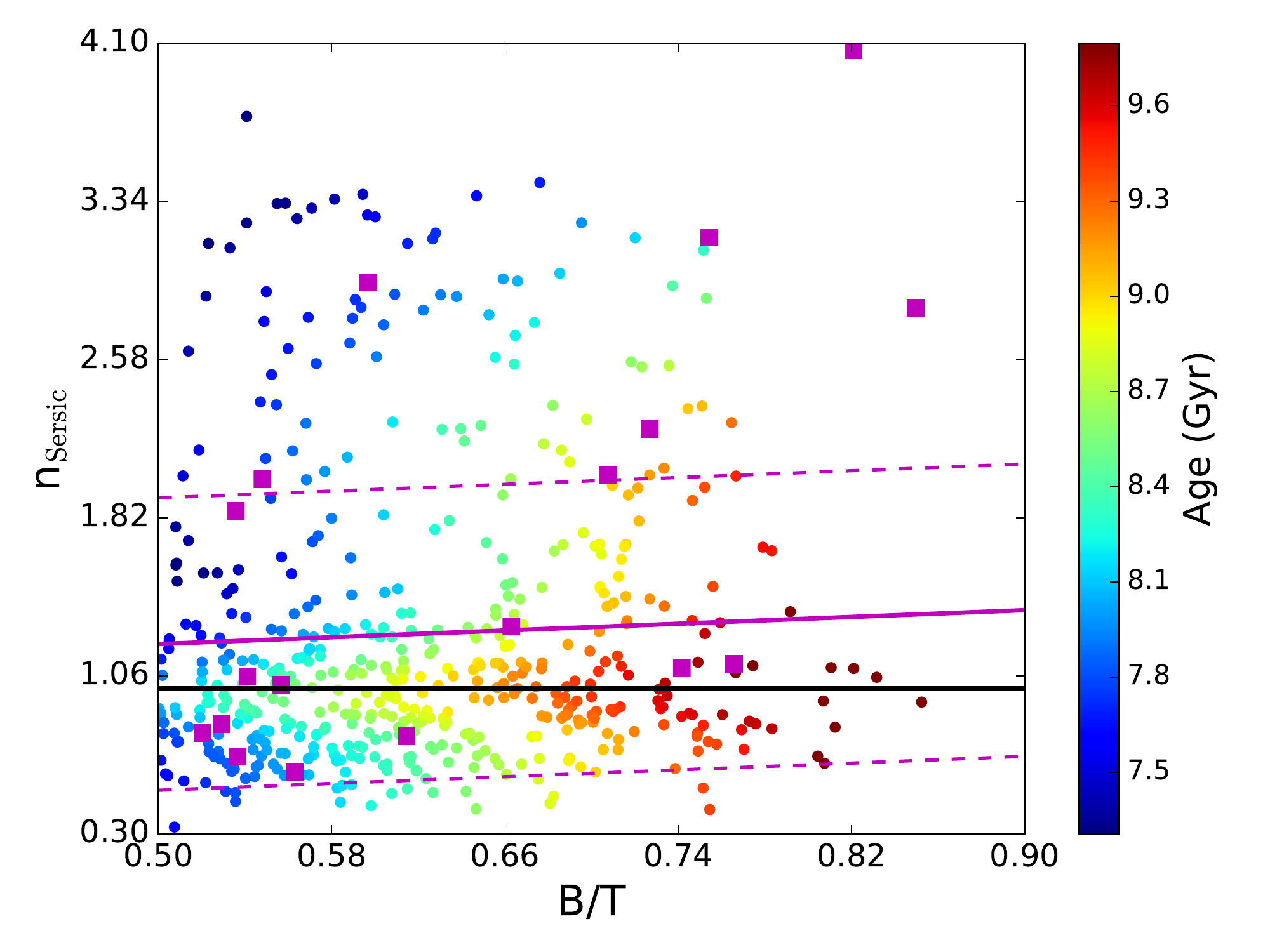}
  \caption{S\'ersic index ($n_{\mathrm{Sersic}}$) as a function of $B/T$ ratio for the simulated E-SDGs, coloured according to the  mass-weighted average
    age of the galaxy stellar mass. A linear regression fit is included (solid magenta line) with its 1$\sigma$ dispersion (dashed magenta line). For comparison, we also include the results by  \citet[][]{Rosito2018} (magenta squares).
    The line $n_{\mathrm{Sersic}}=1$ is depicted in a black line. See Appendix \ref{app:plots} for the non-smoothed distribution.}
  \label{fig:n_frac}
\end{figure}

The E-SDGs show no clear correlation between $B/T$ and $n$ as shown in Fig. \ref{fig:n_frac} (Spearman coefficient 0.09, p-value of 0.03).
From Fig. \ref{fig:n_frac}, we can also  see that those E-SDGs with larger $B/T$ ratios have, on average, the oldest SPs (E-SDGs galaxies are coloured according to mass-weighted age of the total galaxy). 
We note that a fraction  of E-SDGs with the larger $n$ index tend to have 
slightly younger SPs, on average. 
This is because there is a larger fraction of young stars associated to the disc components as we will discuss in more detail in Section \ref{sec:mgh}.

\begin{figure}[!ht]     
  \centering
  \includegraphics[width=0.45\textwidth]{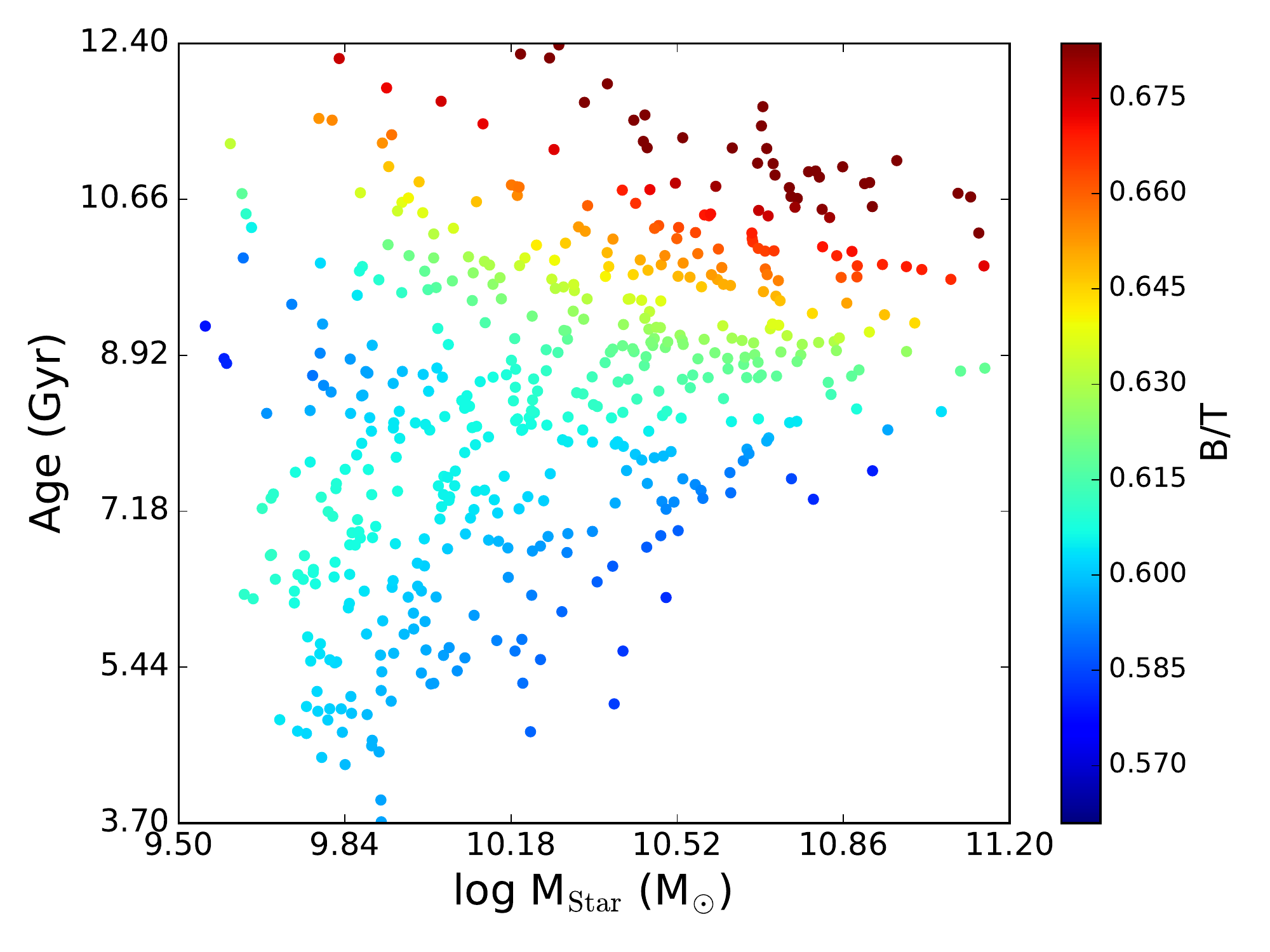}
  \caption{Mass-weighted stellar age of the E-SDGs (i.e. bulge and disc SPs) as a function of stellar mass for
    the simulated galaxies. The symbols are coloured according to the $B/T$ ratios. See Appendix \ref{app:plots} for the non-smoothed distribution. }
  \label{fig:agemstar}
\end{figure}

In Fig. ~\ref{fig:agemstar}, we show the mass-weighted ages as a
function of the stellar mass for the E-SDGs. The colour code denotes the $B/T$
ratios. As can be seen, there is a general trend to have  more massive galaxies  populated by  old
stars, on average, and with larger $B/T$ ratios as expected.
As one moves to more discoidal E-SDGs, the stellar ages are smaller and the
galaxies are less massive. However, at a given stellar mass, there
is a large variety of both morphologies and ages.

\subsubsection{Scaling relations}

In this subsection, we analyse three main scaling relations for the E-SDGs: the size-mass relation,  the FJR \citep{Faber1976} and
the FP \citep{Faber1987, Dressler1987, Davis1987},
and compare them with observations.
To estimate these scaling relations we use the $R_{\mathrm{hm}}$ as the characteristic size for the simulated galaxies.

\subsubsection*{Size-mass relation}
\label{subsect:sm}

\begin{figure}
  \centering
  \includegraphics[width=0.45\textwidth]{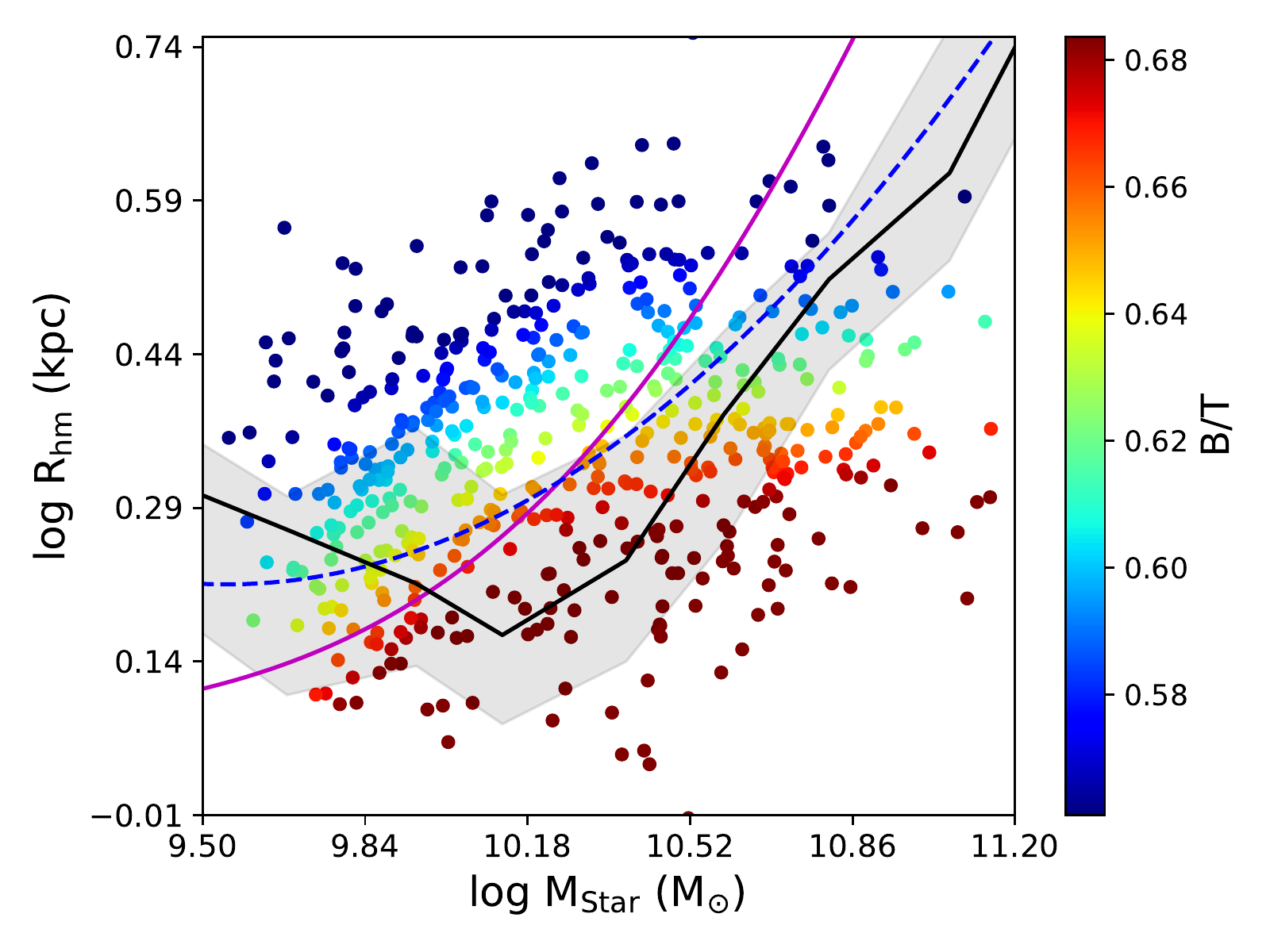}
  \caption{Mass-size relation estimated for E-SDGs. Symbols are coloured according to their $B/T$ ratio.
  The medians of the observations from ATLAS$^{3\mathrm{D}}$ (black line; the first and third quartiles  are shown as shadowed region)
and the observed relations for ETGs reported by \citet[][solid magenta line]{Mosleh13} and \citet[][dashed blue line]{Bernardi2014} are included for comparison. See Appendix \ref{app:plots} for the non-smoothed distribution.}
   \label{fig:MS2}
\end{figure}

 The size-mass relation for the E-SDGs and those obtained from different observational works are shown in Fig. \ref{fig:MS2}.
We compare simulated size-mass relation with the
observational trends reported by \cite{Mosleh13} and
\cite{Bernardi2014}.
For the former, we take those corresponding to ETGs \citep[table 1
in][]{Mosleh13} and for the latter, we choose the case of a single S\'ersic profile (their table
4).
We also compare our results with the observations from ATLAS$^{3\mathrm{D}}$ Project that consists of a sample of 260 nearby ETGs \citep{AtlasI, Cappellari2013}.
 The stellar masses are calculated from the luminosities given by  \citet[][table1]{Cappellari2013} and by  using the mass-to-light ratios of \citet[][table1]{CappellariAtlasXX} for a Salpeter IMF \citep{Salpeter55}.
The corresponding correction to transform Salpeter estimated stellar masses to the Chabrier IMF \citep{Chabrier2003} has been applied.

As can be seen in  Fig. \ref{fig:MS2}, E-SDGs within given $B/T$ ratios are located on particular tracks, showing a correlation in global agreement with  observations
albeit shallower  \citep[see also][]{Rosito2018}. There is a systematic variation of sizes as a function of $B/T$ at a given stellar mass.
E-SDGs with $B/T \ge 0.6$ are in better agreement with the observed relation from ATLAS$^{3\mathrm{D}}$. 
E-SDGs with smaller $B/T$ ratios are displaced to larger $R_{\mathrm{hm}}$; they are more consistent 
with the relation for late-type galaxies \citep[e.g.][]{Mosleh13}.
This is consistent with  the existence of larger disc components in the central region that contribute to radially expand the stellar distributions as shown in Fig. \ref{fig:fracfracs}.

\subsubsection*{The Faber-Jackson relation}

The FJR links kinematic and photometric properties: $L \propto \sigma^{\gamma}$, where $L$ is the luminosity and $\sigma$, the velocity dispersion.
The observational results from \citet{Cappellari2013} are used to confront the simulated relations.
We estimated the simulated FJR by using the r-band luminosity at $z=0.1$ obtained from the EAGLE public database \citep{mcalpine2016} and the velocity dispersion within the projected $R_{\rm hm}$.
The E-SDG FJR has a slope of  $0.30 \pm 0.01$, which is in agreement with that obtained from  ATLAS$^{3\mathrm{D}}$, $0.34 \pm 0.02$  within 2 $\sigma$ (the errors are calculated by a bootstrap method).
As can be seen, E-SDGs follow the expected trend
with brighter galaxies having larger dispersion velocities.

\begin{figure}     
  \centering
  \includegraphics[width=0.45\textwidth]{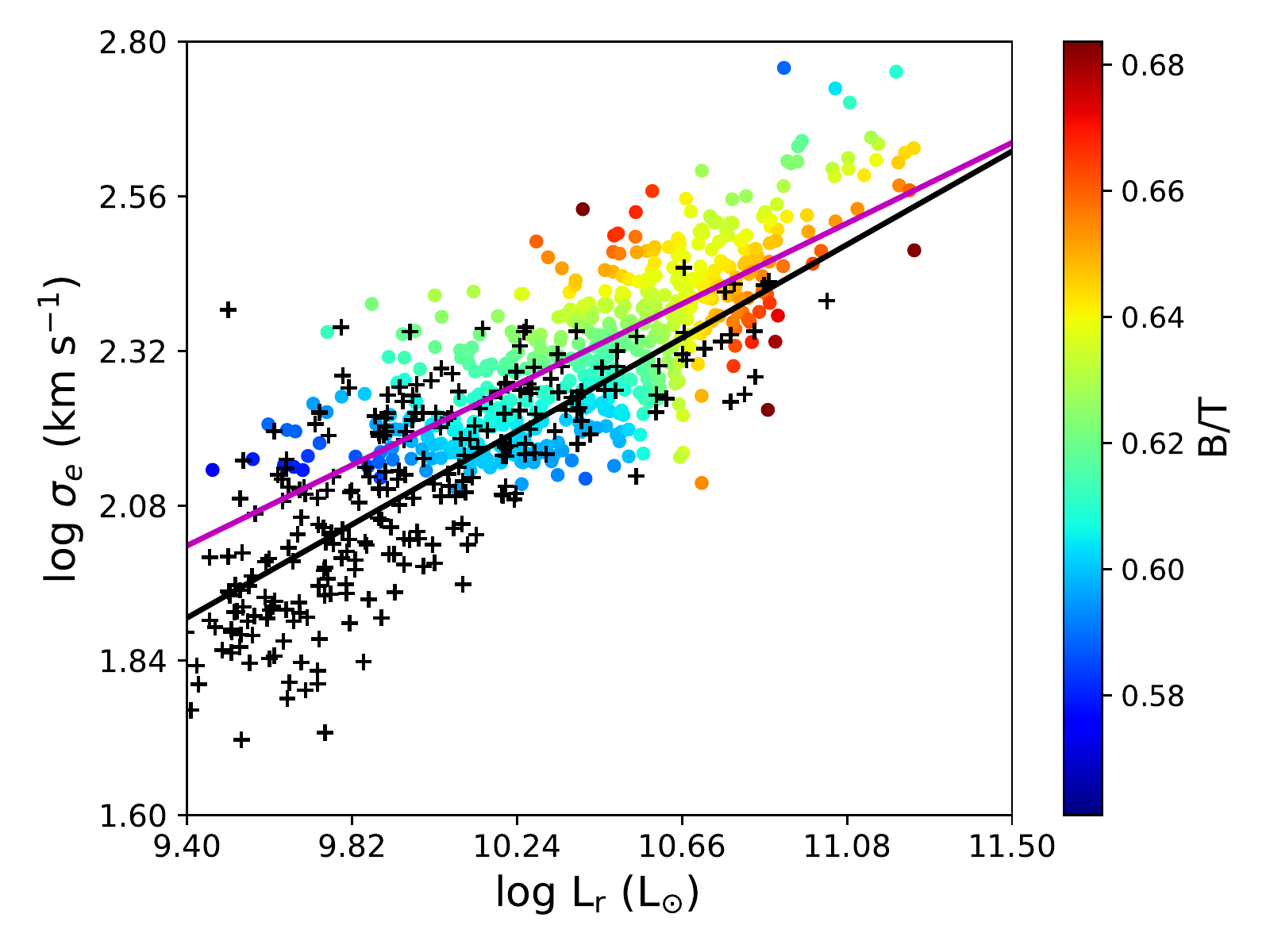}
  \caption{FJR for simulated E-SDGs (filled circles) and observed ETGs from ATLAS$^{3\mathrm{D}}$ (black crosses) galaxies. The symbols are coloured according to their $B/T$ ratios. The least squared regressions are included for simulated and observed data (magenta and black lines, respectively). See Appendix \ref{app:plots} for the non-smoothed distributions.}
  \label{fig:FJ}
\end{figure}

\subsubsection*{The fundamental plane}

The FP \citep{Faber1987, Dressler1987, Davis1987} links the size ($R_{\mathrm{eff}}$) with the luminosity surface density ($\Sigma_e$) and the velocity dispersion ($\sigma_e$):
\begin{equation*}
  R_{\mathrm{eff}} \propto \sigma_e^{\alpha} \Sigma_e^{\beta}
\end{equation*}

\begin{figure}
  \centering
  \includegraphics[width=0.45\textwidth]{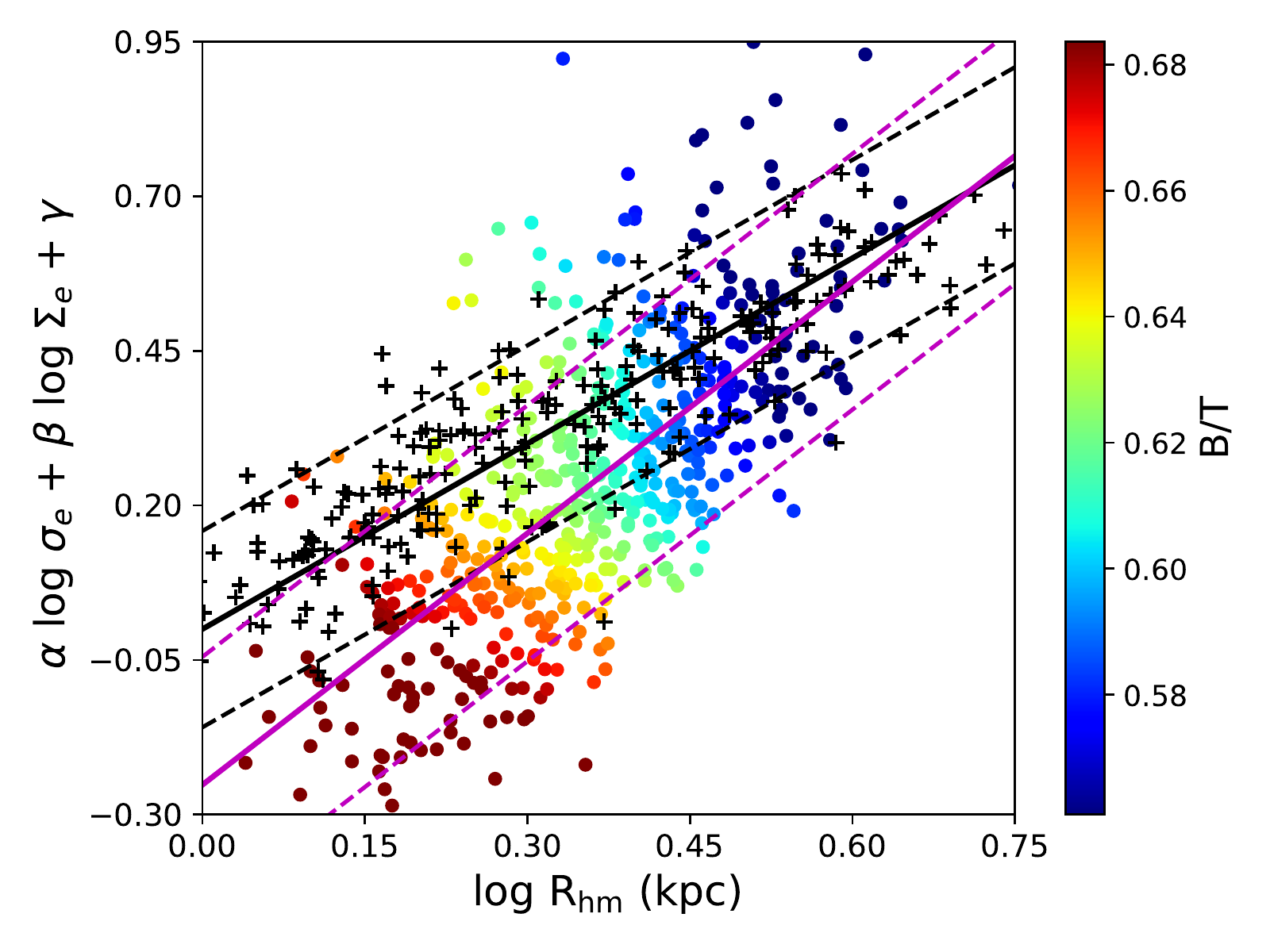}
  \caption{FP for simulated SDGs calculated with the
    parameters estimated for the  ATLAS$^{3\mathrm{D}}$ sample.
   E-SDGs are also depicted according to their  $B/T$ ratio.
     The one-to-one relation (black line) and  the best fit for the E-SDGs (magenta line), together with their corresponding \textit{rms} (dashed lines), are also included for comparison.  See Appendix \ref{app:plots} for the non-smoothed distribution. } 
  \label{fig:FP}
\end{figure}

We confronted the simulated relation with that reported by the ATLAS$^{3\mathrm{D}}$ Project with the observed parameters: $\alpha=0.98$ and $\beta=-0.74$.
In Fig. \ref{fig:FP}, we show the FP for the E-SDGs calculated with those parameters.
As can be appreciated from this figure, there is a reasonable agreement with the one-to-one relation for the E-SDGs: the \textit{rms} is $\sim 0.16$ and $\sim 0.21$ for the least squares regression and the one-to-one relation, respectively.
However, there is a group of E-SDGs that lies above the plane. 
It can be appreciated that, at a
given $R_{\mathrm{hm}}$,  there is a large dispersion. 
The zero point of the observed FP is reproduced by assuming $M/L=1$.

E-SDGs have been coloured according to the $B/T$ ratios. As can be seen from  Fig. \ref{fig:FP}, E-SDGs with larger $B/T$ ratios are
more compact. Considering Fig. \ref{fig:FJ}, they are also more massive.

\section{Shape and kinematics}
\label{sec:sk}

In this section, we focus on the core aspect of our study: the intrinsic properties of the E-SDGs and their history of 
assembly.
In order to analyse the relation between shape and kinematics,  we calculated the $V/\sigma_L$ ratio where $V$ is the average rotational velocity of the SPs 
and $\sigma_{L}$ is the one dimensional velocity dispersion, within the $R^{3D}_{\mathrm{hm}}$. We assume the velocity dispersion to be isotropic in order to derive $\sigma_{L}$ from the 3D $\sigma$.
The ratio $V/\sigma_L$  provides a measure of the relative importance of ordered rotation with respect to velocity dispersion \citep[e.g.][]{Dubois2016}.
For  the analysed E-SDGs,  a maximum $V/\sigma_L$ estimated is $\sim 1.1$
whereas a maximum  $V/\sigma_L$ obtained considering  the whole sample (i.e. E-SDGs and E-DDGs) is $V/\sigma_L \sim 2.4$.

The shape of the E-SDGs are determined by fitting ellipsoides to the 3D
stellar distributions \citep[as described in][]{tissera2010}.
For each galaxy, we estimated the semi-axis of the ellipsoides as $a \geq b \geq c$ within $R^{3D}_{\mathrm{hm}}$.
The ellipticity is hereafter defined  as  $\varepsilon = 1-\frac{b}{a}$. 
Hence, values of $\varepsilon$ close to 0 refer to galaxies that are more oblate.\footnote{This is the intrinsic ellipticity, which is equivalent to
that estimated by \citet{Trayford2019}.}

\begin{figure}
  \centering
  \includegraphics[width=0.45\textwidth]{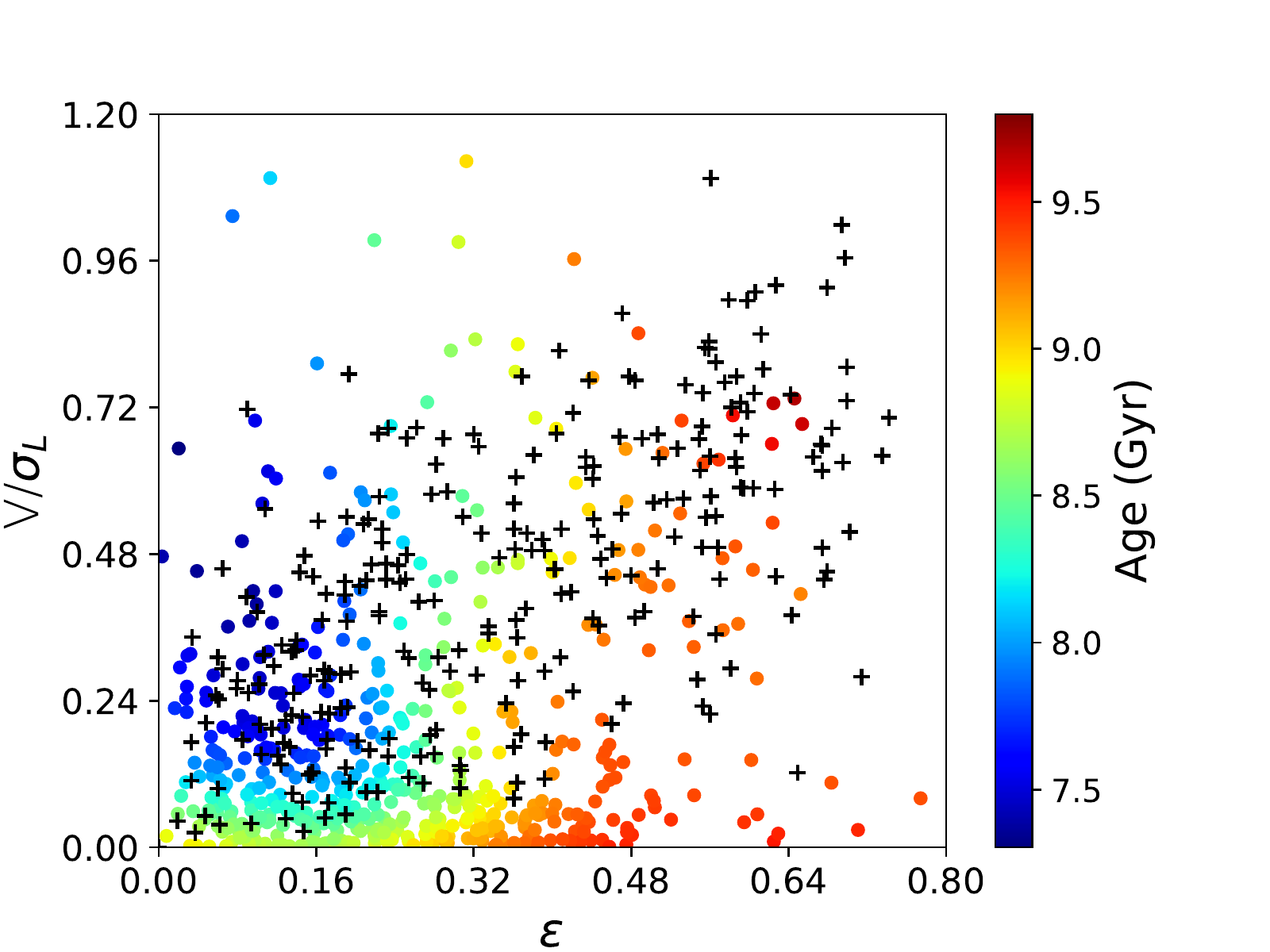} \\
  \includegraphics[width=0.45\textwidth]{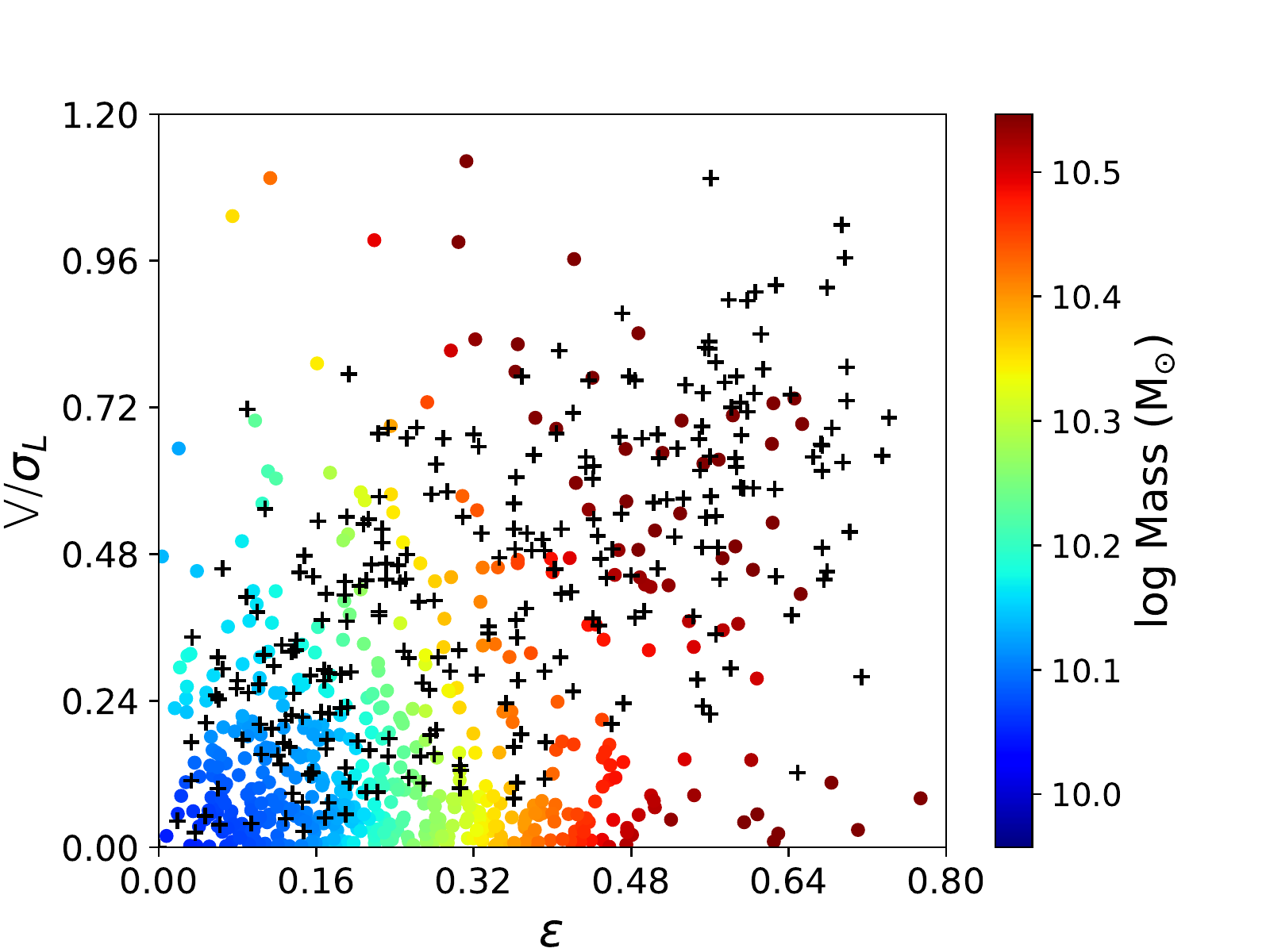} \\
  \includegraphics[width=0.45\textwidth]{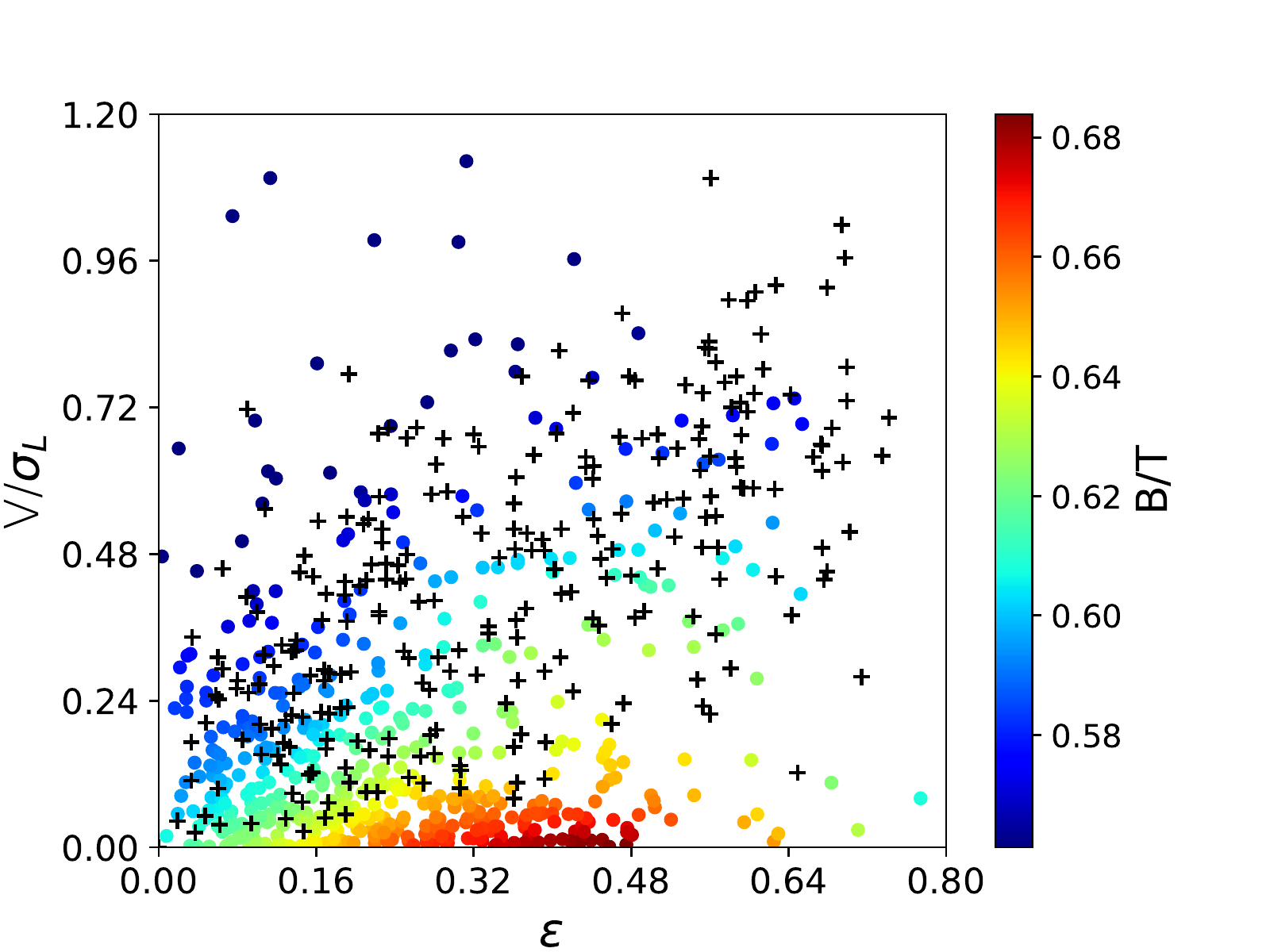}
  \caption{Anisotropy diagram for the E-SDGs. 
  E-SDGs are coloured according to mass-weighted average age (top panel), stellar mass (middle panel) and $B/T$ ratio (bottom panel).
  Observational data from ATLAS$^{3\mathrm{D}}$ are also shown \citep[][black crosses]{Emsellem2011}. 
  See Appendix \ref{app:plots} for the non-smoothed distributions.} 
  \label{fig:vsigma_eps}
\end{figure}

In Fig. \ref{fig:vsigma_eps}, we show the $V/\sigma_L$ versus $\varepsilon$. The observed galaxies from ATLAS$^{3\mathrm{D}}$ \citep{Emsellem2011} and the simulated ones are located in similar regions of this plane. 
From  Fig. \ref{fig:vsigma_eps}, we can see that the galaxies with the
oldest SPs tend to be less oblate and
massive.  This trend is in global agreement with the results reported by \citet{vdSande2018}. We also find  some massive E-SDGs with high $V/\sigma_L$ and old SPs in agreement with \citet{Lagos2018}
and \citet{cappellarireview2016}.
As can be seen in the bottom panel of this figure, galaxies with high values of $B/T$ tend to have low $V/\sigma_L$. 
Therefore, hereafter and motivated by this trend, we will define as slow rotators those galaxies with $B/T > 0.7$ and the rest will be considered fast rotators.

\section{Mass-growth histories}
\label{sec:mgh}

We investigate the global and radial archaeological MGHs of the E-SDGs.
The MGHs were constructed by using the age distributions of the stellar particles at $z \sim 0$. 
We note that because of our selection criteria of the dynamical components, the  stellar haloes are not included in our calculations.
Three radial bins are defined $[0,0.5]$, $[0.5,1]$ and $[1,1.5]$ of the $R^{3D}_{\rm hm}$, following the observational study of MaNGA galaxies by  \citet{IbarraMedel2016}.
We estimated the MGH of each of the 508 E-SDGs. 

For this analysis, the E-SDGs are divided in six subsamples.
First, we distinguished them according to their stellar masses, defining three
subsamples with similar number of galaxies ($\sim 33$ per cent of the total in each
one).  The mass ranges of the subsamples are given in Table \ref{table:t70table}.
Within each mass subsample, $B/T$, ratios are used  to group them according to their morphology:  $0.5
< B/T < 0.7$ (hereafter, fast rotators) and $B/T > 0.7$ (hereafter, slow rotators) consistently with Fig. \ref{fig:vsigma_eps} (bottom panel).

\begin{figure*}
  \centering
  \includegraphics[width=0.45\textwidth]{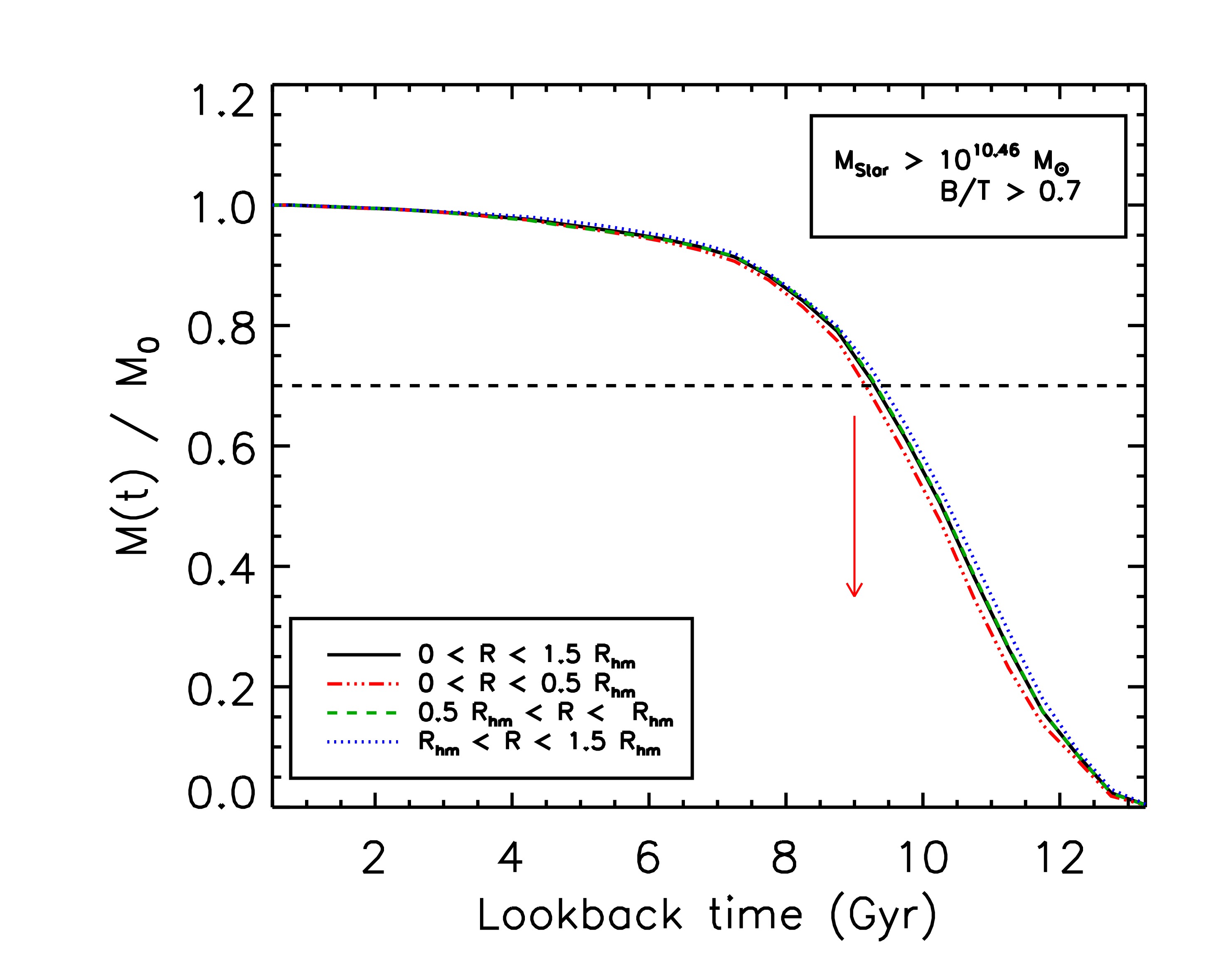}
  \includegraphics[width=0.45\textwidth]{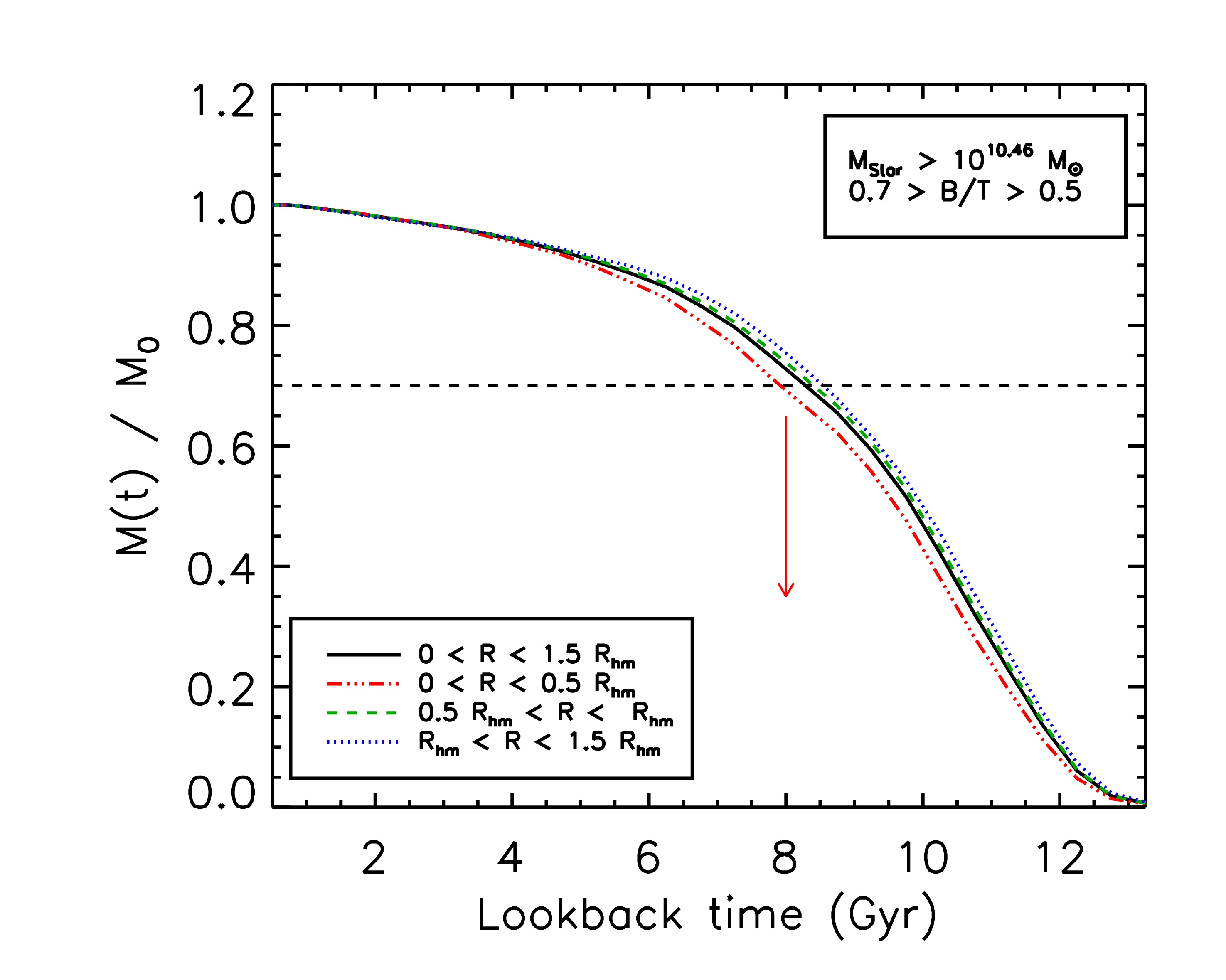} \\  
  \includegraphics[width=0.45\textwidth]{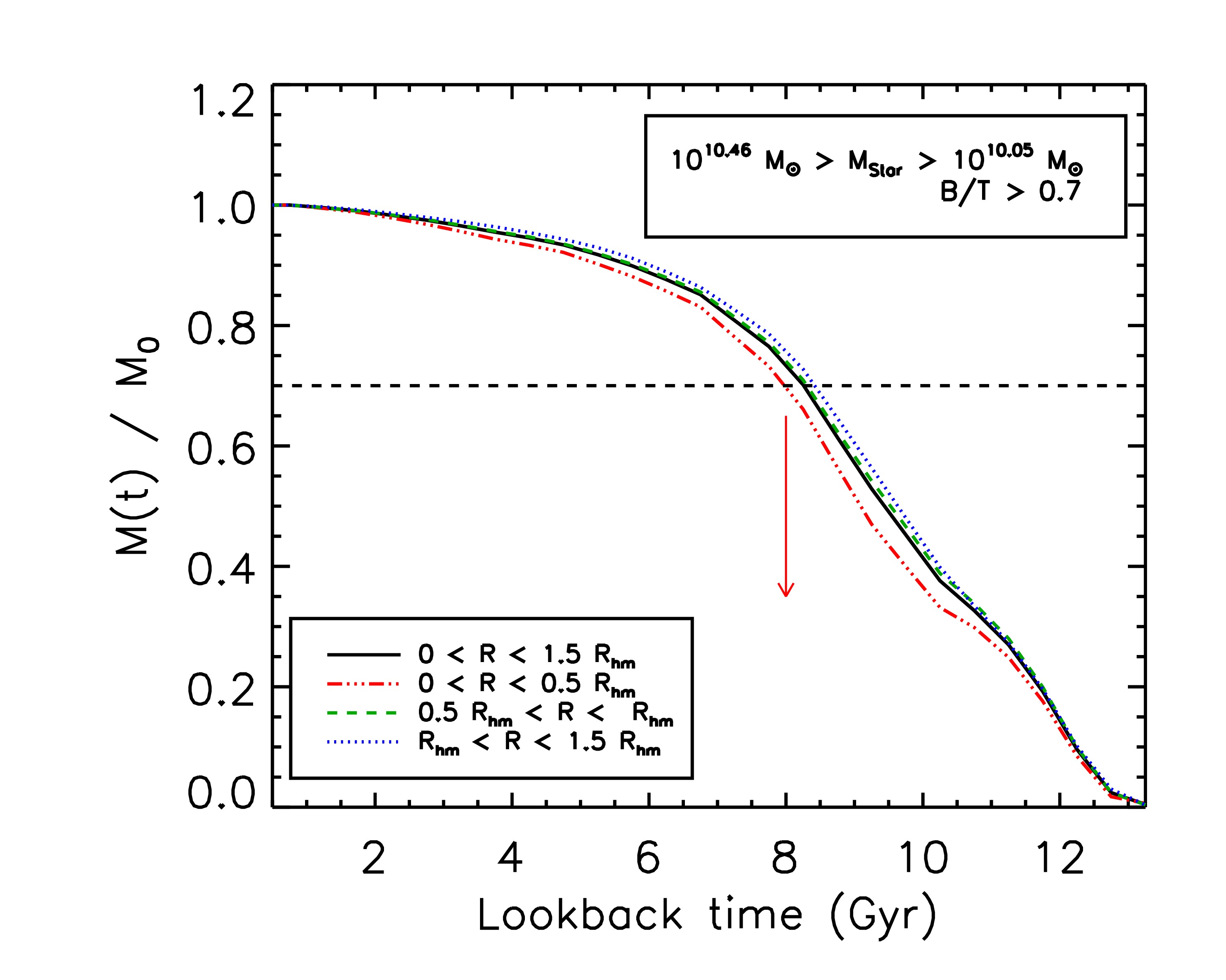}
  \includegraphics[width=0.45\textwidth]{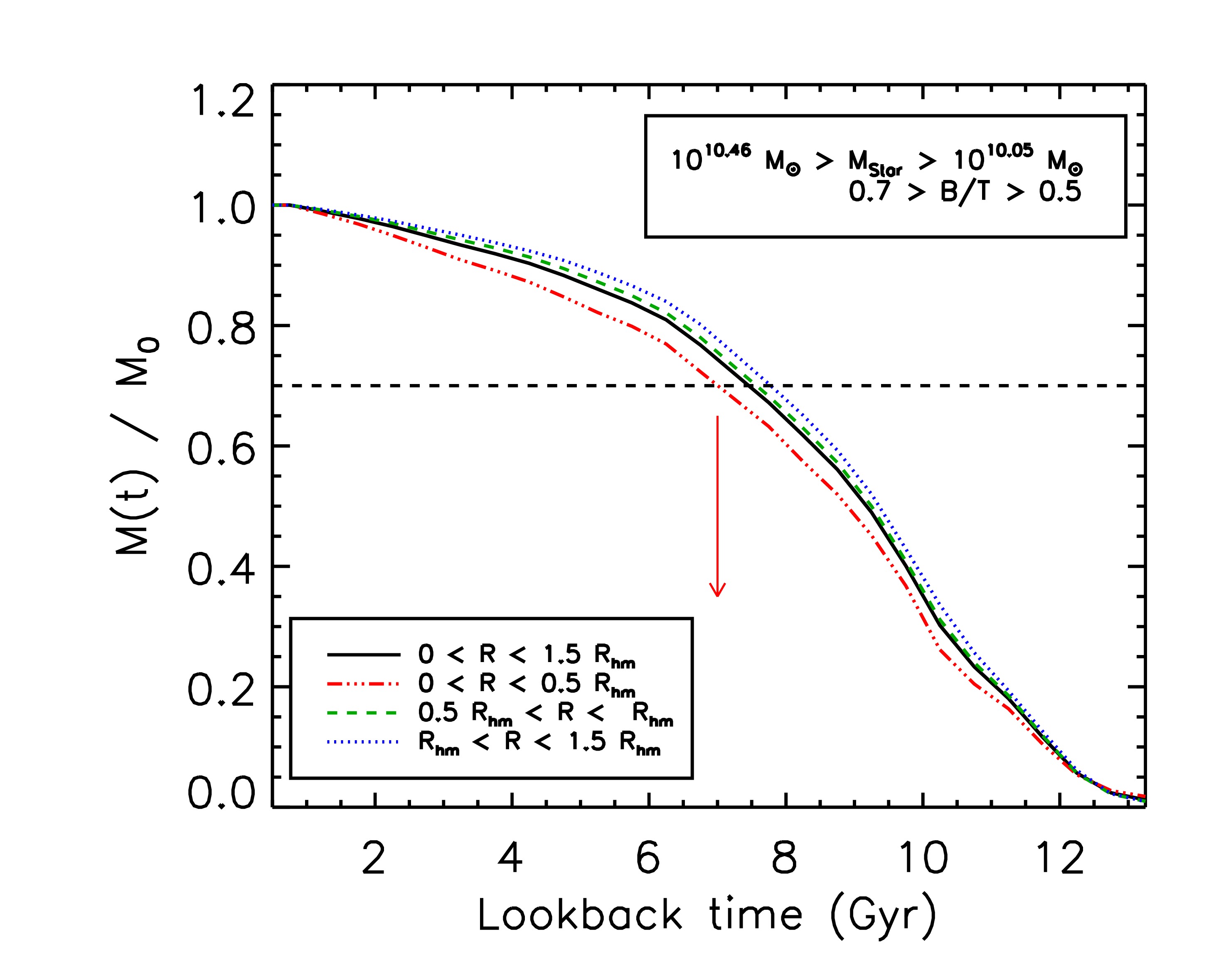}  \\
\includegraphics[width=0.45\textwidth]{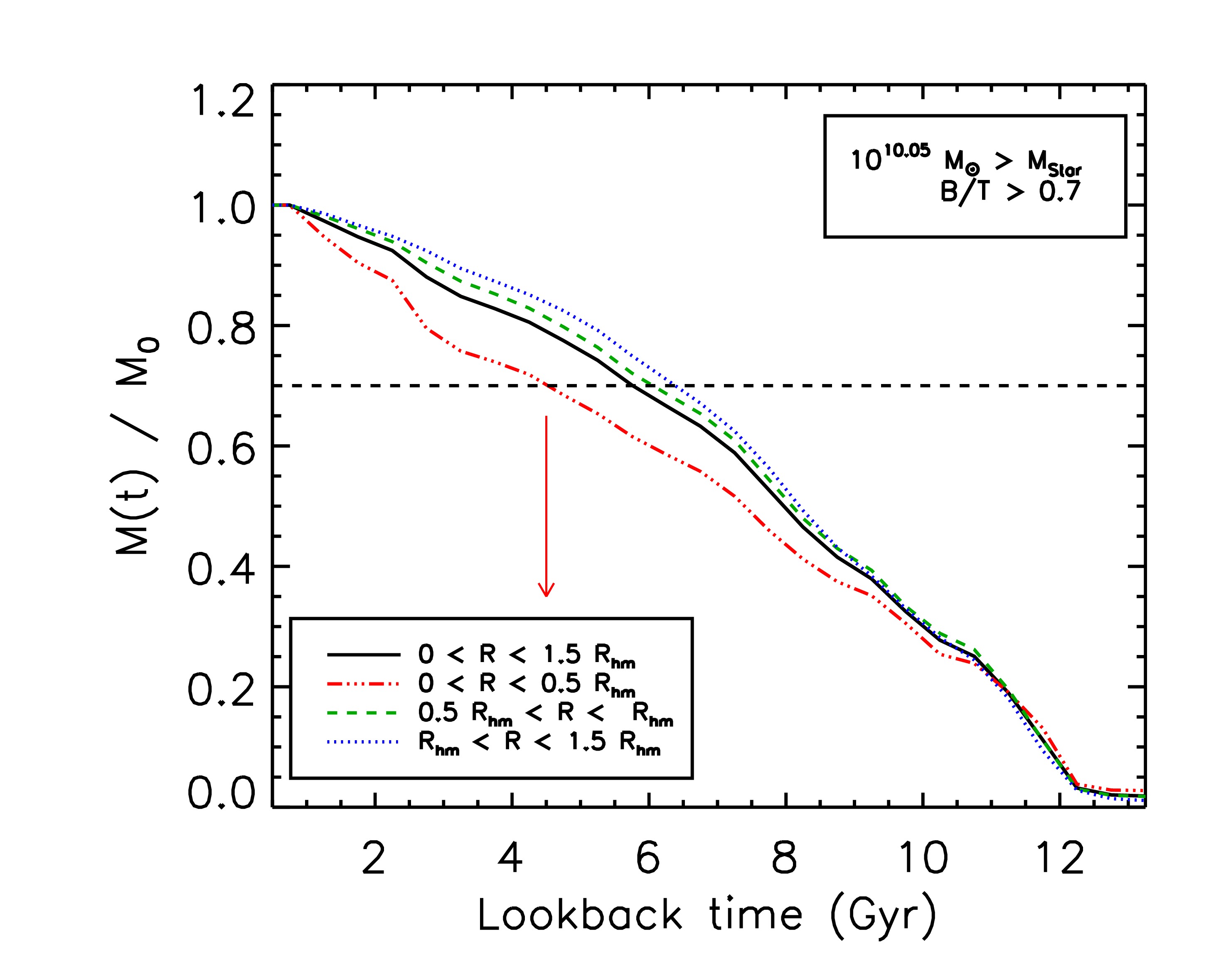}
  \includegraphics[width=0.45\textwidth]{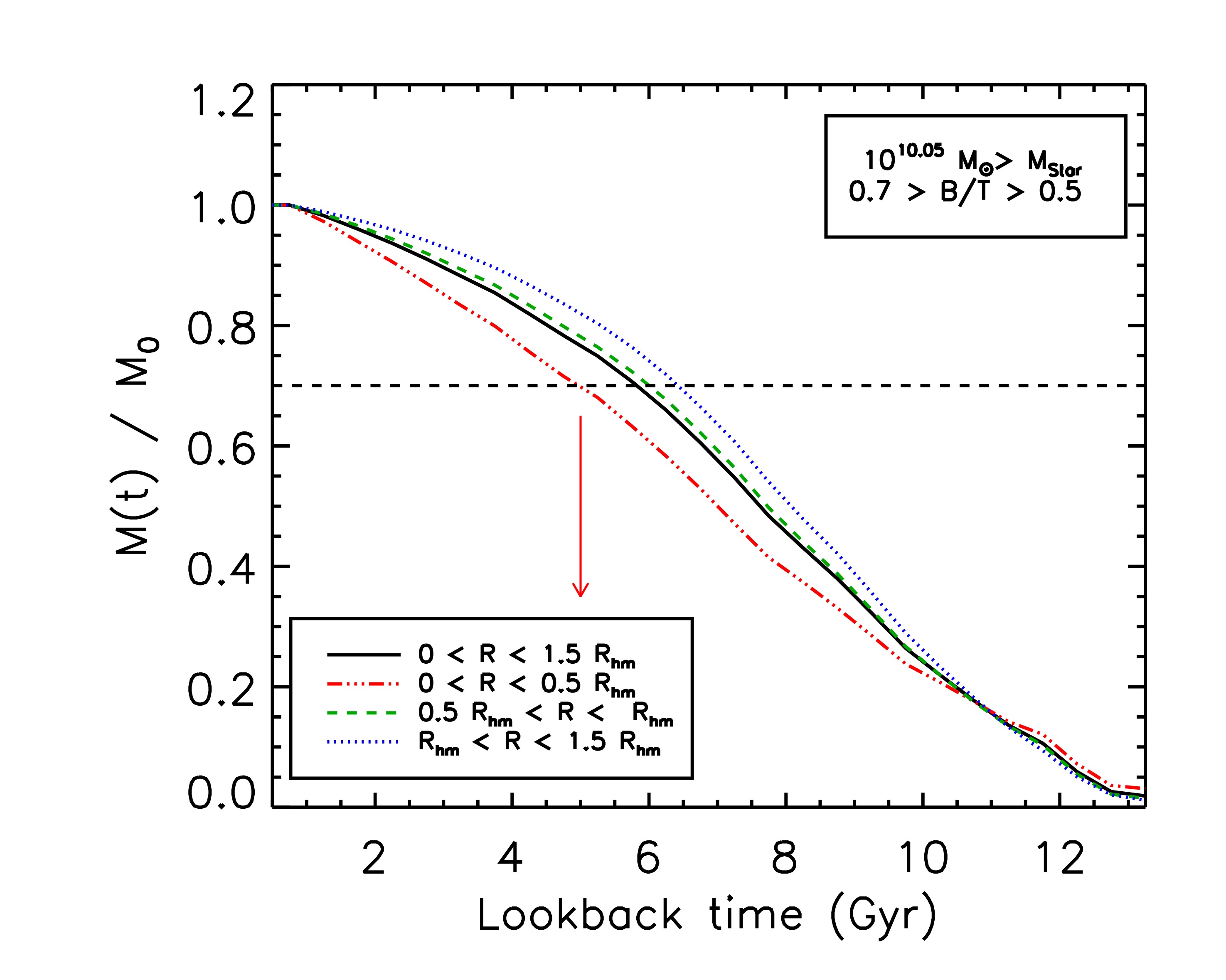} 
  \caption{Average global and radial normalised MGH for the E-SDGs: massive galaxies (top panels), intermediate mass galaxies (middle panels), and  low mass galaxies (lower panels).
 Each group is  subdivided according to the $B/T$ ratio: fast rotators (left panels) and slow rotators (right panels).
  The red arrows point out the lookback time at which 70 per cent of the stellar mass in the inner bin is attained.}
  \label{fig:mghs}
\end{figure*}

We calculated the average global and radial MGHs for each group as done in \cite{Rosito2018}. These averaged MGHs are normalised to the total mass in each bin ($M_0$) at a lookback time of 0.5 Gyr in order to have a better comparison with observations \citep{IbarraMedel2016}. They are shown in Fig. \ref{fig:mghs}.
As can be seen from this figure, massive galaxies (upper panels) show very similar MGHs for each radial interval, implying a coeval 
formation history, on average. However, there is a slight trend for the stars in the inner radial bin to be slightly younger. The MGHs of massive galaxies are in agreement with results by \citet{IbarraMedel2016}.
In order to quantify these trends, we defined the lookback time $T^{70}$ at
which 70 per cent of the stellar mass is already formed. 
We calculated  $T^{70}$ for each individual galaxy and, in order to understand the global trends, we estimated an average $T^{70}$ under two different interpretations. 
On one hand, and for consistency with \cite{Rosito2018} we consider the lookback time at which the average (stacked) MGH for each radial bin reaches the value 0.7 (Table~\ref{table:t70table}). 
On the other hand, we calculated the mean value of $T^{70}$ of each bin, considering the individual galaxies grouped as mentioned before. By applying a bootstrap method, we estimated the dispersion of these $T^{70}$ (Table \ref{table:t70table_all}). Both approaches suggest coeval stellar populations for massive galaxies and those with $B/T >0.7$. 

For  less massive galaxies, the signals for an overall outside-in
formation become stronger.  
If we consider the stacked $T^{70}$ as in Table \ref{table:t70table}, the outside-in trend is present regardless the $B/T$ ratio.
By analysing Table \ref{table:t70table_all}, we cannot ensure this behaviour within the bootstrap errors for galaxies with high $B/T$ ratios. However, it must be noticed that there are few galaxies with $B/T>0.7$ and the dispersion is quite large.
In spite of this, we find that $T^{70}$ moves systematically
to earlier values with smaller distances.

\begin{table*}
\caption{Average stacked $T^{70}$ (Gyr) for SPs in each radial bin for galaxies grouped according to their stellar masses $M_{\mathrm{Star}} $ (in units of M$_{\odot})$.}    
\label{table:t70table}      
\centering                          
\begin{tabular}{ccccccc} 
\hline\hline   
  & \multicolumn{2}{c}{$10^{10.46} <M_{\mathrm{Star}} < 10^{11.15}$}  & \multicolumn{2}{c}{$10^{10.05 }< M_{\mathrm{Star}} < 10^{10.46}$} & \multicolumn{2}{c}{$10^{9.56}<M_{\mathrm{Star}} < 10^{10.05} $}\\
  & $B/T > 0.7$ &  $0.5 < B/T < 0.7$  & $ B/T >0.7 $  & $0.5 < B/T < 0.7$  & $ B/T > 0.7$  & $0.5 < B/T < 0.7$ \\
  \hline                        
$0 < R < 0.5 R_{\mathrm{hm}}$ & $9.0$ & $8.0$ & $8.0$ & $7.0$ & $4.5$ & $5.0$ \\
$0.5 R_{\mathrm{hm}}  < R < R_{\mathrm{hm}}$ & $9.5$ & $8.5$ & $8.5$ & $7.5$ & $6.0$ & $6.0$ \\
$R_{\mathrm{hm}}  < R < 1. 5 R_{\mathrm{hm}}$ & $9.5$ & $8.5$ & $8.5$ & $8.0$ & $6.5$ & $6.5$ \\
  \hline 
\end{tabular} \\
\end{table*}

\begin{table*}
\caption{Mean $T^{70}$ (Gyr) values for SPs in each radial bin for galaxies grouped according to their stellar masses $M_{\mathrm{Star}} $ (in units of M$_{\odot})$. The errors are calculated using a bootstrap method.}    
\label{table:t70table_all}      
\centering                          
\begin{tabular}{ccccccc} 
\hline\hline   
  & \multicolumn{2}{c}{$10^{10.46} <M_{\mathrm{Star}} < 10^{11.15}$}  & \multicolumn{2}{c}{$10^{10.05 }< M_{\mathrm{Star}} < 10^{10.46}$} & \multicolumn{2}{c}{$10^{9.56}<M_{\mathrm{Star}} < 10^{10.05} $}\\
  & $B/T > 0.7$ &  $0.5 < B/T < 0.7$  & $ B/T >0.7 $  & $0.5 < B/T < 0.7$  & $ B/T > 0.7$  & $0.5 < B/T < 0.7$ \\
  \hline                        
$0 < R < 0.5 R_{\mathrm{hm}}$ & $9.75 \pm 0.23$ & $8.46 \pm 0.19$ & $8.58 \pm 0.44$ & $7.17 \pm 0.25$ & $6.30 \pm 0.80$ & $5.94 \pm 0.27$ \\
$0.5 R_{\mathrm{hm}}  < R < R_{\mathrm{hm}}$ & $9.86 \pm 0.21$ & $8.59 \pm 0.15$ & $8.74 \pm 0.41$ & $7.73 \pm 0.20$ & $6.64 \pm 0.74$ & $6.28 \pm 0.22$ \\
$R_{\mathrm{hm}}  < R < 1. 5 R_{\mathrm{hm}}$ & $9.68 \pm 0.21$ & $8.52 \pm 0.15$ & $8.83 \pm 0.34$ & $7.96 \pm 0.17$ & $6.92 \pm 0.63$ & $6.67 \pm 0.19$ \\
  \hline 
\end{tabular} \\
\end{table*}

The younger populations identified in the central regions could be
associated to secular evolution or/and wet minor merger of   smaller
systems. The MGHs of galaxies with $M_{\mathrm{Star}}<10^{10.46} \ {\rm M_{\odot}}$ show
features (i.e. bumps) that indicates important sudden contributions of SF which might be due to processes such as mergers or gas inflows.

The outside-in formation trends are at odds  with the results by \citet{IbarraMedel2016} for small stellar mass galaxies\footnote{In the Appendix, we show that DDGs selected
from the same simulation  have, on average, an inside-out formation
(Fig \ref{fig:mgh_d}) as reported by previous work
\citep{zavala2016,tissera2019}.}. We note, however, that
there are larger uncertainties in the observational determination of
ages. This is particularly important for SPs older
than $\sim 7$ Gyr. Previous results by  \cite{Rosito2018},  using   galaxies
selected from Fenix project \citep{Pedrosa2015}, show an overall inside-out
formation for SDGs in agreement with the results of
\citet{IbarraMedel2016}.
However, we note that our sample of E-SGDs are significantly older, in general, that those observed in recent works, in agreement with \cite{vdSande2019} where simulations, including EAGLE, are confronted with observations. The youngest galaxy, considering E-SDGs and E-DDGs has a mass-weighted average age of $\sim 3$ Gyr.

Our findings suggest the existence of discs, particularly, in low-mass E-SDGs. These discs could have formed as a results of 
low efficient transformation of gas into stars at higher redshift that could have left a gas reservoir for late disc formation. 
The  available gas could set on disc structures and  feed new SF in the central region via secular evolution. Mergers could also bring in gas that trigger new SF activity.
In that case, we expect the younger stars that rejuvenate the central regions of E-SDGs to be associated to the  disc components.

\begin{table*}
\caption{ Mean $T^{70}$ (Gyr) values for SPs for bulge and disc components in each radial bin for galaxies grouped according to their stellar masses $M_{\mathrm{Star}} $ (in units of M$_{\odot})$. We consider only two radial bins for the bulge. The errors are calculated using a bootstrap method.}    
\label{table:t70table_comp}      
\centering                          
\begin{tabular}{ccccccc} 
\hline\hline   
  & \multicolumn{2}{c}{$10^{10.46} <M_{\mathrm{Star}} < 10^{11.15}$}  & \multicolumn{2}{c}{$10^{10.05 }< M_{\mathrm{Star}} < 10^{10.46}$} & \multicolumn{2}{c}{$10^{9.56}<M_{\mathrm{Star}} < 10^{10.05} $}\\
  & $B/T > 0.7$ &  $0.5 < B/T < 0.7$  & $ B/T >0.7 $  & $0.5 < B/T < 0.7$  & $ B/T > 0.7$  & $0.5 < B/T < 0.7$ \\
  \hline 
  Bulge & & & & & & \\ 
  \hline        
  $0 < R < 0.5 R_{\mathrm{hm}}$ & $9.75 \pm 0.23$ & $8.47 \pm 0.19$ & $8.58 \pm 0.43$ & $7.19 \pm 0.25$ & $6.31 \pm 0.82$ & $5.96 \pm 0.27$ \\
  $0.5 R_{\mathrm{hm}}  < R < R_{\mathrm{hm}}$ & $10.34 \pm 0.21$ & $8.85 \pm 0.19$ & $9.09 \pm 0.40$ & $7.74 \pm 0.24$ & $6.93 \pm 0.80$ & $6.50 \pm 0.26$ \\
  \hline 
  Disc & & & & & & \\
  \hline                        
  $0 < R < 0.5 R_{\mathrm{hm}}$ & $9.68 \pm 0.26$ & $8.15 \pm 0.22$ & $8.67 \pm 0.43$ & $6.95 \pm 0.26$ & $6.31 \pm 0.78$ & $5.78 \pm 0.28$ \\
  $0.5 R_{\mathrm{hm}}  < R < R_{\mathrm{hm}}$ & $9.67 \pm 0.26$ & $8.13 \pm 0.20$ & $8.80 \pm 0.42$ & $7.35 \pm 0.23$ & $6.85 \pm 0.81$ & $5.97 \pm 0.24$ \\
  $R_{\mathrm{hm}}  < R < 1. 5 R_{\mathrm{hm}}$ & $9.47 \pm 0.30$ & $7.69 \pm 0.23$ & $8.85 \pm 0.40$ & $7.34 \pm 0.22$ & $7.11 \pm 0.59$ & $6.37 \pm 0.22$ \\
  \hline
\end{tabular} \\
\end{table*}

\begin{figure*}
  \centering
  \includegraphics[width=0.45\textwidth]{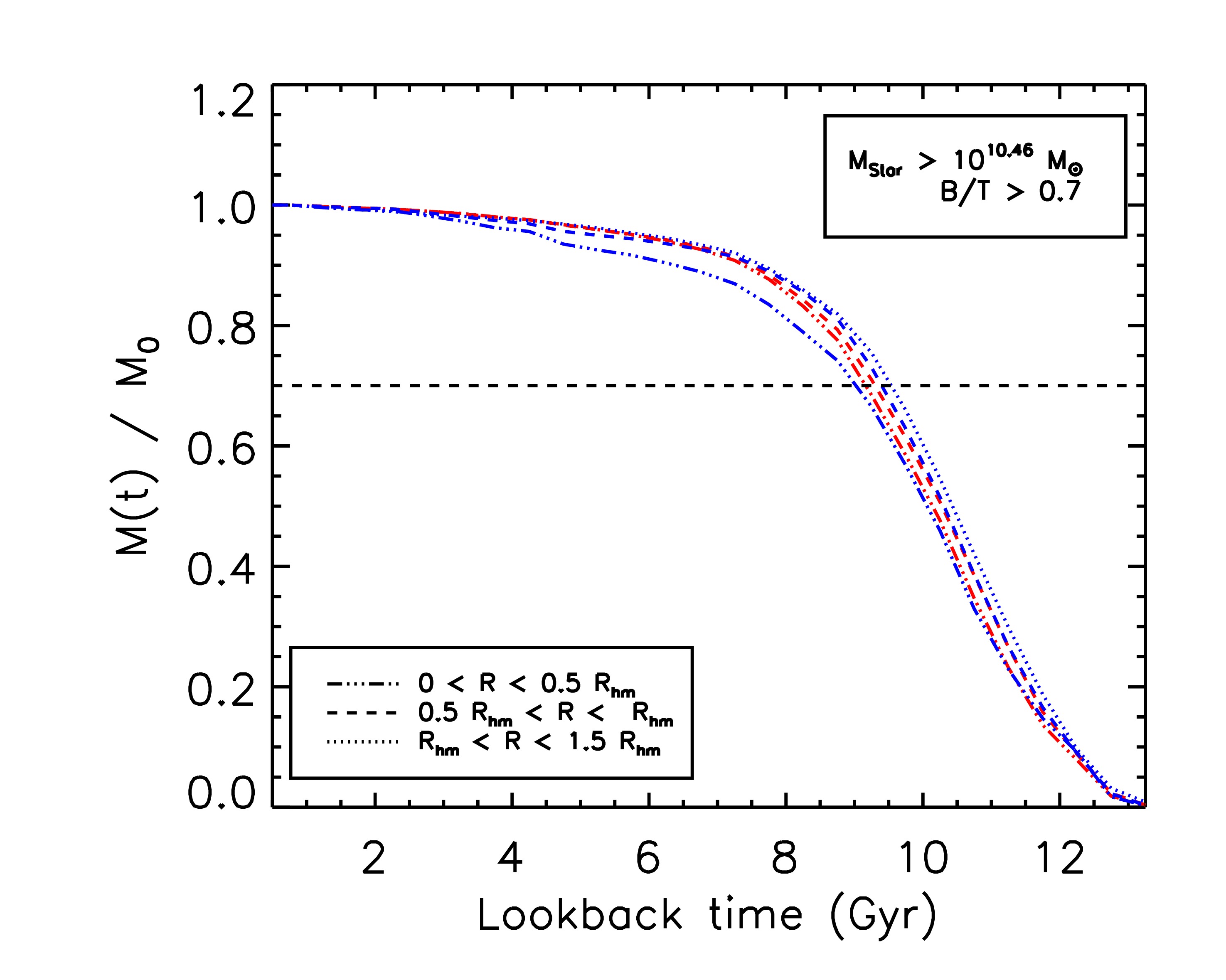}
  \includegraphics[width=0.45\textwidth]{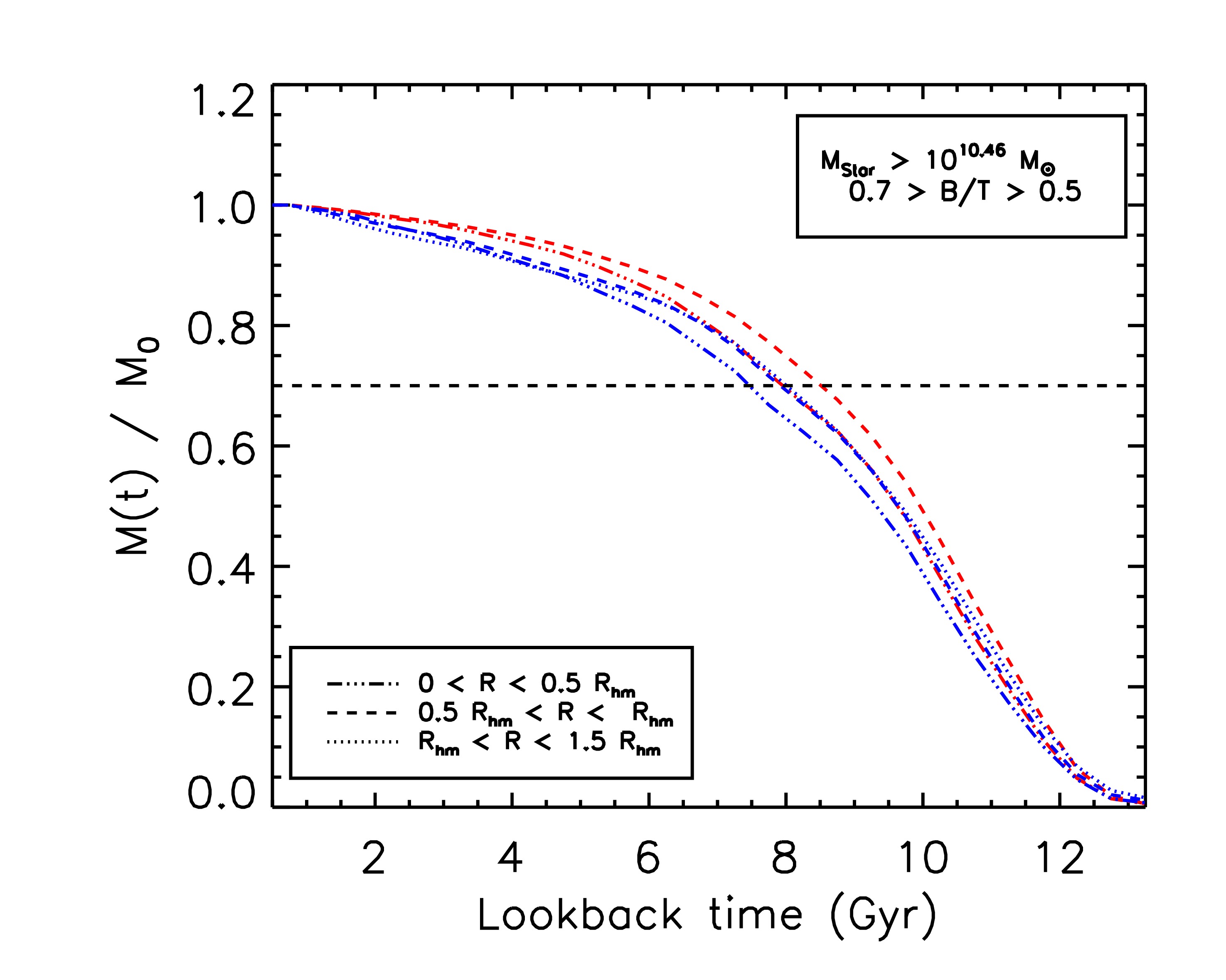} \\  
  \includegraphics[width=0.45\textwidth]{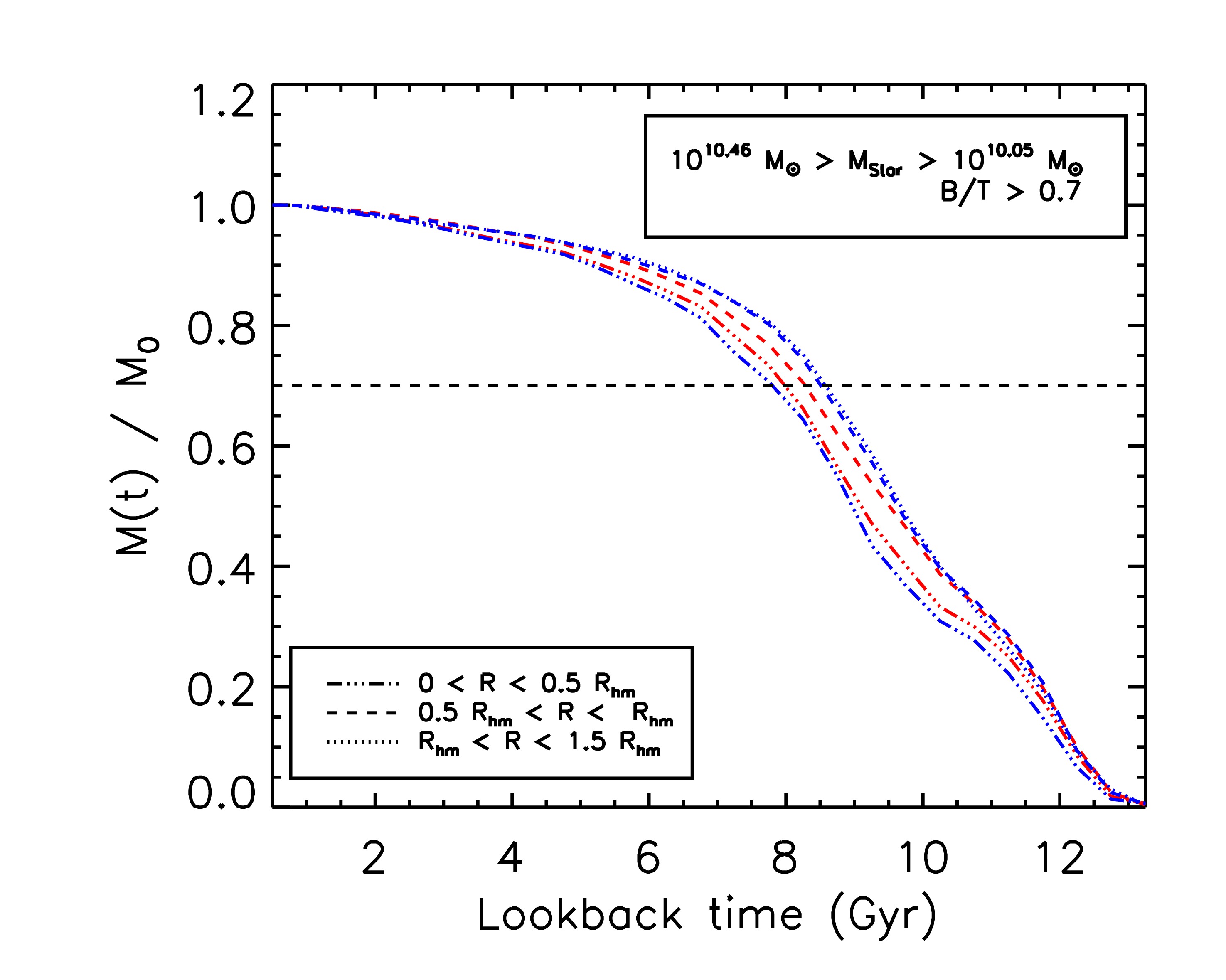}
  \includegraphics[width=0.45\textwidth]{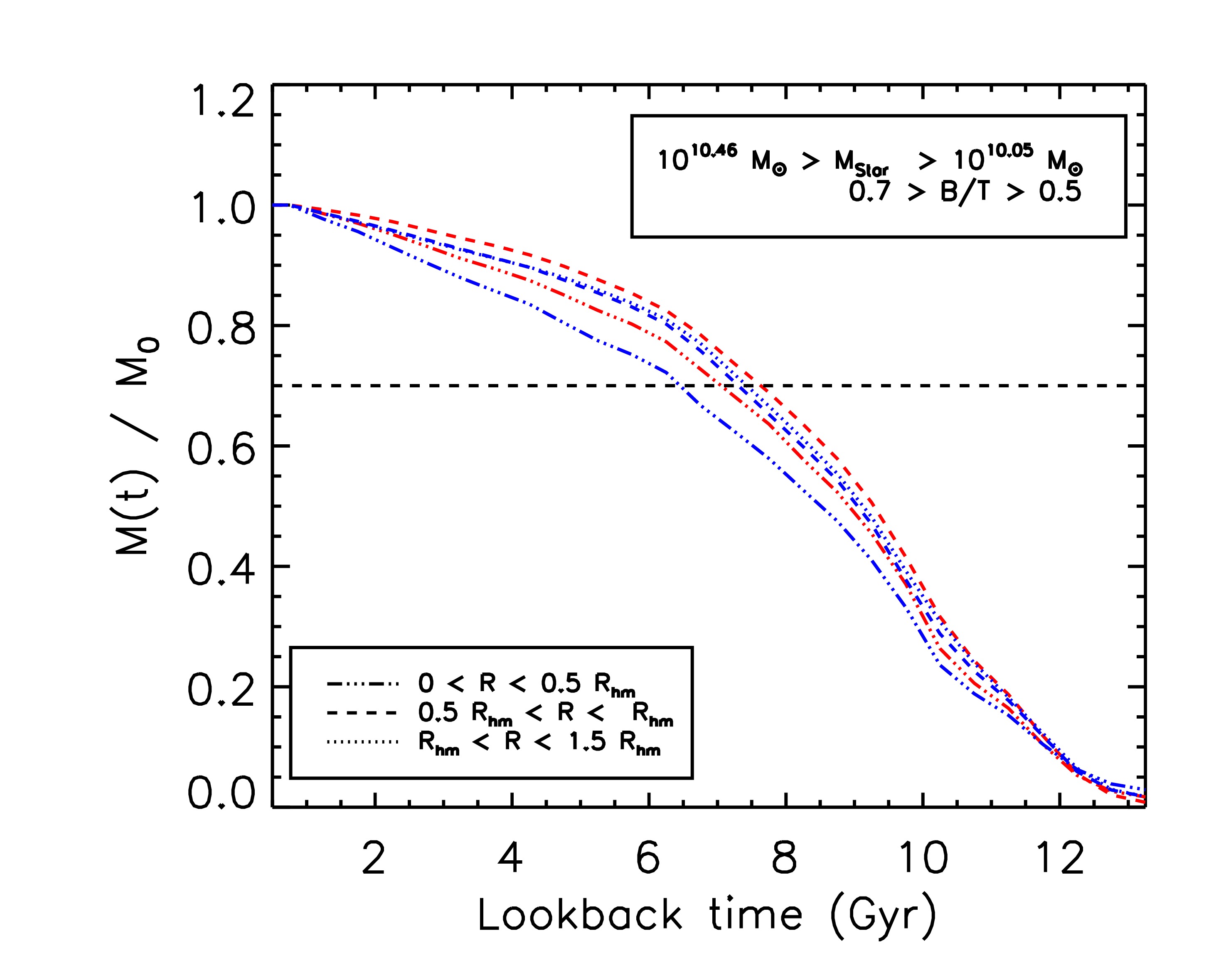}  \\
\includegraphics[width=0.45\textwidth]{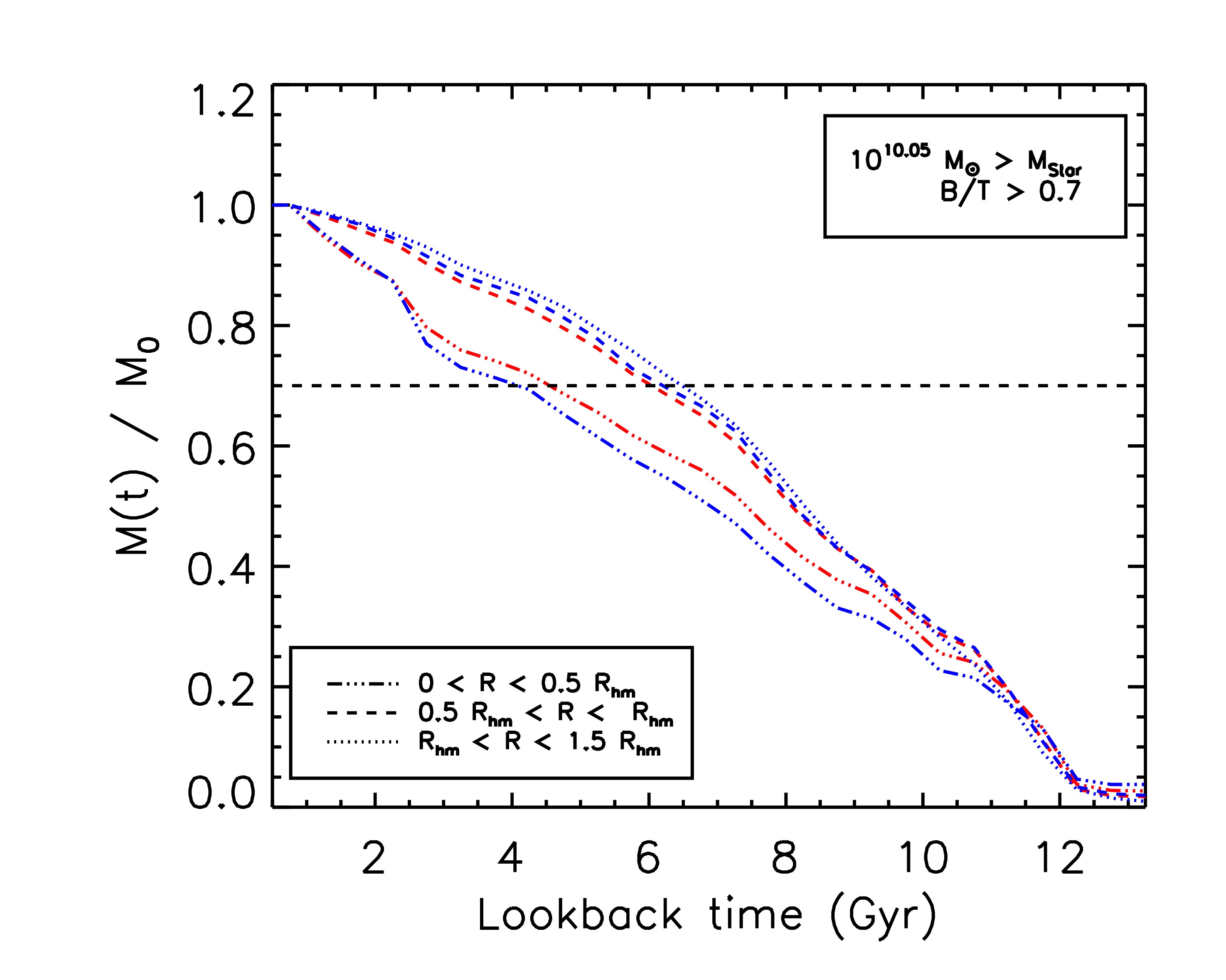}
  \includegraphics[width=0.45\textwidth]{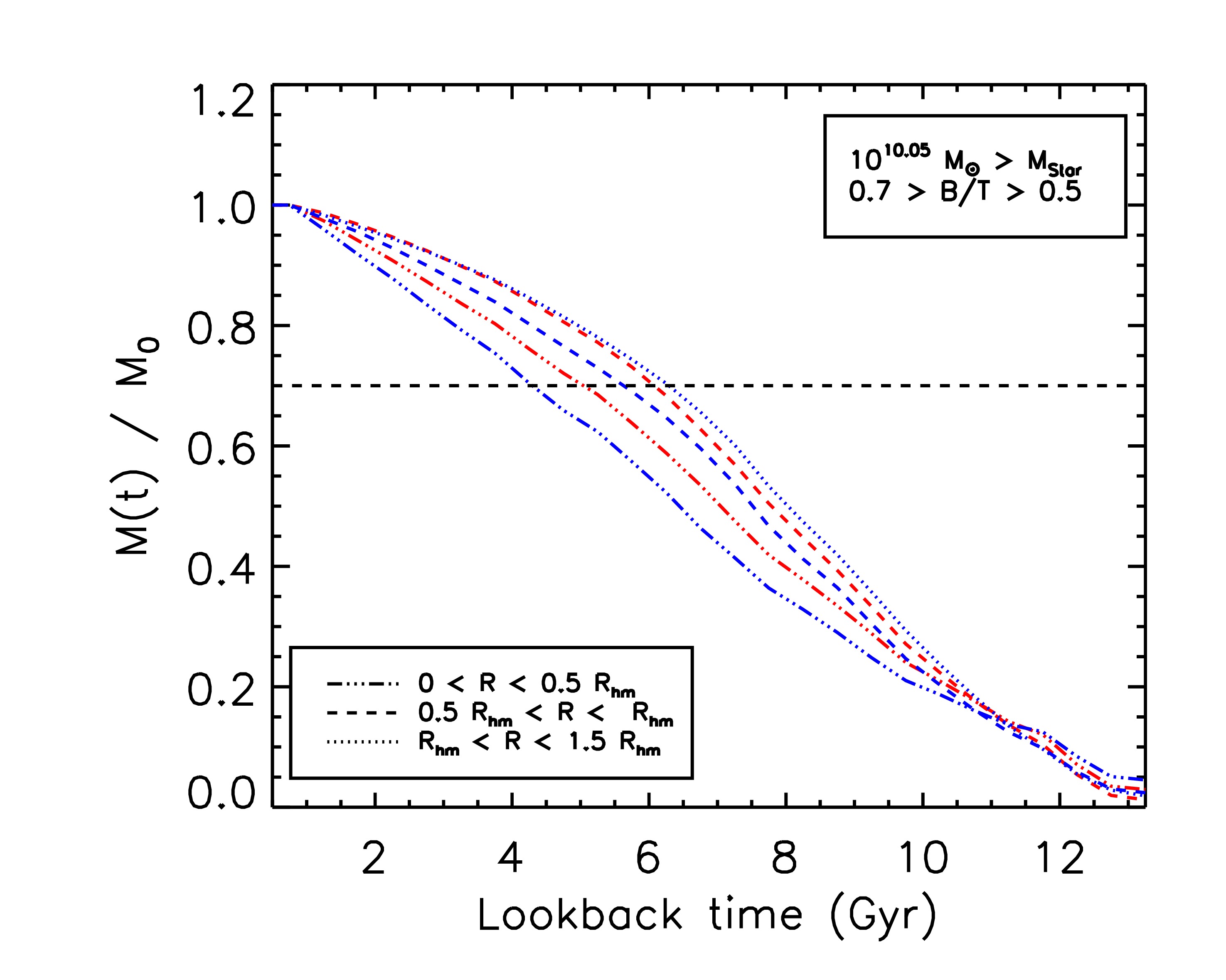} 
  \caption{Radial MGHs for both the bulge (red lines) and the disc (blue lines) components of the E-SDGs: 
massive galaxies (top panels), intermediate mass galaxies (middle panels), and  low mass galaxies (lower panels).
 Each group is  subdivided according to the $B/T$ ratio: fast rotators (left panels) and slow rotators (right panels).}
  \label{fig:mghs_comp}
\end{figure*}

To test this, the MGHs for the bulge and disc components are estimated separately as shown in  Fig. ~\ref{fig:mghs_comp}.
As bulges are smaller, we considered only two radial bins whereas for the disc components, we use the same bins mentioned above.

As shown in Table \ref{table:t70table_all}, we calculated the values of $T^{70}$ from which it can be appreciated that discs tend to be younger than bulges (Table \ref{table:t70table_comp}).
For  bulges and discs, the SPs are younger for lower mass galaxies as expected.
As can be seen from  Fig, ~\ref{fig:mghs_comp}, regardless of mass or $B/T$ ratios, the inner regions of the disc components of the  E-SDG are younger, on average. Our results agree with recent findings by \citet{Trayford2019} who show that there is SF activity associated to the disc components in ETGs in the EAGLE simulations. We go further in this analysis and identify that this SF activity is located mainly in the central regions contributing to rejuvenate  the SP and to determine an  outside-in formation history of the E-SDGs. We note that E-SDGs with younger SPs tend to have $B/T <0.7$ and that
there is a large spread in the averaged ages, which is due to large variety of MGHs.

\begin{figure*}
  \centering
  \includegraphics[width=0.45\textwidth]{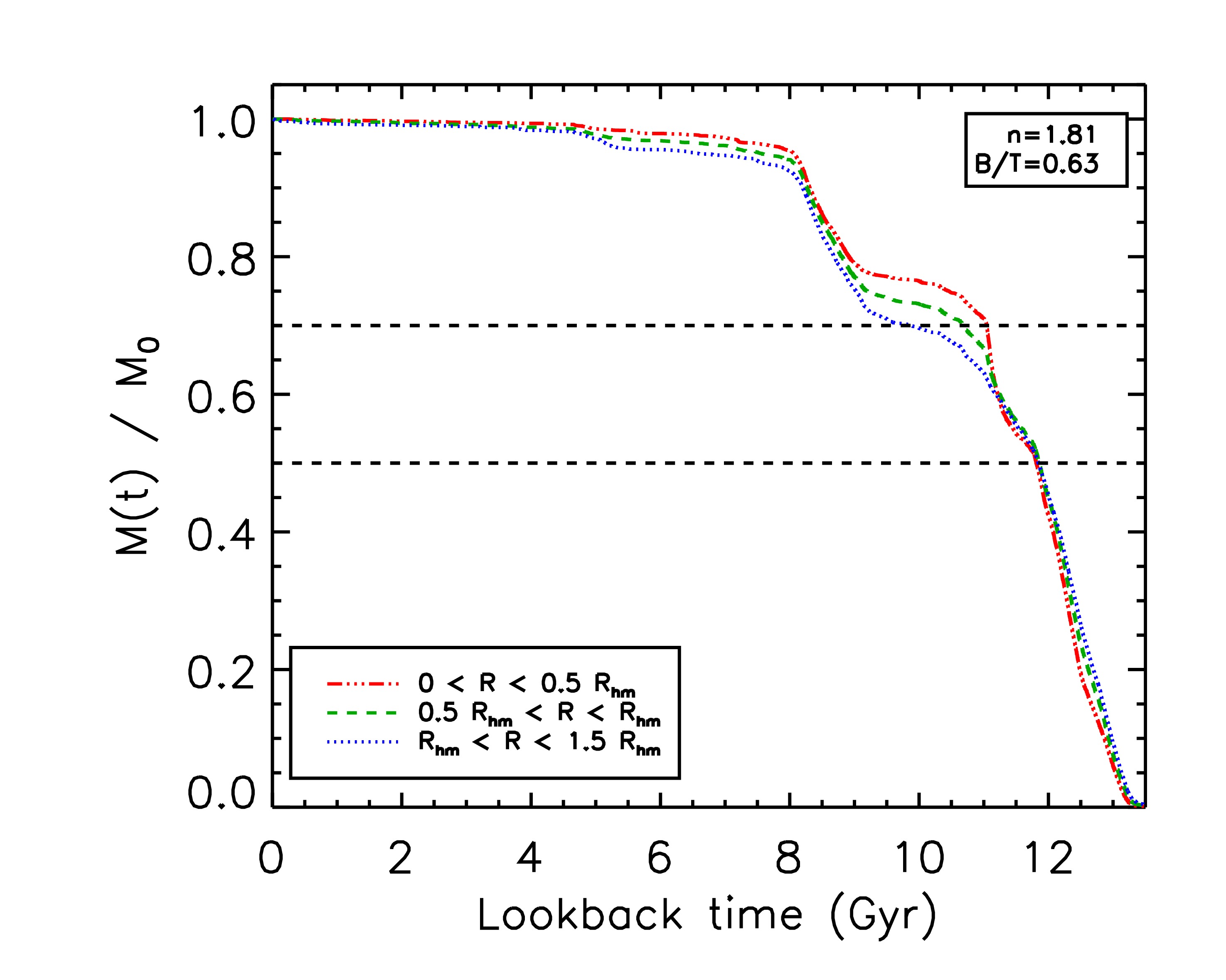}
  \includegraphics[width=0.45\textwidth]{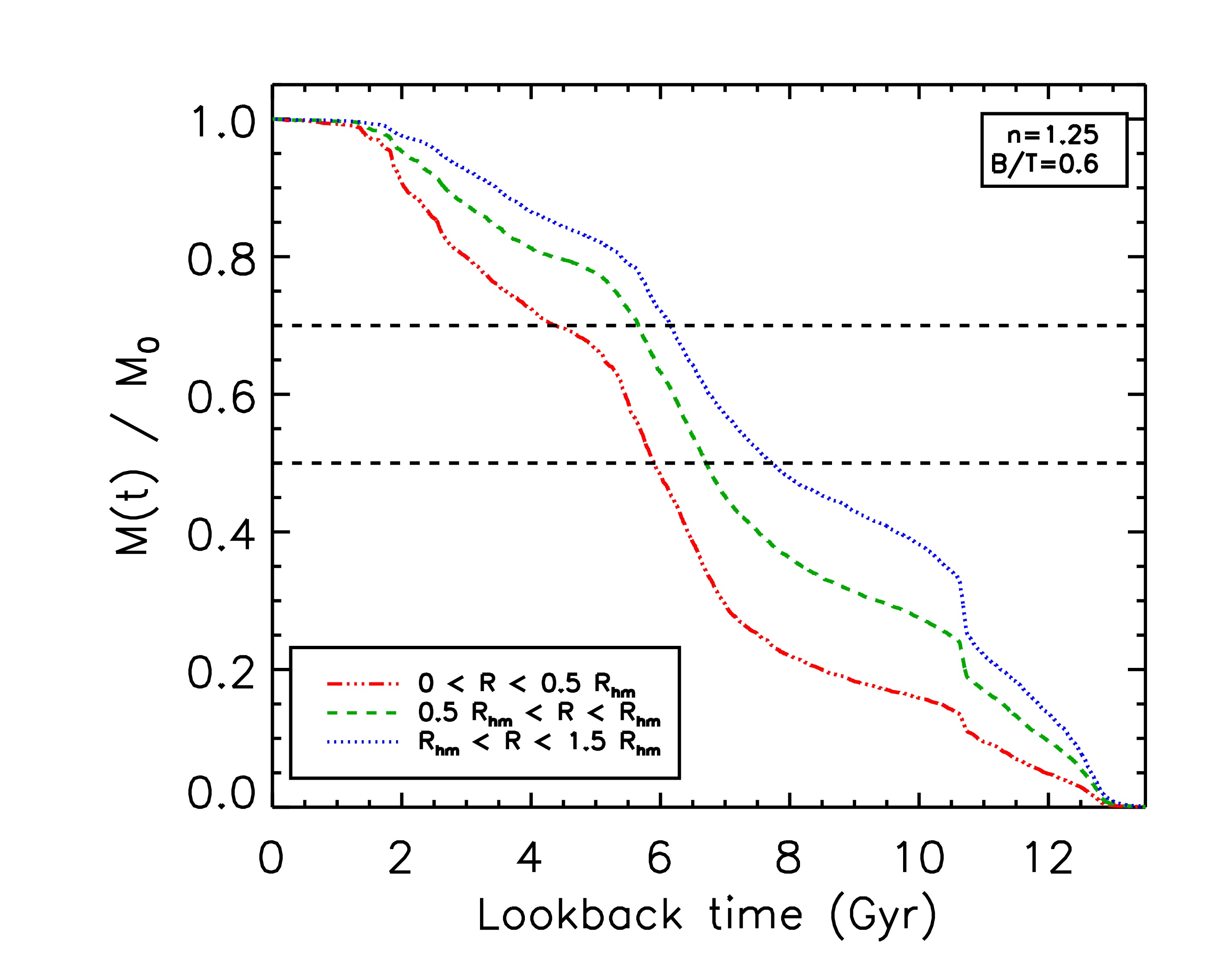}\\
  \caption{Top panels: MGHs for two E-SDGs showing different assembly
    histories: an inside-out (left panel) and an outside-in (right panel).
  These E-SDGs have similar $B/T$ ratios.
} 
  \label{fig:mghs_ind}
\end{figure*}

\begin{figure*}
  \centering
  \includegraphics[width=0.45\textwidth]{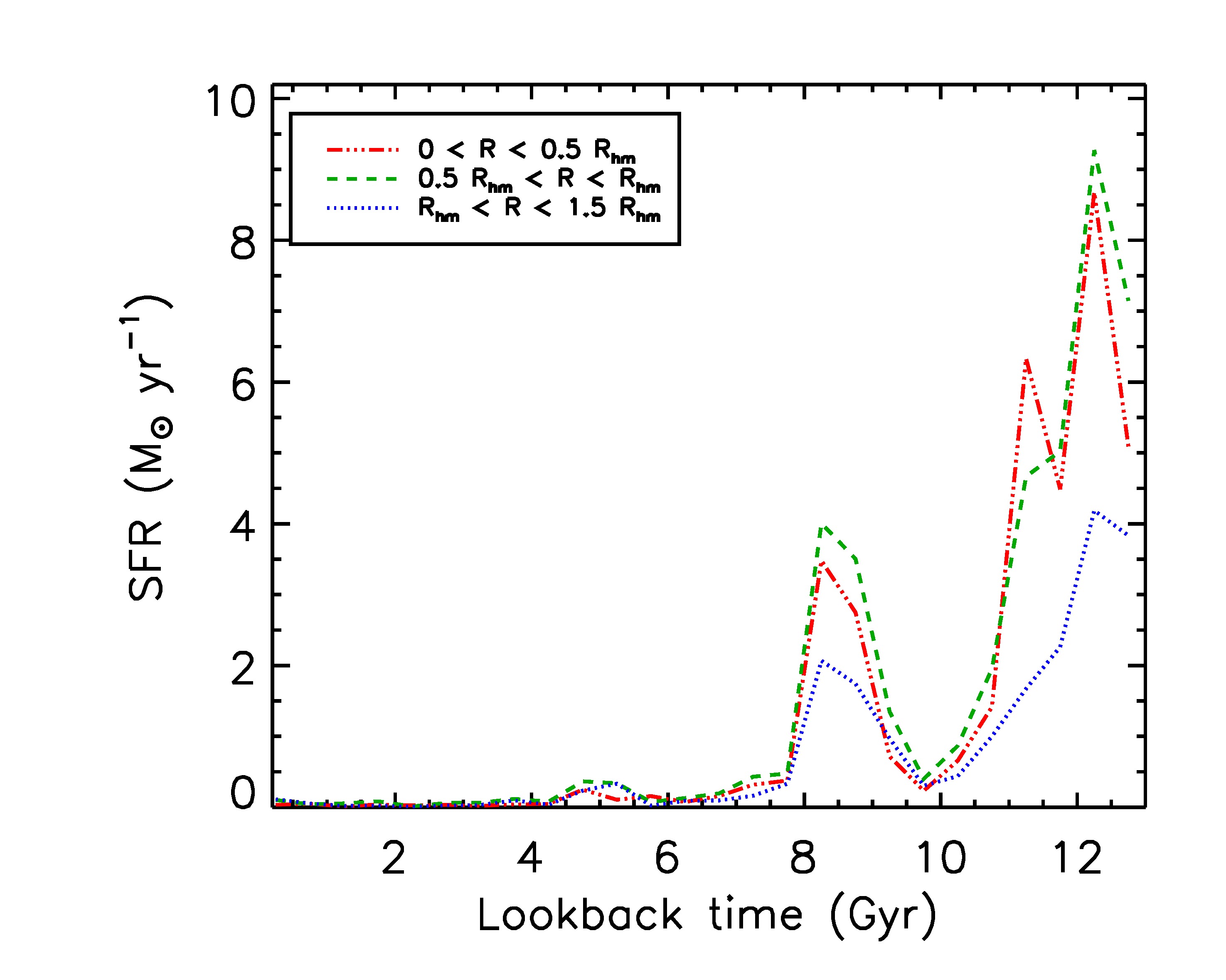}
  \includegraphics[width=0.45\textwidth]{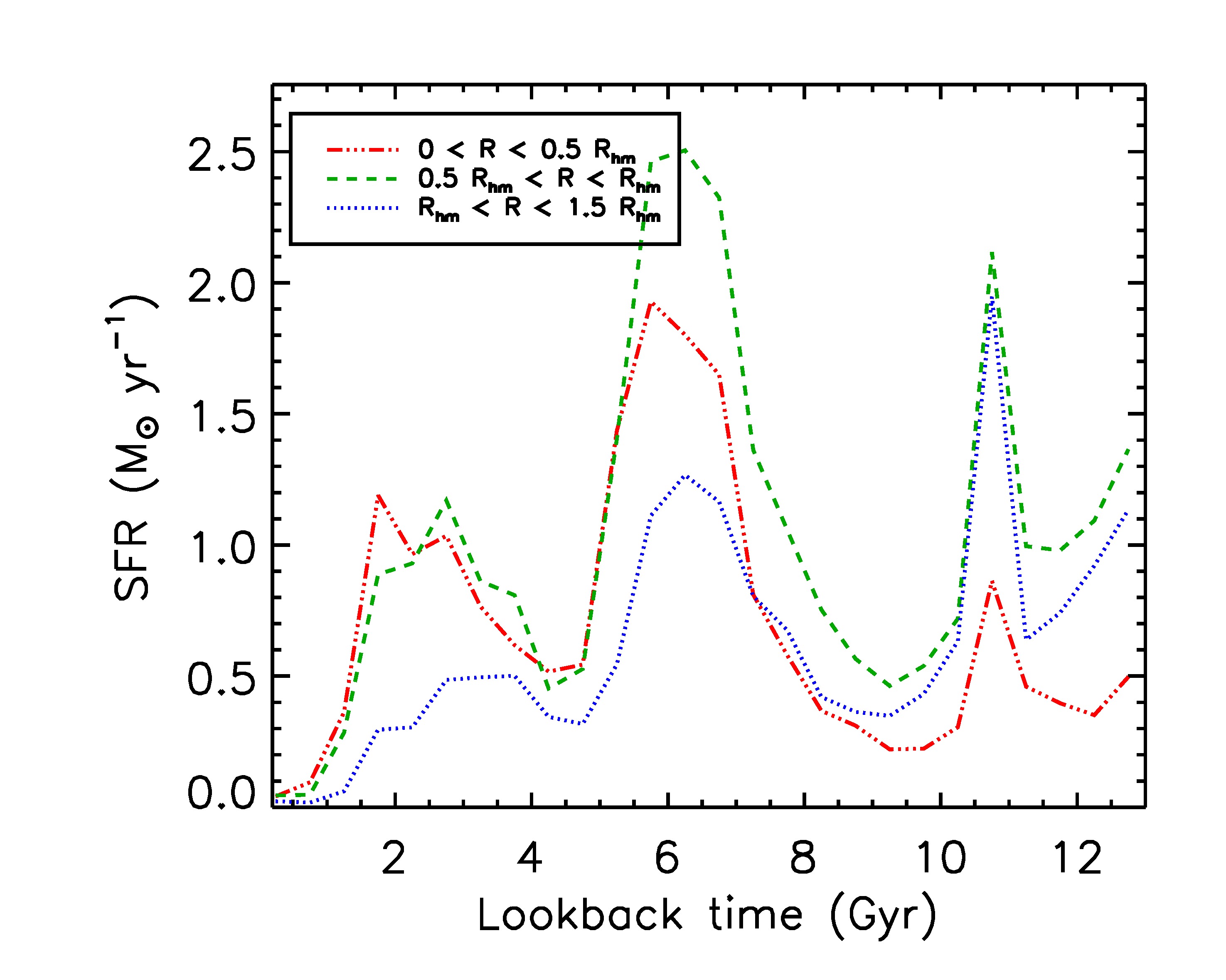}\\
  \caption{Top panels: MGHs for two E-SDGs showing different assembly
    histories: an inside-out (left panel) and an outside-in (right panel).
  These E-SDGs have similar $B/T$ ratios.
} 
  \label{fig:sfr_ind}
\end{figure*}

The complexity of the formation processes of spheroidal galaxies is reflected in the diversity of features and assembly histories
\citep[see e.g.][for recent reviews]{Brooks+2016, Kormendy2016}. 
To illustrate this, Fig.~\ref{fig:mghs_ind} shows the MGHs of two typical E-SDGs   with an inside-out (left panel) and an outside-in (right  panel) growth.
These two galaxies have similar $B/T$ ratios but the first one is more massive (i.e. $M_{\rm Star} \sim 8.2\times 10^{10}$M$_{\odot}$ and $M_{\rm Star} \sim 4.5\times 10^{10}$M$_{\odot}$, respectively).
The galaxy that formed outside-in shows a larger inner disc fraction ($\sim 0.15$ for the outside-in and $\sim 0.07$ for the inside-out galaxy).
In general, galaxies with an outside-in behaviour tend to have slightly larger inner discs fractions: the first and third quartiles of these fractions are 0.11 and 0.18 respectively, whereas the corresponding ones for inside-out galaxies are 0.11 and 0.15, respectively.
Both MGHs show signature of strong starbursts (i.e. sudden changes in the stellar mass in the MGHs). 
From
this figure, we can also see that  bumps are present at all
defined radial intervals, suggesting that mergers could be a
channel of gas fuelling since they can affect the inner and the outer parts of
galaxies by inducing bars and/or accreting material from the incoming systems \citep{sillero2017}. 
In Fig. \ref{fig:sfr_ind}, we show the SFR as a function of the look back time for the inside-out galaxy (left panel) and the outside-in one (right panel).  
It can be seen that in the first one, the main starburst is placed in the inner region at high redshift, whereas in the one with outside-in behaviour, the central region is formed at later epochs. 
We must be aware that we are studying archaeological MGHs, as done in observations, and we therefore cannot determine only by this analysis whether stars were formed in-situ or they were accreted from satellites \citep[see][for an ex-situ and in-situ discussion]{Trayford2019}.

The exploration of the merger trees shows that both have the last mergers event at about $8-8.5$ Gyr ago but the galaxy with the outside-in formation had a massive merger (1:4) while the galaxy with the
inside-out formation had a minor merger (1:10).  \citet{Alpine2018} show that, in the EAGLE simulations massive mergers are efficient mechanisms to trigger the rapidly growth of the central black hole and the
triggering of AGN feedback. This result is also reported in previous numerical works \citep[e.g.][]{Springel+2005,Hopkins+2005,HopkinsQ2010,Capelo2015}. The outside-in formation that we detect some of the E-SDGs, particularly for low mass galaxies might suggest that the SN feedback is not efficient enough
to quench the SF and/or  that AGN feedback might be needed even in galaxies of about $10^{10}$M$_{\odot}$ \citep{Argudo2018}. Moreover, \cite{Manzano2019} provide new observational evidence of AGN activity that would quench SF in low mass galaxies. This is a controversial area of active discussion. In fact  previous numerical works suggest that this activity is mainly suppressed by SN feedback in this kind of galaxies \citep{Dubois2015, Habouzit2017, Bower2017, Angles2017}.

\subsection{Bulge and disc global properties}

\begin{figure}
  \centering
   \includegraphics[width=0.45\textwidth]{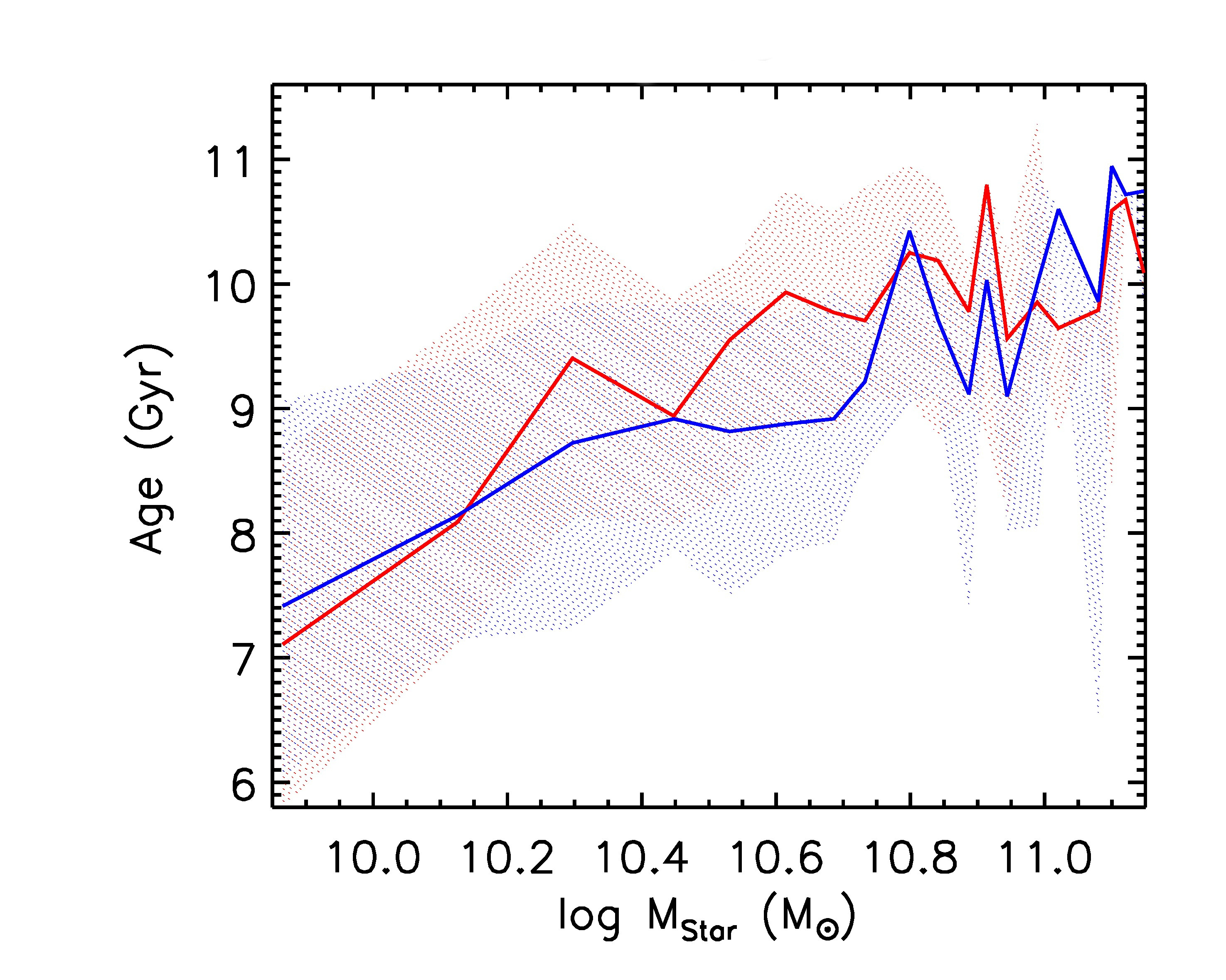}
  \caption{Median stellar ages for bulge (red lines)  and disc
    (blue lines) components as a function of galaxy stellar mass for
    E-SDGs. The shaded regions represent the first and third quartiles.} 
  \label{fig:ages}
\end{figure}

To gain more insight in the assembly process, we explore the global properties, ages and SF efficiency of the bulge and disc components of these galaxies separately.
In Fig~\ref{fig:ages} we show the mean stellar ages for the spheroid and
the disc components in E-SDGs.
As can be seen, the bulge and the disc have similar median ages. For both of
them there is a correlation with stellar mass so that the more massive
galaxies are  dominated by the older SPs (conversely, the
DDGs show a clear age difference between the bulge and the discs, so
that the SPs in the bulges are systematically older by
at least 1.5-2 Gyr in comparison  to those of the discs, Fig.~\ref{fig:agesDDG} as shown in the Appendix).

\begin{figure}
  \centering
  \includegraphics[width=0.45\textwidth]{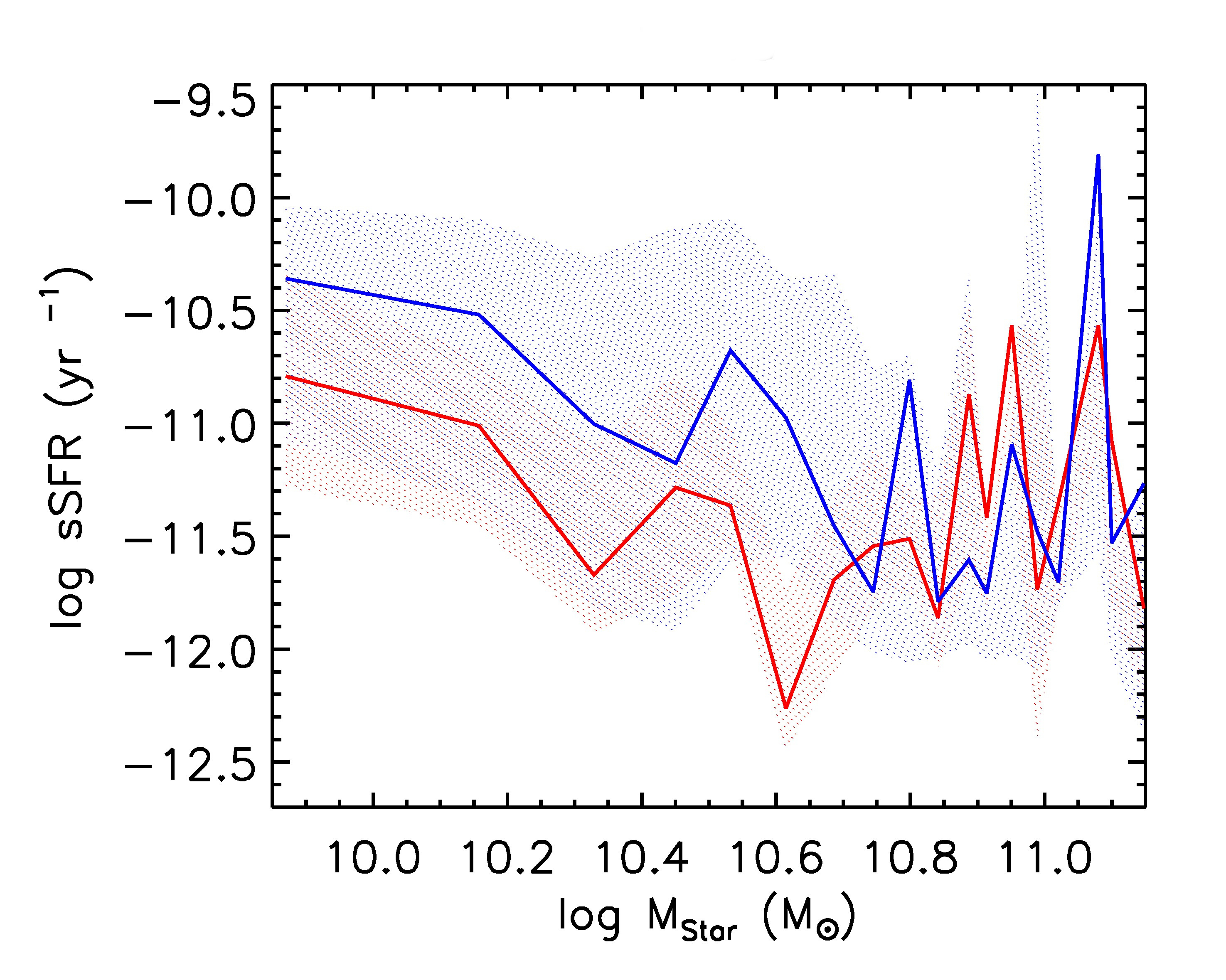}
  \caption{Median sSFR for the bulge (red line) and the disc (blue line) as a function of galaxy stellar mass for the E-SDGs. 
  Shaded regions depict the first and third quartiles. }
   \label{fig:ssfr}
\end{figure}

The SF efficiency of the bulge and the disc components
show similar trends with galaxy mass as  shown in 
Fig.~\ref{fig:ssfr}, although the disc components show systematically larger sSFR as expected.
Smaller galaxies have slightly stronger SF activity. These global trends over the whole disc and bulge
components are in agreement with the analysis of the MGHs.

\subsection{Chemical abundances}

\begin{figure}
  \centering
  \includegraphics[width=0.45\textwidth]{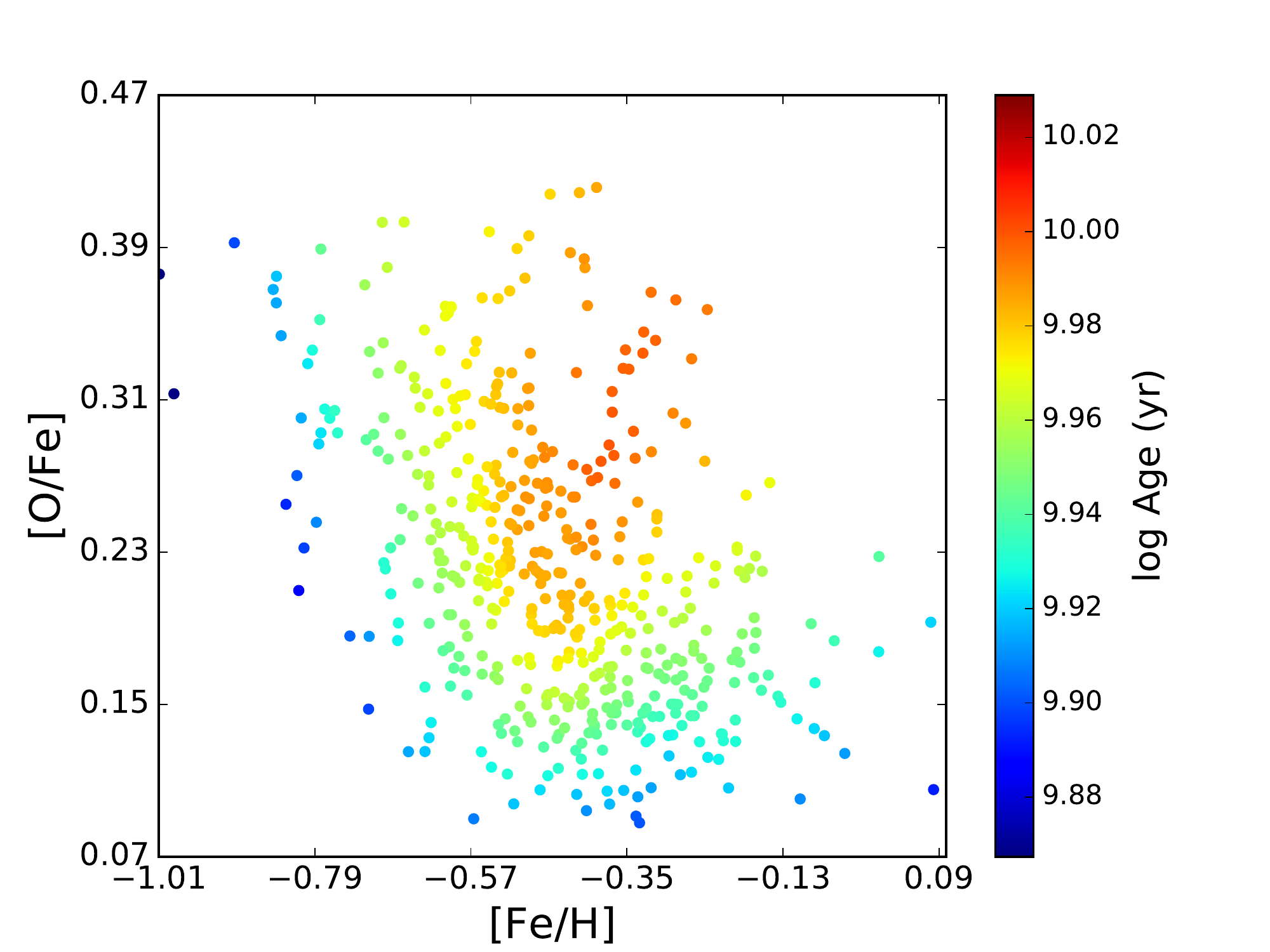}
  \caption{Distribution of median [O/Fe] and median [Fe/H] for the E-SDGs coloured according to median ages of the total SPs. See Appendix \ref{app:plots} for the non-smoothed distribution.} 
  \label{fig:ofefeh2}
\end{figure}

Our analysis indicated that the outside-in behaviour  is triggered by stars that are formed in the inner region at later time.
If these stars are created from enriched gas by  earlier SPs, then they could have different chemical abundances.
Hence, we investigate their chemical abundances at different radial distances and in relation with their ages.
In Fig.~\ref{fig:ofefeh2}, the distribution of the median [O/Fe] and median [Fe/H] for the SPs of the E-SDGs as a
function of median age of their SPs  are shown.

The median abundances for each E-SDGs are estimated by using the individual values for
each single SP (i.e. represented by a star particle) within
the $R^{3D}_{\mathrm{hm}}$.  Hence, the SPs
in the very outer regions are not included.

Overall, stars in the E-SDGs are
$\alpha$-enhanced as expected. The [O/Fe] decreases with increasing [Fe/H].
As can be seen from this figure, massive galaxies dominated by old stars have [Fe/H] in the range $[-0.6,-0.2]$ dex and are $\alpha$-enriched \citep[see also][]{Vincenzo2018}. This is consistent with the fact that more massive
galaxies  tend to form stars in  strong starbursts that also quenched the subsequent SF (via stellar or AGN feedback). E-SDGs with
younger stars are distributed in a type of U-shape: those with higher
[O/Fe] are located at low metallicities while poor-$\alpha$ systems move
to higher metallicities. These trends are consistent with a larger variety in
assembly histories for  smaller galaxies and SF activity (with more than one burst). 

\begin{figure}
  \centering
  \includegraphics[width=0.45\textwidth]{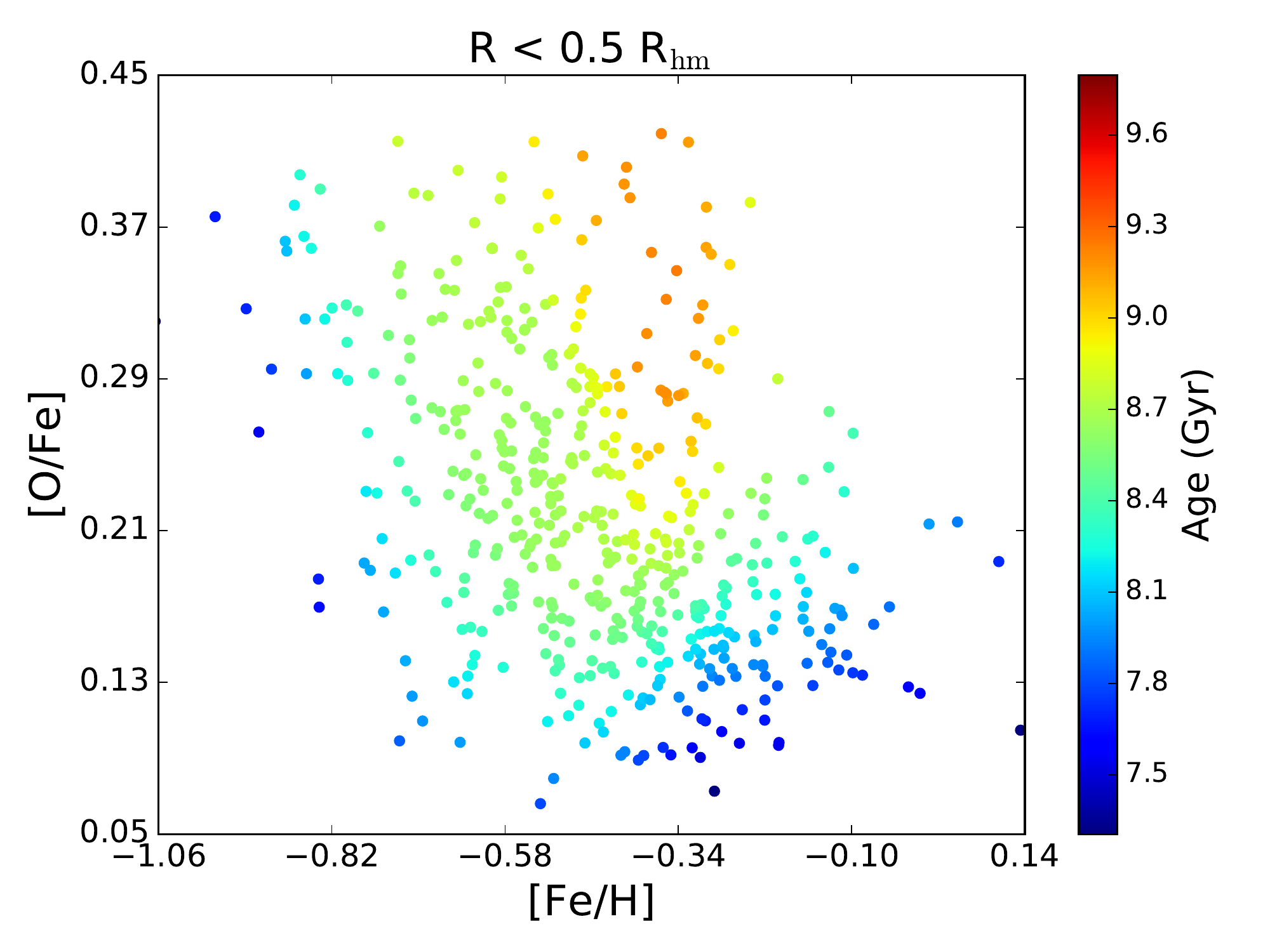} \\
  \includegraphics[width=0.45\textwidth]{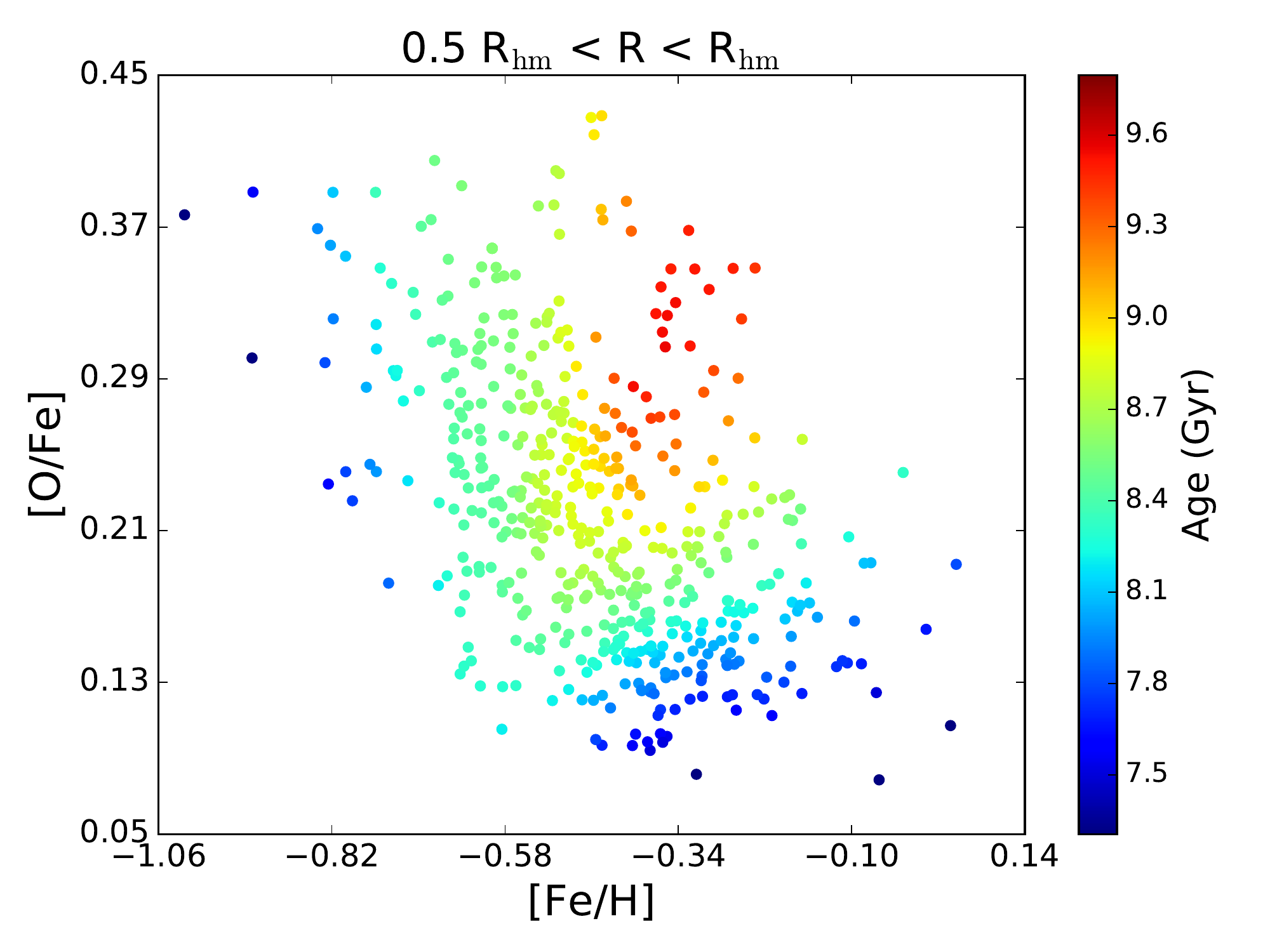} \\  
  \includegraphics[width=0.45\textwidth]{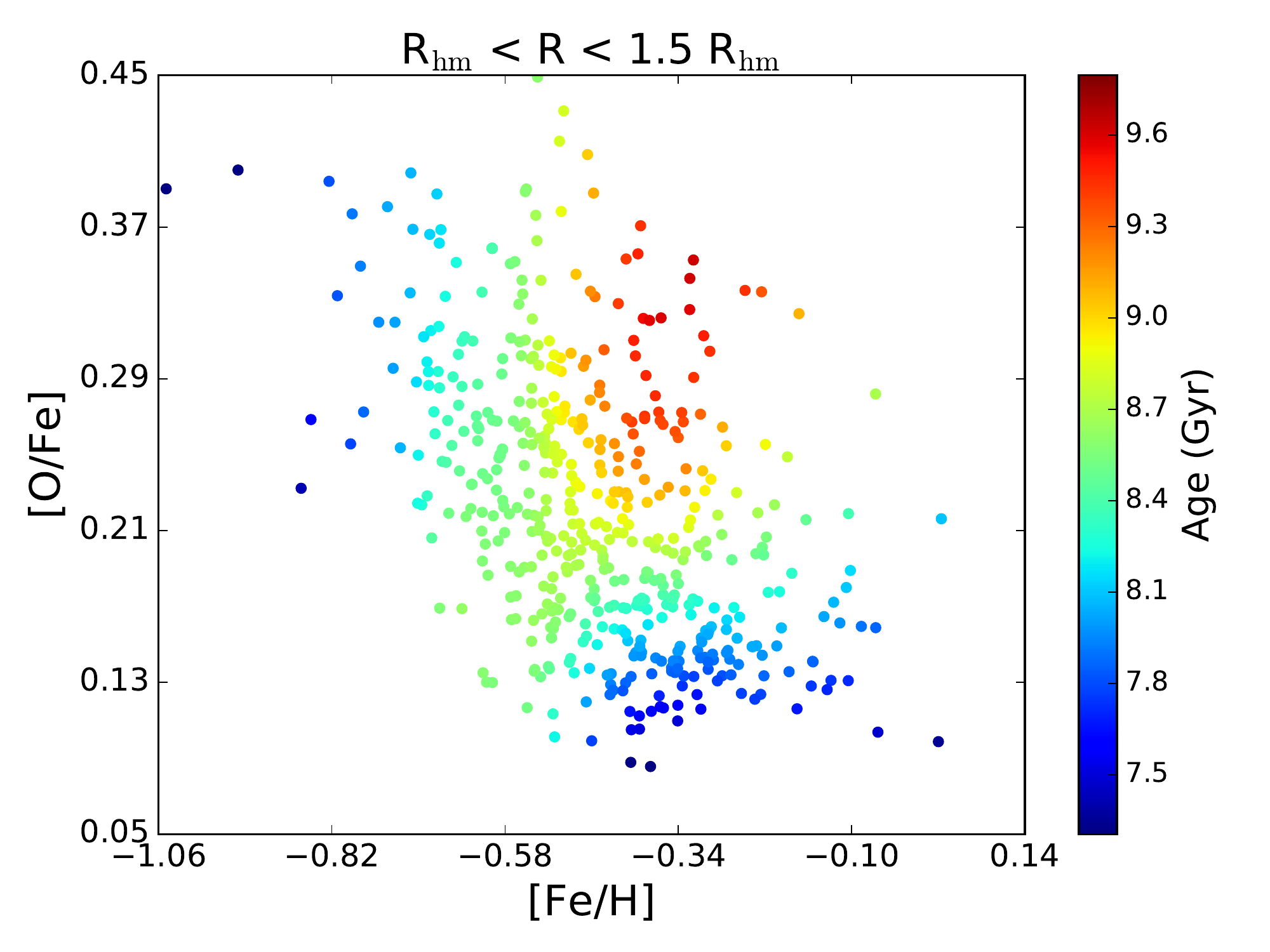}
  \caption{Distribution of  median [O/Fe] and median [Fe/H] within each radial bin. Symbols are coloured according to mean ages. See Appendix \ref{app:plots} for the non-smoothed distribution.}
  \label{fig:ofefeh_bin}
\end{figure}

In Fig. \ref{fig:ofefeh_bin} we show the distributions of [O/Fe] and [Fe/H] calculated within each radial bin considered in Fig. \ref{fig:mghs}.
There is a slight trend for lower  [O/Fe] at low [Fe/H] for SPs in the inner radial bin. This suggests that a fraction of the inner
stars formed from pre-enriched material compared to the outer parts. From this plot, we can also see that the outer radial regions have a larger
contribution of old and $\alpha$-rich stars.
These behaviours could be explained by the later accretion of gas into the central regions via secular evolution and/or the contribution of
old, high $\alpha$-enriched stars acquired by accretion of small satellites \citep{Clauwens2018}. 
We acknowledge the fact that there is a large dispersion in the abundance relations that prevents a robust conclusion. This analysis is intended to
provide a description of the chemical abundance distributions which can serve as basis for future improvements.

\section{Conclusions}
\label{sec:conclusion}

We have performed a comprehensive study of a sample of well-resolved numerical spheroid dominated galaxies from the EAGLE simulation.
The features and behaviours of these systems are confronted with previous results.
A galaxy is considered a SDG if $B/T>0.5$.

We find the following results:

\begin{enumerate}[(i)]

  \item{In agreement with \cite{Rosito2018}, all E-SDGs are found to host   disc component. Part of these discs co-exists with the bulge.
  The mass fraction of inner disc over the bulge anti-correlates with $B/T$ ratio. }
 
  \item{Most of E-SDGs have S\'ersic indexes $n< 2$ that increases with $B/T$ ratio in agreement with observations \citep{Fisher2008}.
  Furthermore, we find that galaxies with larger $B/T$ ratios are more massive and have older SPs.}

\item{E-SDGs follow  scaling relations such as the FJR and FP,  in agreement with observations \citep{Cappellari2013, CappellariAtlasXX}.
  Regarding the size-mass relation, there is also a reasonable agreement with observations, although E-SDGs with lower $B/T$ ratios tend to be more extended.
An exhaustive analysis of the size-mass relation is given by Rosito et al. (2019, accepted for publication).}
  
  \item{By analysing the relation between shape and kinematics, we find trend that agree with results by \cite{Emsellem2011}, who report that a significant fraction of  ETGs tend to be  fast rotators. 
  The relations between shape and kinematics indicators are consistent with observations from ATLAS$^{3{\mathrm{D}}}$. We find a trend for less oblate E-SDGs to have older SPs and to be slower rotators in global agreement with observations. The median stellar ages estimated for the E-SDGs are within the range [3.4, 12.7] Gyr with the median value at $\sim 8.7$ Gyr.
These ages are older than reported by observations in agrement with the results found by \citet{vdSande2019}.}
  
  \item{ The MGHs of the E-SDGs suggest different assembly histories according to the stellar mass of the galaxies in global agreement with the picture described by \citet{Clauwens2018} for the complete sample of EAGLE galaxies. 
  In general, there is a large variety of MGHs for the galaxies analysed in this work and in many of them there are signals of mergers and starbursts.
E-SDGs with stellar masses in the range  $10^{10.5} \mathrm{M}_{\odot} <M_{\mathrm{Star}} < 10^{11.2}  \mathrm{M}_{\odot}$ have MGHs  consistent with almost coeval SPs, with a slight trend for the inner regions to be $\sim 0.5$ Gyr younger
than the outskirts. E-SDGs with $B/T>0.7$ are consistent to have coeval SPs that are systematically older than those in systems with $B/T<0.7$ in the same mass range. 
For less massive galaxies, there is an increasing contribution of younger SPs within the  $0.5 \ R_{\mathrm{hm}}$, that translates into an age gap between
the median age of the inner SPs and that of the outer SPs. The age gap increases from $\sim 0.5$ Gyr for the most massive analysed E-SDGs to $\sim 2$ Gyr for the smallest ones. The outside-in assembly of ETGs with 
$M_{\mathrm{Star}} < 10^{10.5} {\rm M_{\odot}}$ is in tension with current observational findings.}

\item {We find that, at a given median [Fe/H],   more massive and older E-SDGs have higher [O/Fe] abundances. E-SDGs with
$M_{\mathrm{Star}} < 10^{10} \mathrm{M}_{\odot}$ determine a U-shape relation showing a larger variety of chemical abundances in [Fe/H] vs [O/Fe]. 
 The outside-in scenario is consistent with the chemical abundances estimated for SPs at different radial distances.
The larger values of [O/Fe] in the outer regions are consistent with stars formed in short starbursts so that the pristine gas could not have been enriched with iron due to SNIa. This trend supports the suggestion that part of the
outer stellar enveloped is formed by accreted material  \citep[e.g.][]{Clauwens2018}. On the other hand the slightly younger stars with lower  [O/Fe] in the inner regions indicates that they might have formed from pre-enriched gas.  }

\item{As mentioned before, all selected E-SDGs have a disc component, even if it represents a small fraction of the total mass. The disc concentrates most of the SF activity which is preferentially located in the central regions.
This is not the case for those galaxies dominated by discs. We find that E-DDG  (i.e. $B/T<0.5$) to  show a clear ordered MGHs consistent with an inside-out scenario. Hence, galaxies that reach $z=0$ as spiral galaxies have managed to grow orderly in an inside-out fashion (see Appendix \ref{app:all}).
We speculate that the following mechanisms could have played a role at modulating the assembly of ETGs.  On one hand, the existence of discs in the  E-SDGs makes them more susceptible to the effects of secular evolution that can drive gas inflows and feed SF in the central region. Stronger SN feedback could help to solve this problem.
AGN feedback might be also  needed to act in smaller galaxies as suggested by recent observational results \citep{Argudo2018, Manzano2019}. However,
a number of numerical results claimed for  less important role of AGN feedback at low mass \citep[e.g.][]{Dubois2015, Angles2017}. Hence, this is still a controversial issue
in galaxy formation.
On the other hand, considering the results of \citet{Clauwens2018} that showed a contribution of ex-situ stars at $r > 5 $ kpc, ranging from 10 to 60 per cent, depending on galaxy mass, it might be possible that the accretion of satellites in the outer regions works to enhance the outside-in formation trend as they would contribute with old stars. These minor mergers could also play a role at destabilising an existing disc component. 
This is a very interesting aspect to explore in more detail in a future work.}

\end{enumerate}

\begin{acknowledgements}
PBT acknowledges partial support from Fondecyt 1150334 (Conicyt). The use of
RAGNAR cluster funded by Fondecyt 1150334 and UNAB is acknowledged.

This project has received funding from the European Union Horizon 2020
Research and Innovation Programme under the Marie Sklodowska-Curie
grant agreement No 734374.

This work used the DiRAC Data Centric system at Durham University, operated by the Institute for Computational Cosmology on behalf of the STFC DiRAC HPC Facility (www.dirac.ac.uk). This equipment was funded by BIS National E-infrastructure capital grant ST/K00042X/1, STFC capital grants ST/H008519/1 and ST/K00087X/1, STFC DiRAC Operations grant ST/K003267/1 and Durham University. DiRAC is part of the National E-Infrastructure. We acknowledge PRACE for awarding us access to the Curie machine based in France at TGCC, CEA, Bruyeres-le-Chatel.
\end{acknowledgements}

\bibliographystyle{aa}

\def\apj{ApJ}
\def\apjl{ApJ}
\def\aj{AJ}
\def\mnras{MNRAS}
\def\aa{A\&A}
\def\nat{Nature}
\def\araa{ARA\&A}
\def\aap{A\&A}

\bibliography{eagle}

\begin{appendix} 

\section{Comparison with the disc-dominated galaxies}
\label{app:all}

For comparison purposes, we also analyse some of the relations studied in the main section for all galaxies resolved with
more than 10000 particles. 
This sample comprises 70 per cent of E-DDGs and 30 per cent of E-SDGs.

\begin{figure}
  \centering
  \includegraphics[width=0.45\textwidth]{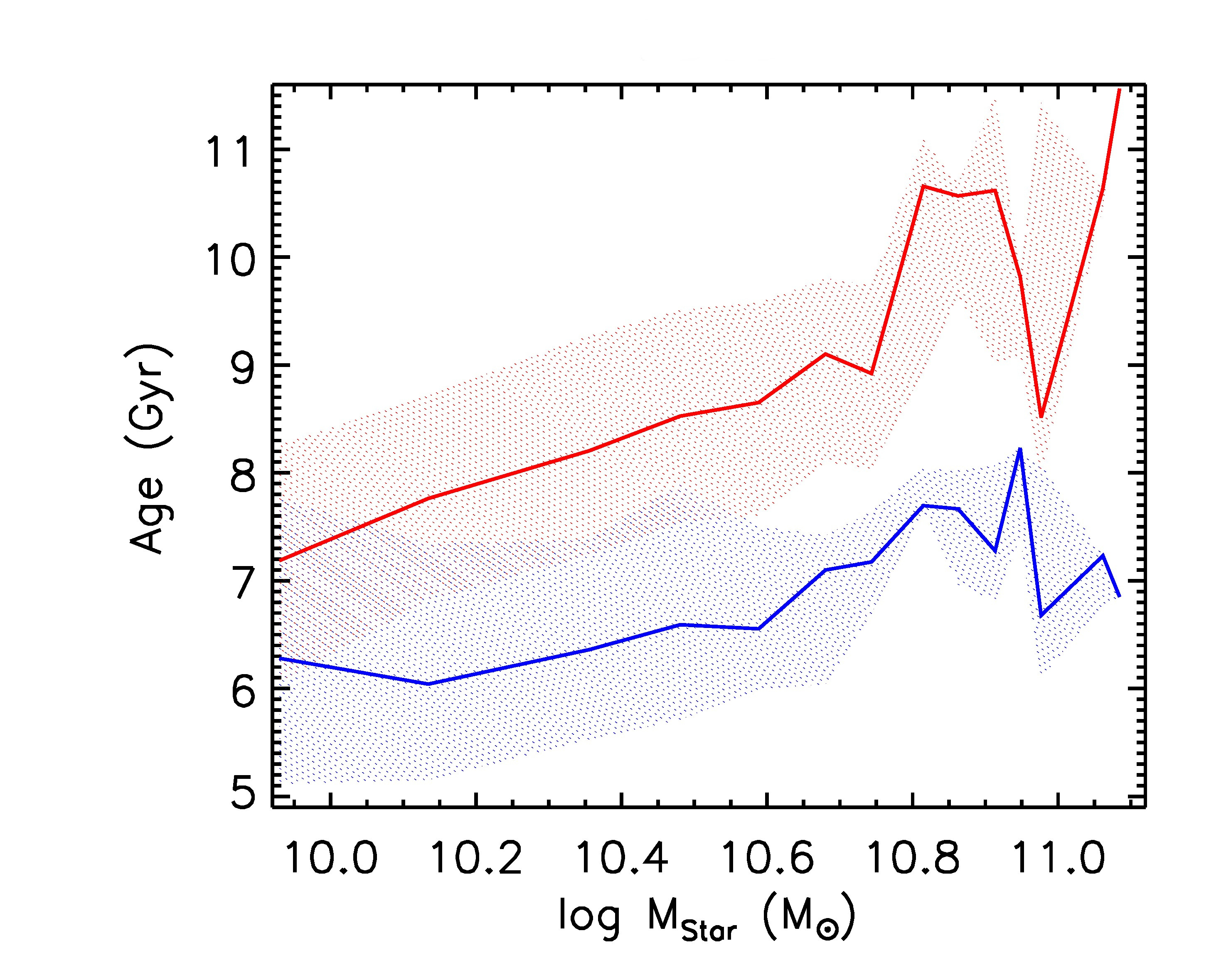}
  \caption{Median stellar ages for the bulge (red lines)  and disc
    (blue lines) components as a function of galaxy stellar mass for
   DDGs. The shadowed regions represent the first and third quartiles.} 
  \label{fig:agesDDG}
\end{figure}

\begin{figure}
  \centering
  \includegraphics[width=0.45\textwidth]{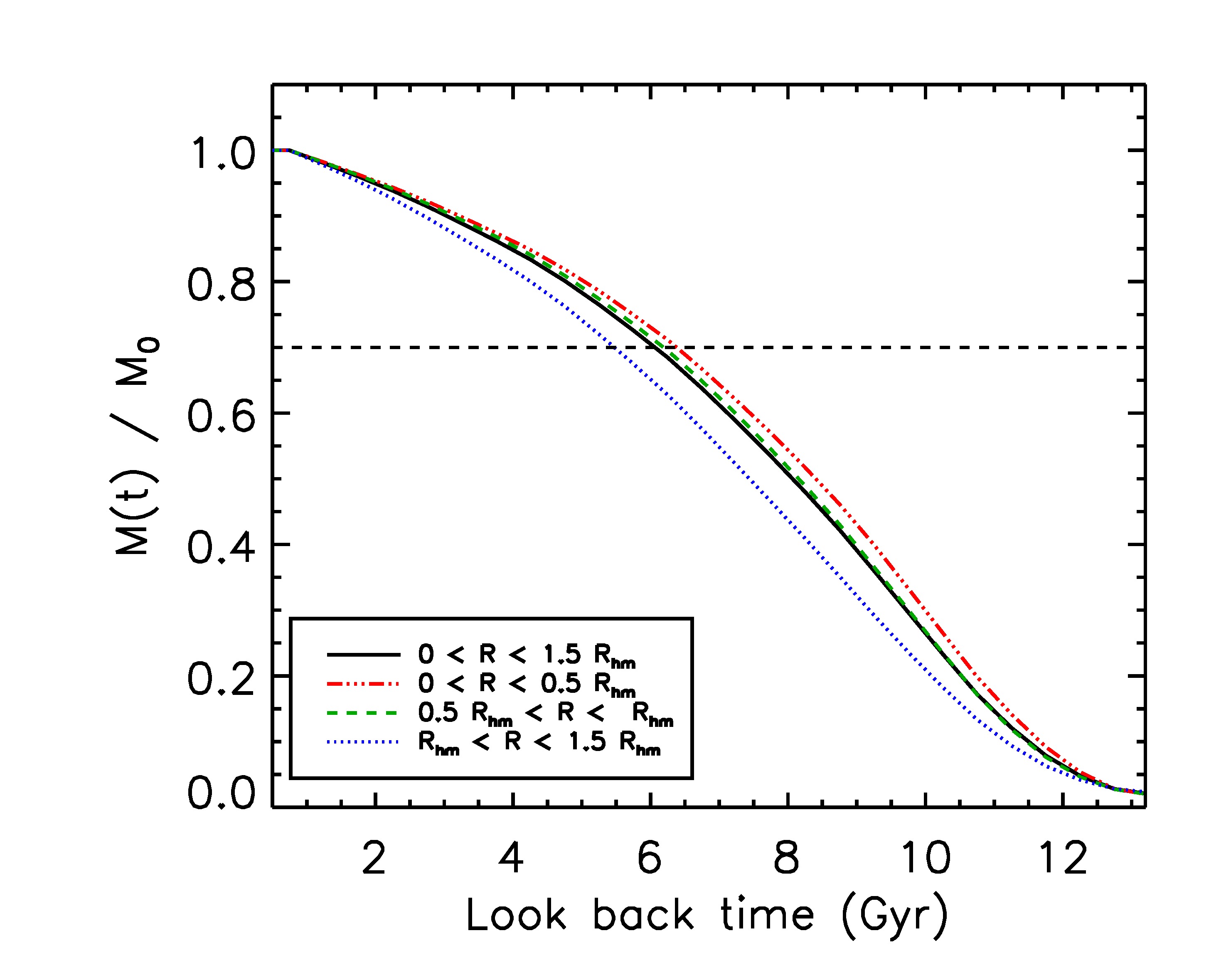}
  \caption{Average global and radial normalised MGH for the DDGs. It presents a clear inside-out growth.}
   \label{fig:mgh_d}
\end{figure}

In Section \ref{sec:mgh}, we emphasise the fact that SDGs in the EAGLE
simulation tend to show an outside-in growth. However as expected this is not the case for
DDGs. 
The inside-out scenario is consistent with previous analysis of the
EAGLE simulations \citep{zavala2016,tissera2019}.
While  no significant  differences between the mean
ages of the bulge and disc components of E-SDGs are found (Section \ref{sec:mgh}),
in the DDGs
bulges are older than discs by about $\sim 2$ Gyr (Fig. \ref{fig:agesDDG}).
The average global and radial MGHs for DDGs can be seen in Fig. \ref{fig:mgh_d}.
The 70 per cent of the stellar mass in the inner, intermediate and
outer radial intervals is attained at
lookback times of 6.5 Gyr, 6 Gyr and 5.5 Gyr, respectively.

\begin{figure}
  \centering
  \includegraphics[width=0.45\textwidth]{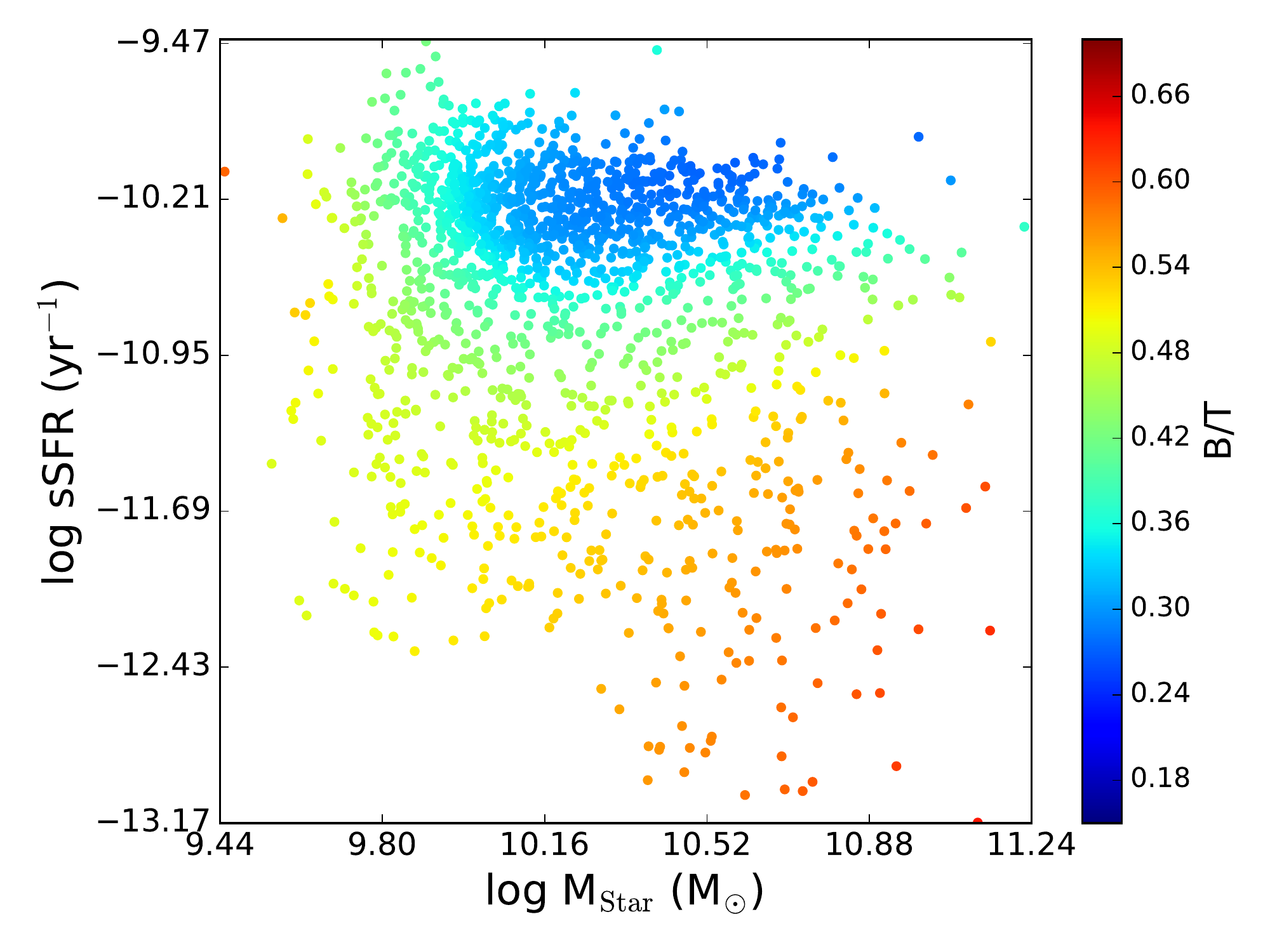}
  \caption{sSFR as a function of stellar mass of the total analysed
    sample. Symbols are coloured according to the $B/T$ ratio. See Appendix \ref{app:plots} for the non-smoothed distribution.}
   \label{fig:ssfr2}
\end{figure}

In Section \ref{sec:mgh}  we find the sSFR of disc and bulge components of E-SDGs are similar. 
However,  when all galaxies in the sample regardless of the $B/T$ ratios
are considered a trend for  higher $B/T$ ratio to have the lower sSFR
as shown in Fig. \ref{fig:ssfr2}.

We remark that in these plots we did not fix the colour-bar limits to the first and third quartiles of the variable according to which we colour the symbols. Instead we use a wider range. 
The reason for this is that most of the galaxies in our sample are DDGs, and therefore have $B/T$ lower than 0.5.

\section{Non-smoothed distributions}
\label{app:plots}

In this Appendix we show the scatter plots of the figures analysed in Sections 3 and 4 without applying the smoothing method of \citet{CappellariAtlasXX} in order to provide
a mean to assess the real distributions.

\begin{figure}[!ht]     
  \centering
  \includegraphics[width=0.45\textwidth]{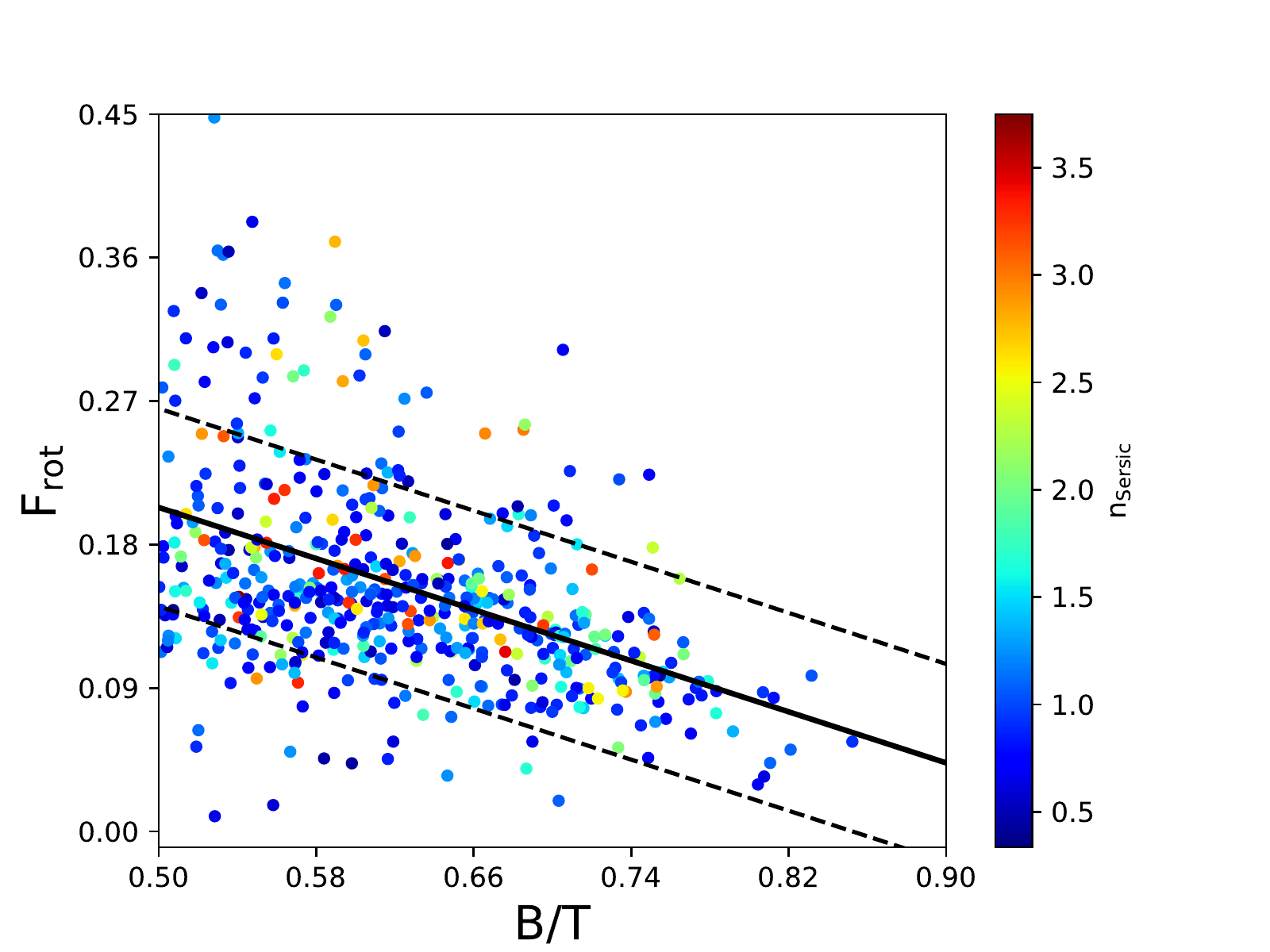}
  \caption{Stellar-mass fractions $F_{\rm rot}$ of the discs that co-exist with
      the spheroidal components as a function of the $B/T$ ratio. 
      Symbols are coloured according to $n$. A linear
    regression fit is included (solid black line) along with its 1$\sigma$ dispersion (dashed black lines).
    In this figure, we do not fix the colour-bar limits.}
  \label{fig:fracfrac}
\end{figure}

\begin{figure}
  \centering
  \includegraphics[width=0.45\textwidth]{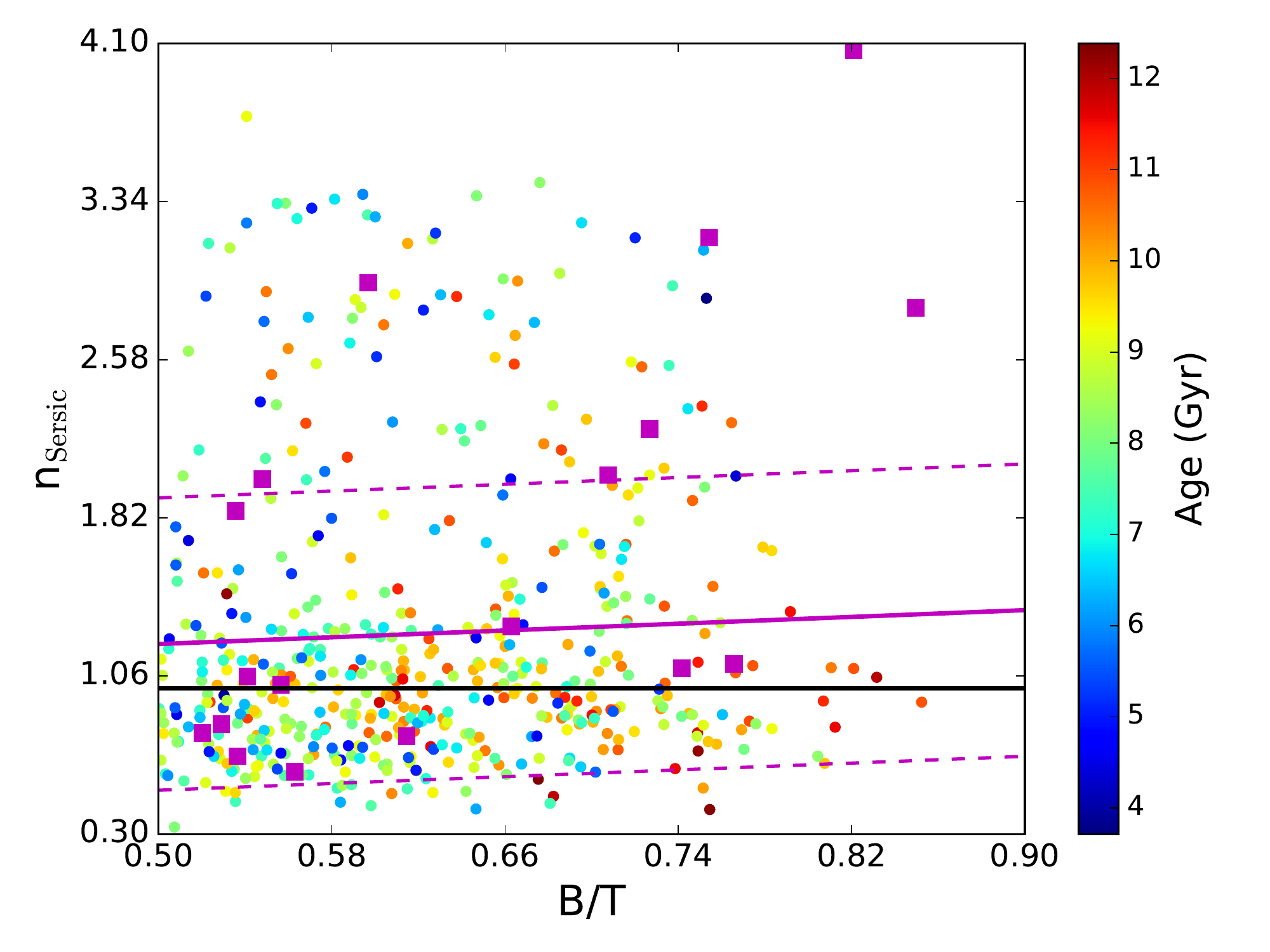}
  \caption{S\'ersic index ($n_{\mathrm{Sersic}}$) as a function of $B/T$ ratio for the simulated E-SDGs, coloured according to the  mass-weighted average
    age of the galaxy stellar mass. A linear regression fit is included (solid magenta line) with its 1$\sigma$ dispersion (dashed magenta line). For comparison, we also include the results by  \citet[][]{Rosito2018} (magenta squares).
    The line $n_{\mathrm{Sersic}}=1$ is depicted in a black line.}
   \label{fig:cuatro}
\end{figure}

\begin{figure}
  \centering
  \includegraphics[width=0.45\textwidth]{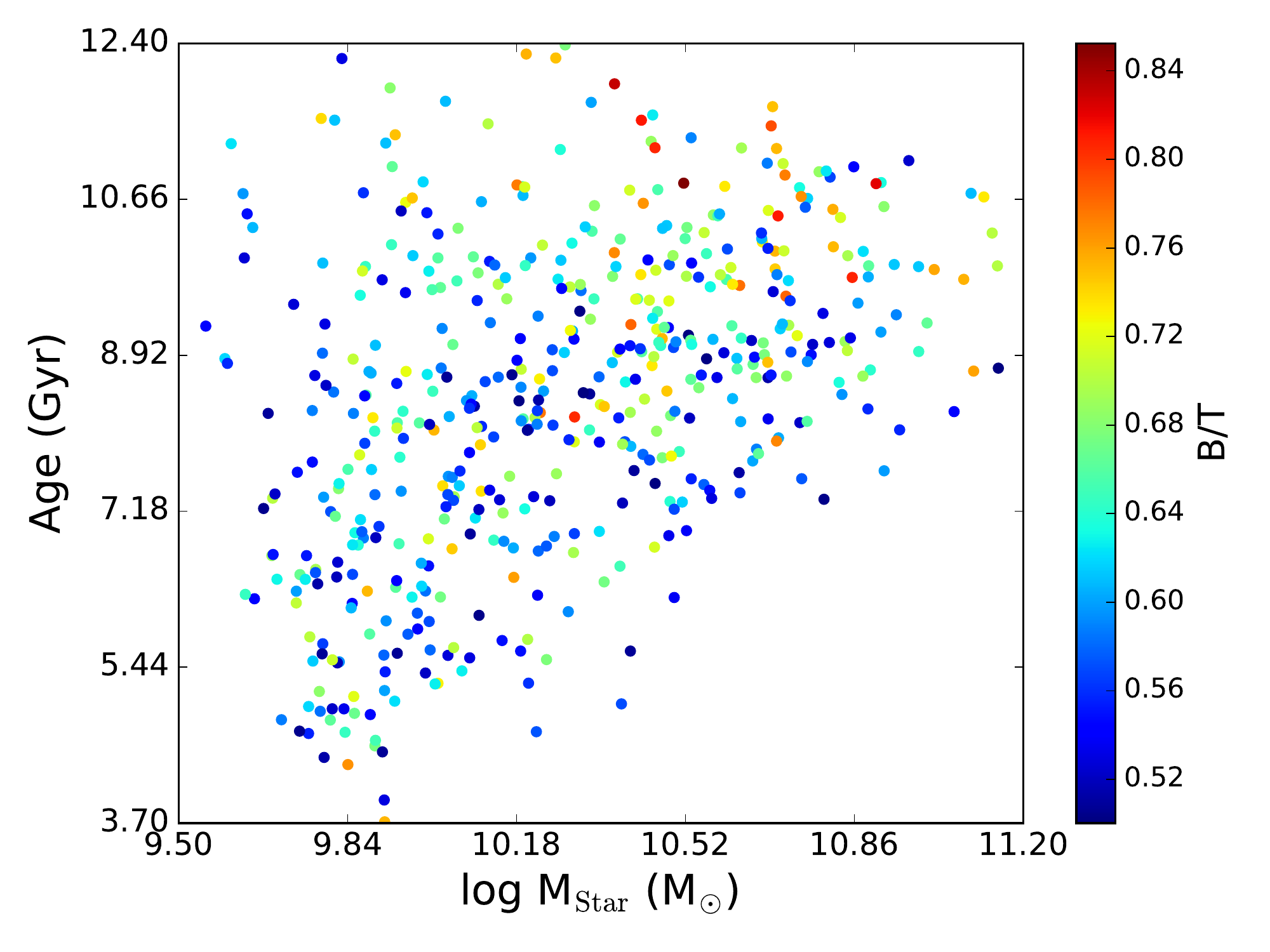}
  \caption{Mass-weighted average stellar age of the E-SDGs (i.e. bulge and disc SPs) as a function of stellar mass for
    the simulated galaxies. The symbols are coloured according to the $B/T$ ratios. }
   \label{fig:cinco}
\end{figure}

\begin{figure}
  \centering
  \includegraphics[width=0.45\textwidth]{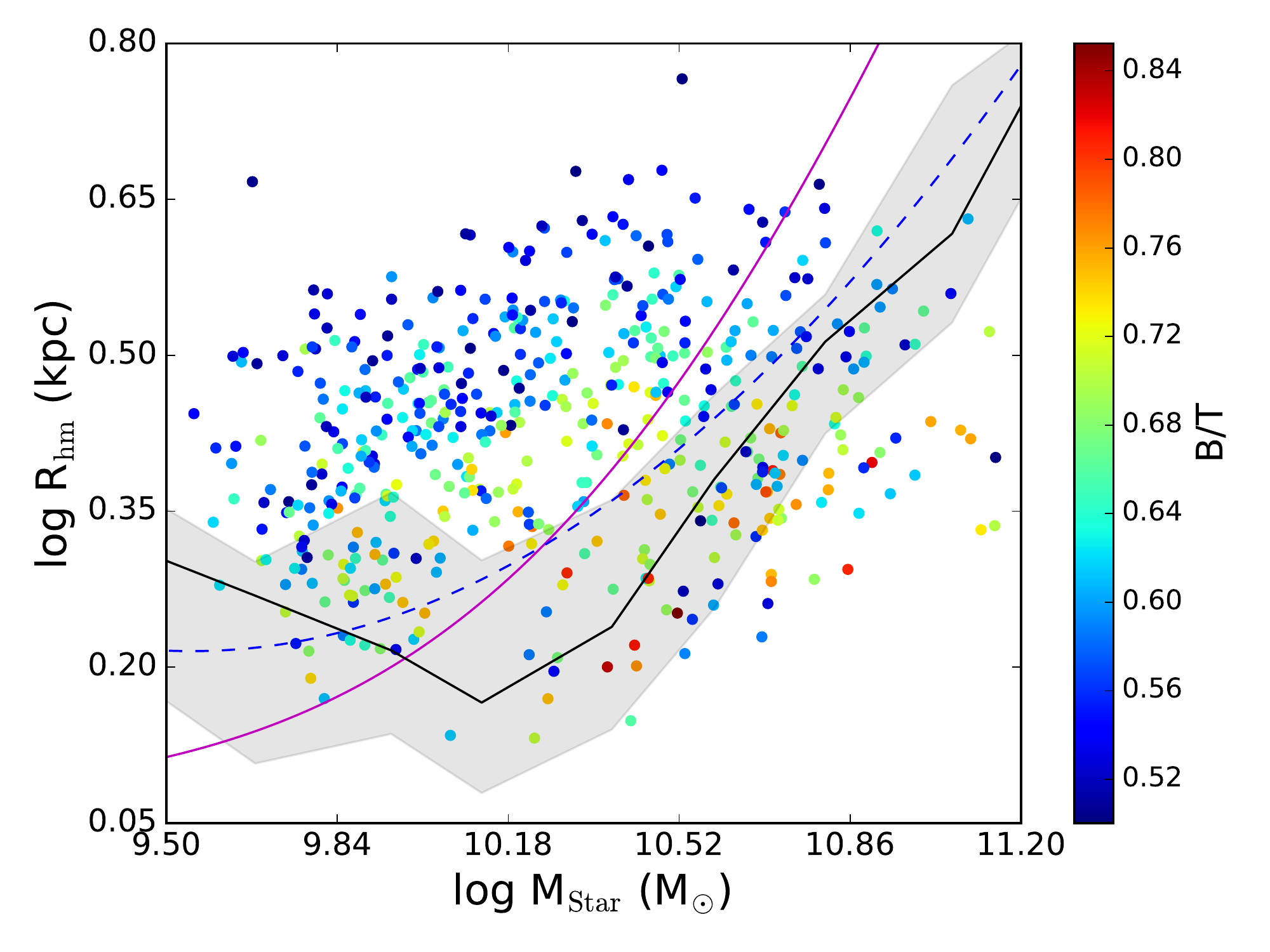}
  \caption{Mass-size relation estimated for  the  E-SDGs. Symbols are coloured according to their $B/T$ ratio.
  The median of the observations from ATLAS$^{3\mathrm{D}}$ (black line; the first and third quartiles  are shown as shadowed region)
and the observed relations for ETGs reported by \citet[][solid magenta line]{Mosleh13} and \citet[][dashed blue line]{Bernardi2014} are included for comparison. }
   \label{fig:seis}
\end{figure}

\begin{figure}
  \centering
  \includegraphics[width=0.45\textwidth]{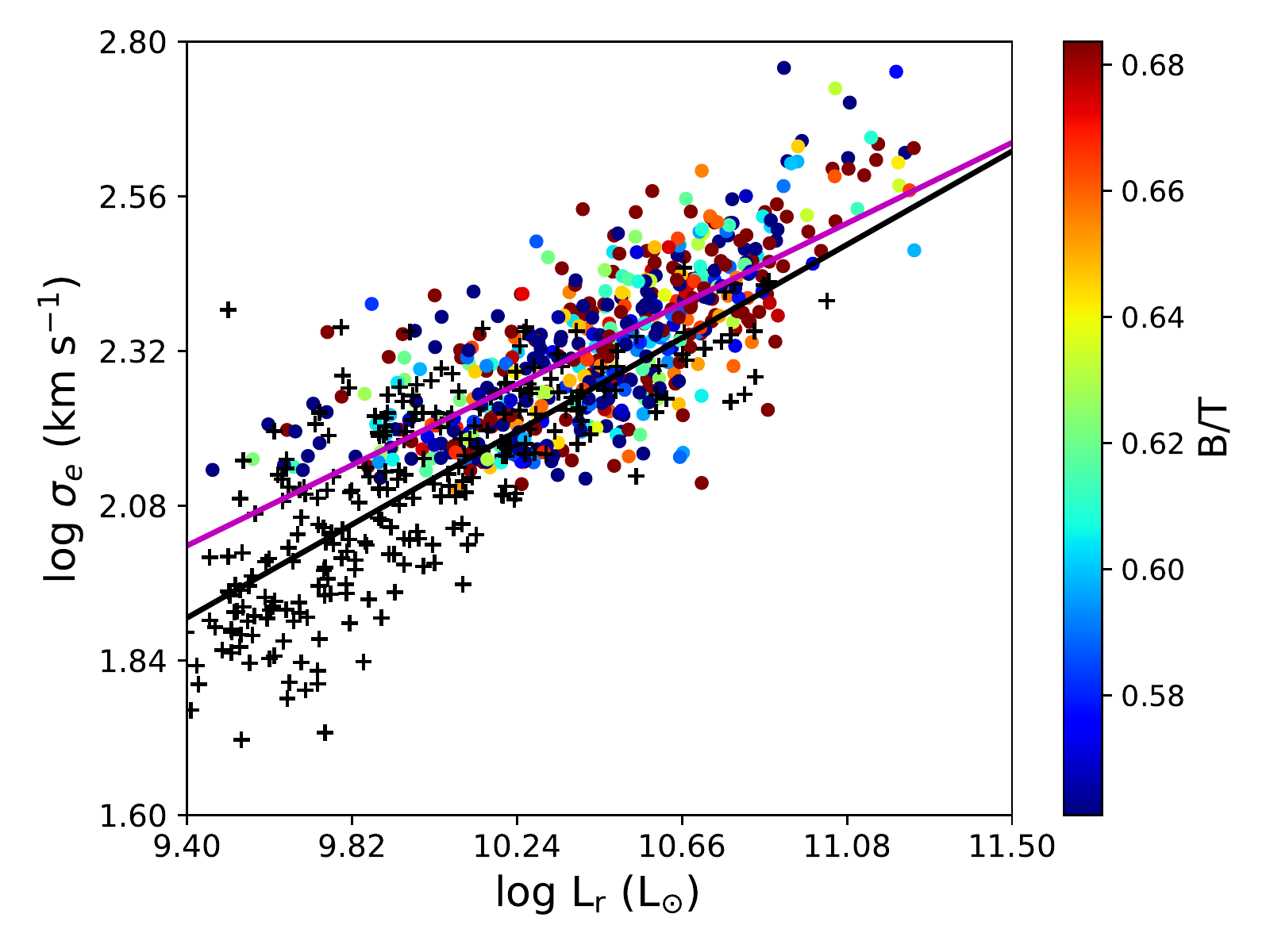} 
  \caption{FJR for simulated E-SDGs (filled circles) and observed ETGs from ATLAS$^{3\mathrm{D}}$ (black crosses) galaxies. The least squared regression lines are included in magenta and black line for simulated and observed data, respectively.}
   \label{fig:siete}
\end{figure}

\begin{figure}
  \centering
  \includegraphics[width=0.45\textwidth]{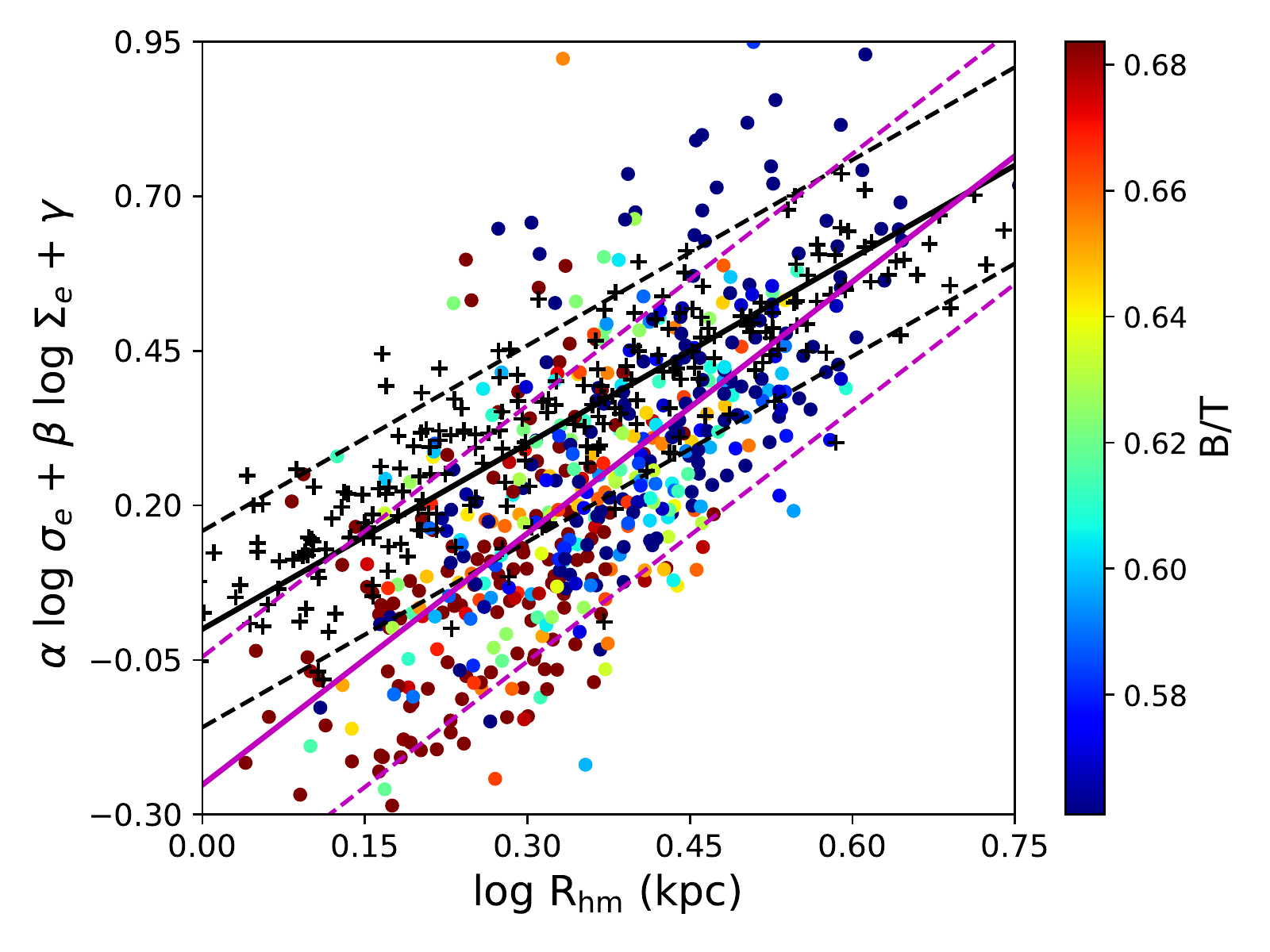}
  \caption{FP for the simulated SDGs calculated with the
    parameters estimated for the  ATLAS$^{3\mathrm{D}}$ sample.
   E-SDGs are also depicted according to their  $B/T$ ratio.
    The black line denotes the one-to-one relation and the magenta line represents the best fit for the E-SDGs.
    In dashed lines we show the \textit{rms} corresponding to the least squared regression and the one-to-one relation.}
   \label{fig:ocho}
\end{figure}

\begin{figure}
  \centering
  \includegraphics[width=0.45\textwidth]{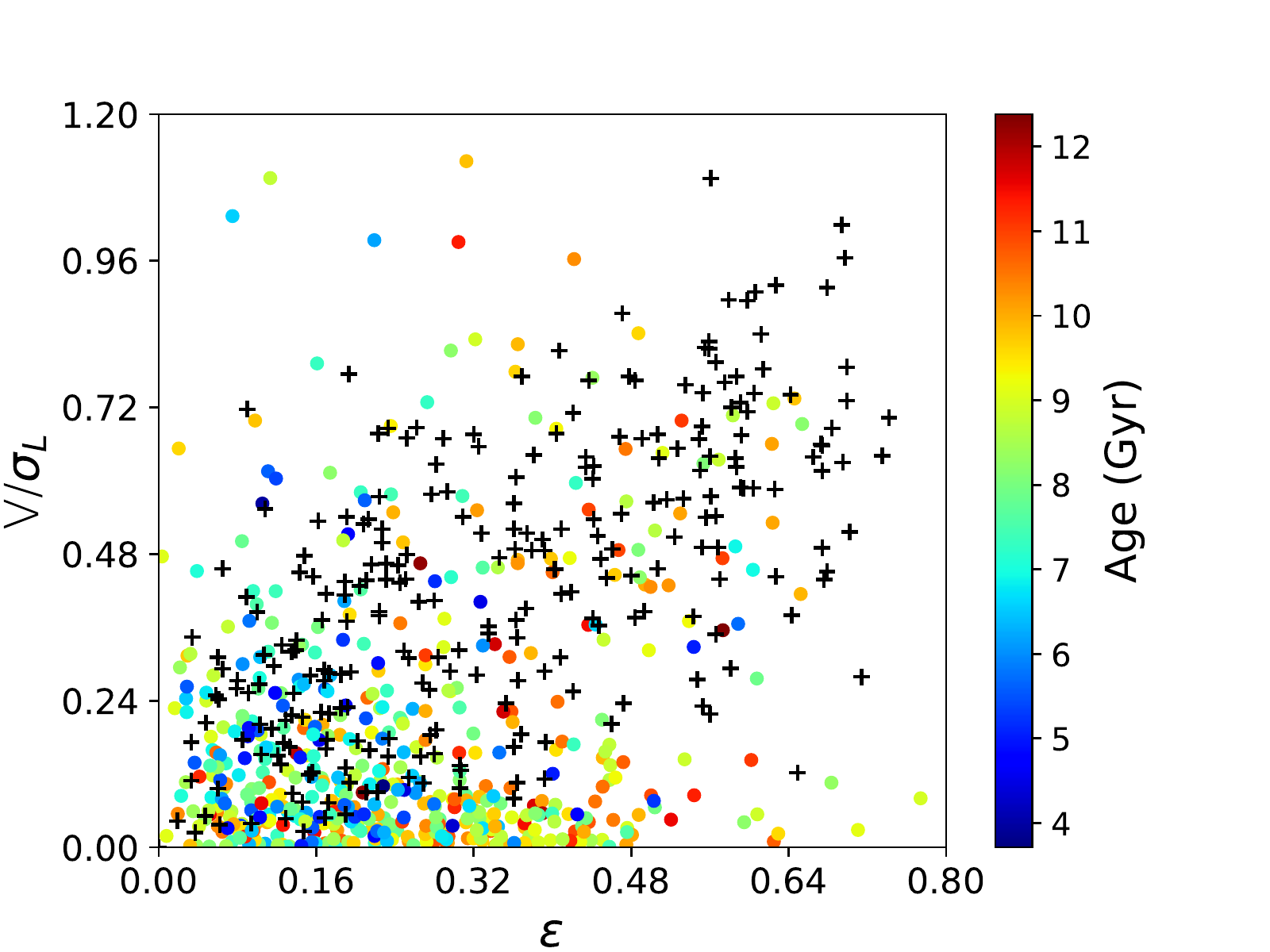} \\
  \includegraphics[width=0.45\textwidth]{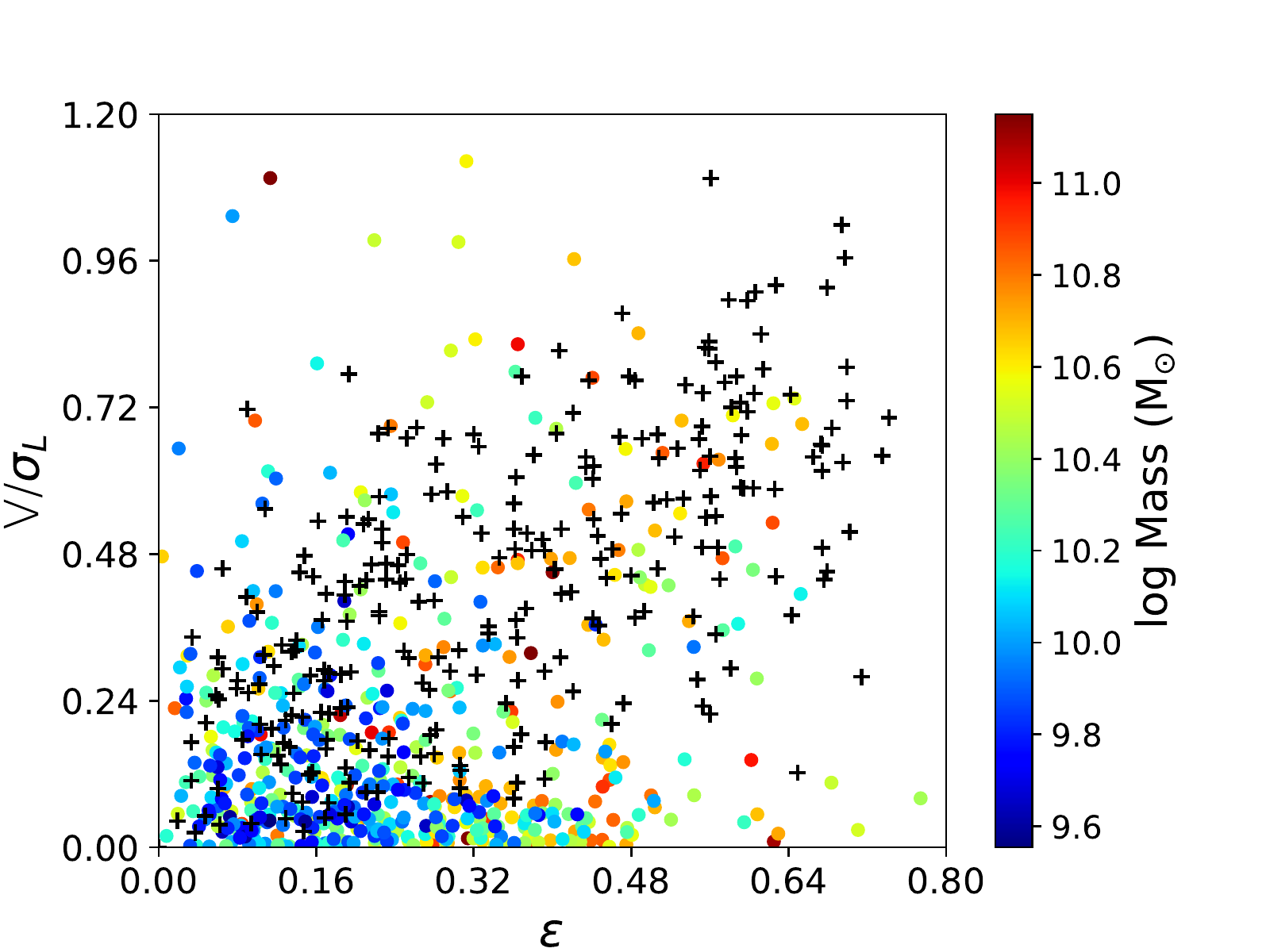} \\
  \includegraphics[width=0.45\textwidth]{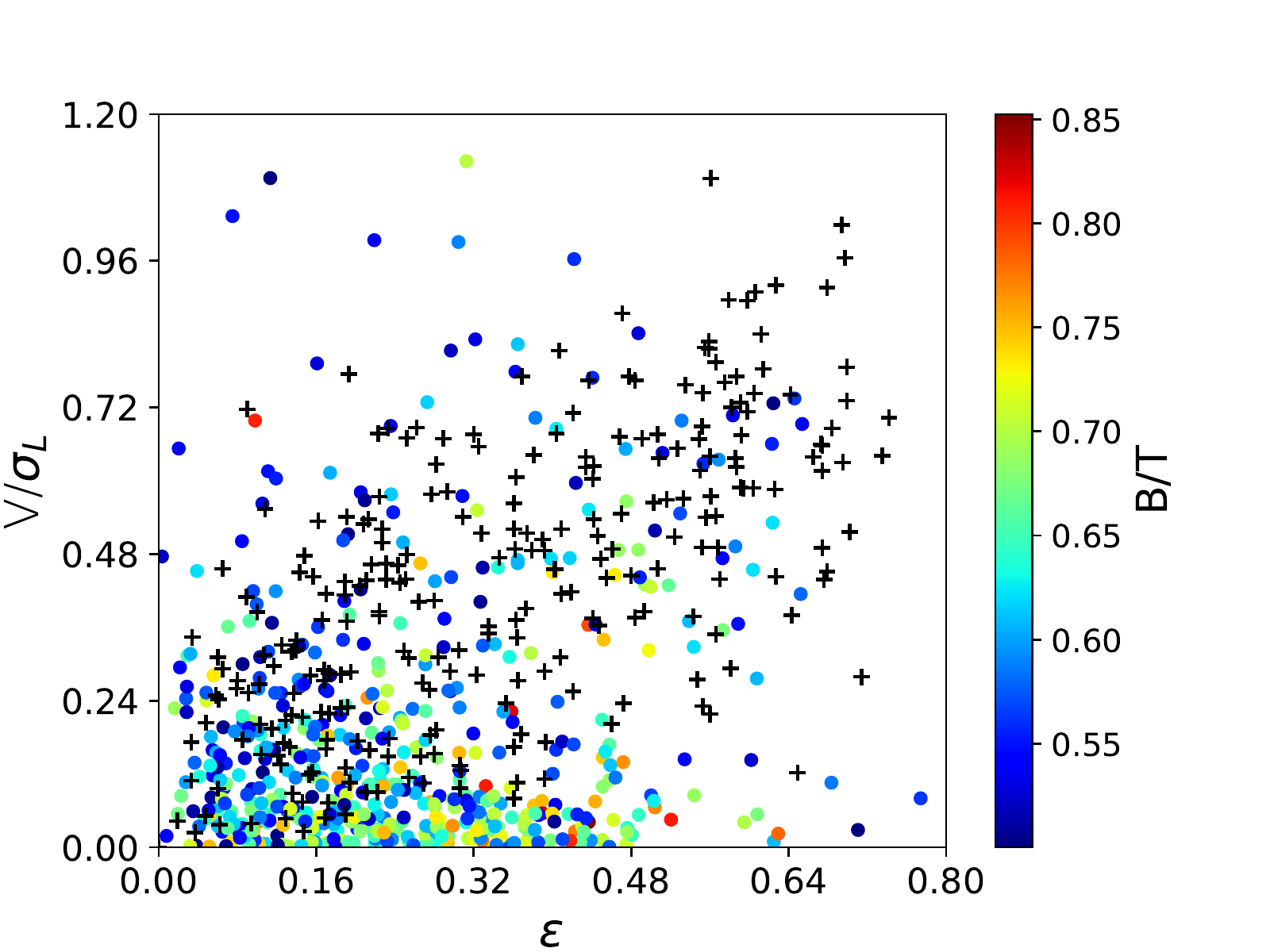}
  \caption{Anisotropy diagram for the E-SDGs. 
  E-SDGs are coloured according to mass-weighted average age (top panel), stellar mass (middle panel) and $B/T$ ratio (bottom panel).
  Observational data from ATLAS$^{3\mathrm{D}}$ are also shown \citep[][black crosses]{Emsellem2011}.  }
   \label{fig:siete}
\end{figure}

\begin{figure}
  \centering
  \includegraphics[width=0.45\textwidth]{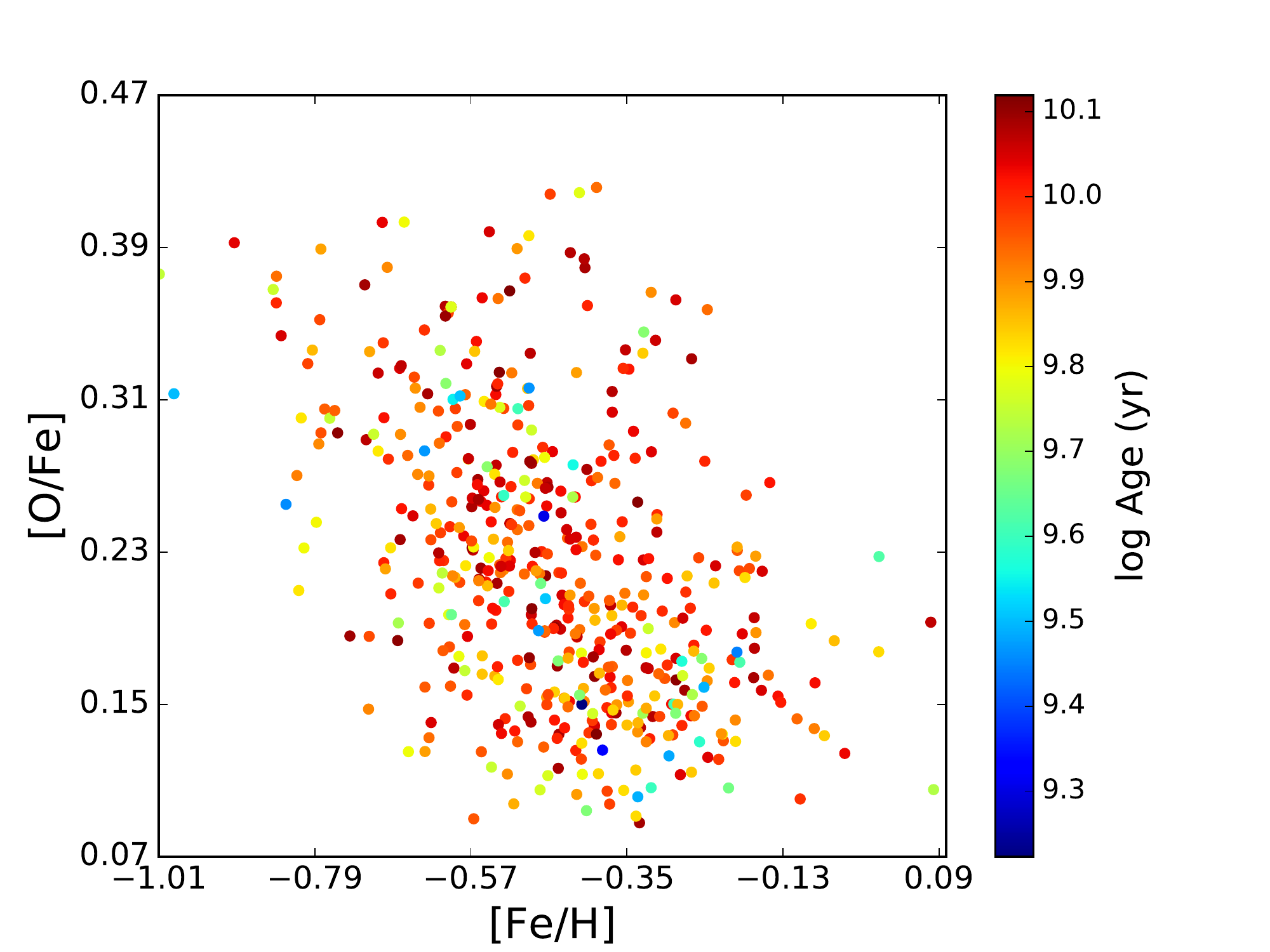}
  \caption{Distribution of [O/Fe] and [Fe/H] for the E-SDGs coloured according to the median ages of the total SPs.}
   \label{fig:dieciseis}
\end{figure}

\begin{figure}
  \centering
  \includegraphics[width=0.45\textwidth]{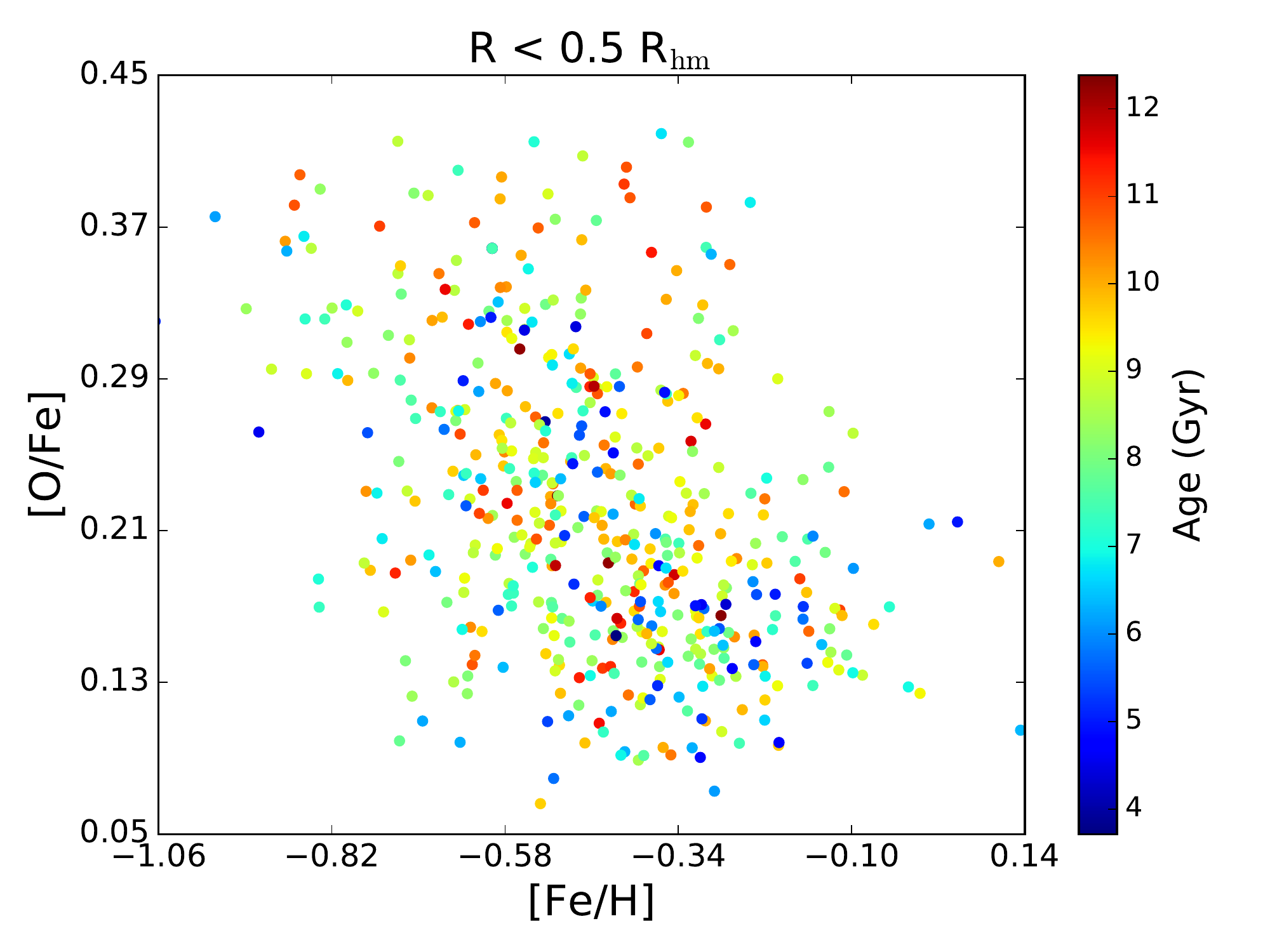} \\
  \includegraphics[width=0.45\textwidth]{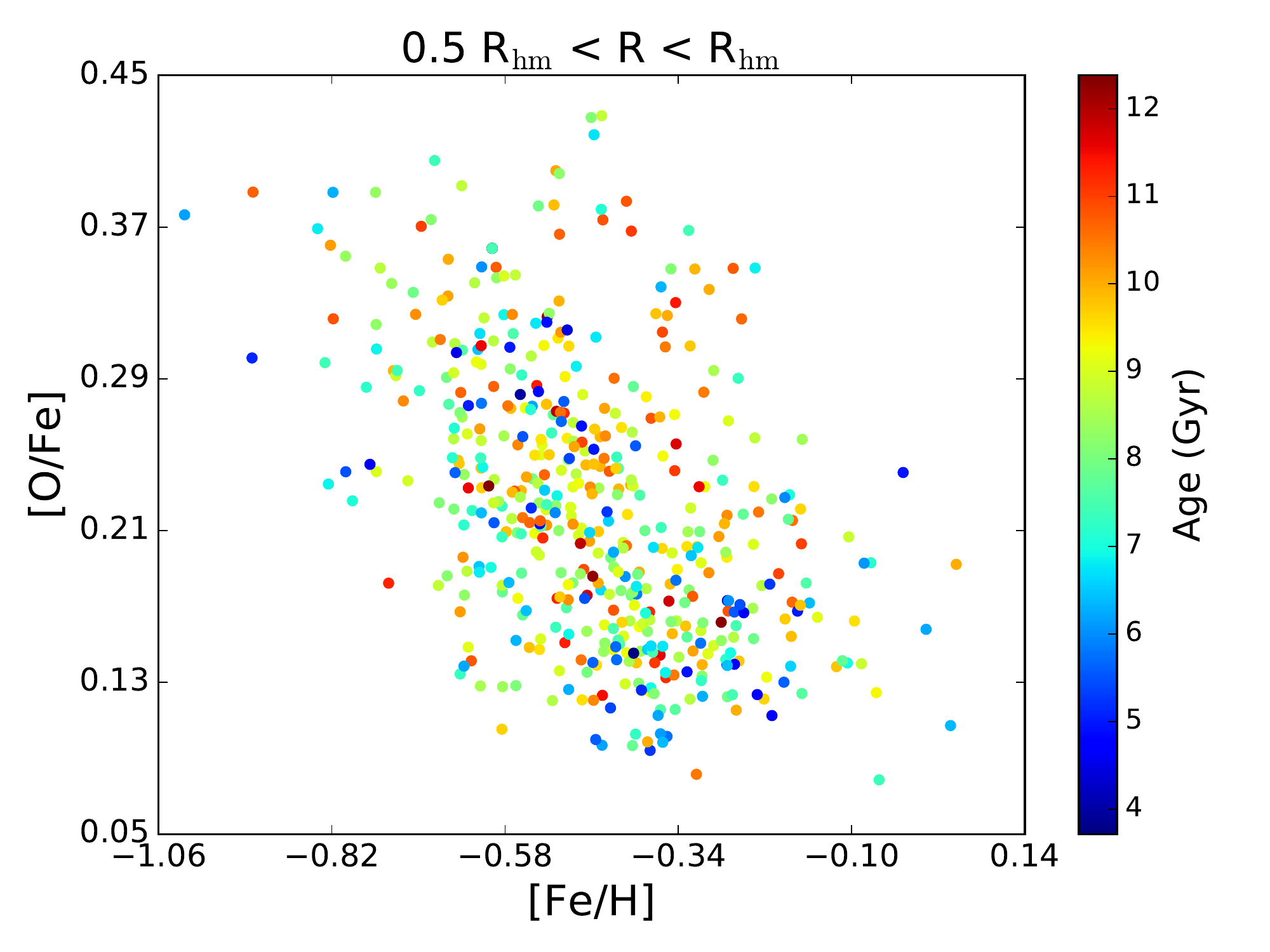} \\
  \includegraphics[width=0.45\textwidth]{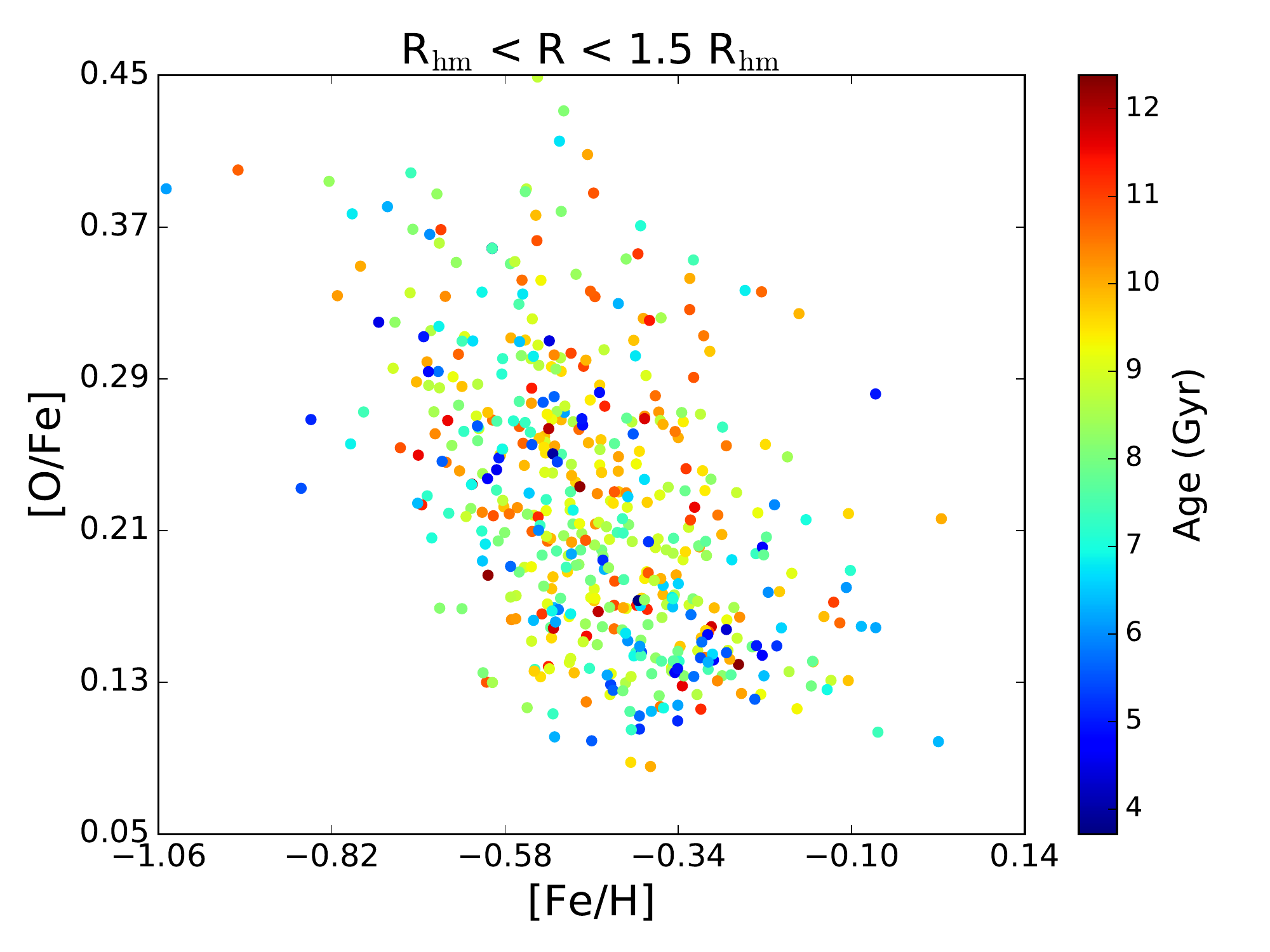}
  \caption{Distribution of [O/Fe] and [Fe/H] within each radial bin. Symbols are coloured according to mean ages. }
   \label{fig:diecisiete}
\end{figure}

\begin{figure}
  \centering
  \includegraphics[width=0.45\textwidth]{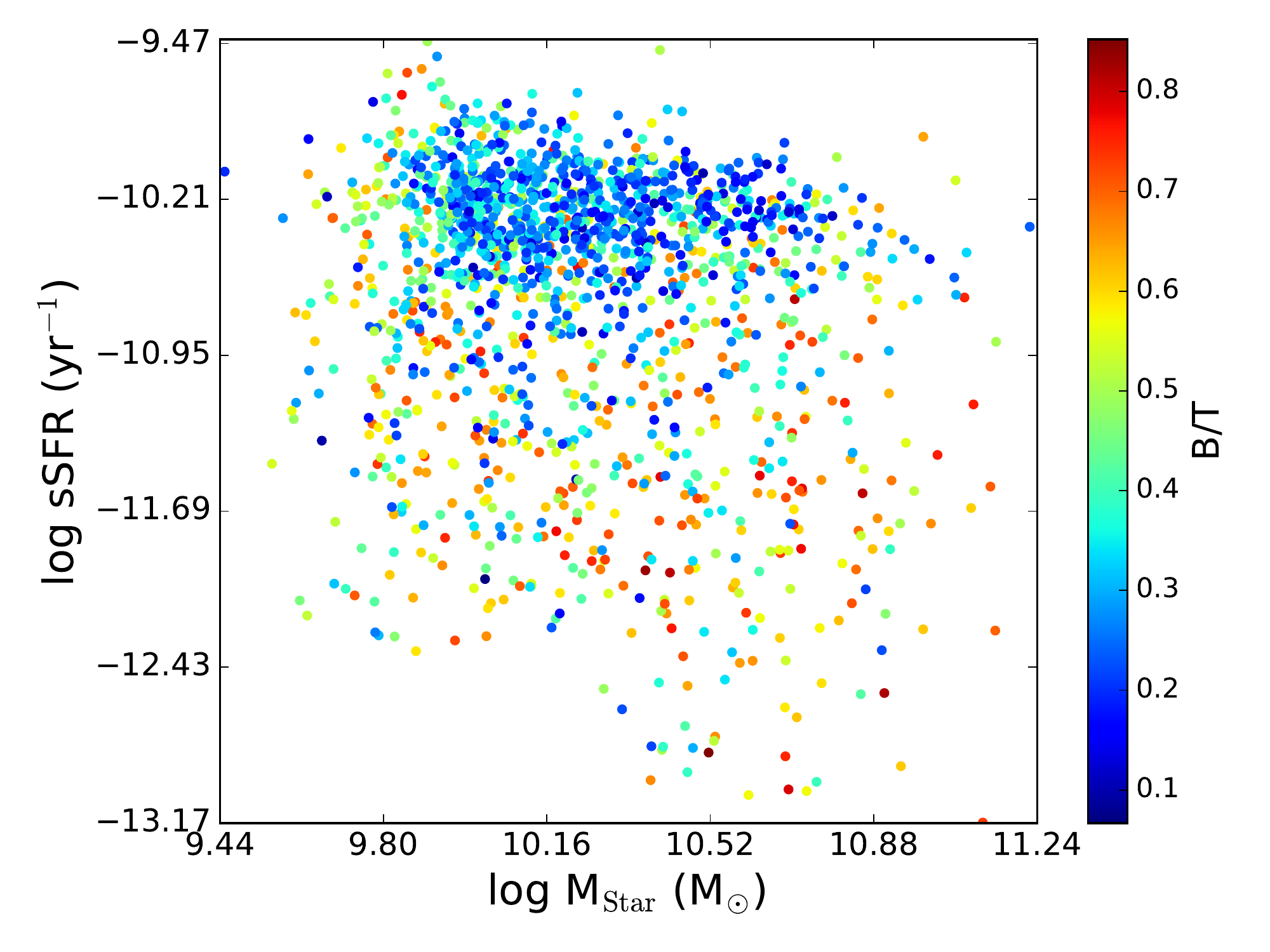}
  \caption{sSFR as a function of stellar mass of the total analysed
    sample. Symbols are coloured according to $B/T$ ratio.}
   \label{fig:a3}
\end{figure}

\end{appendix}

\end{document}